\documentclass[apjl]{emulateapj}

\usepackage{amsmath}
\usepackage{multirow}
\usepackage{relsize}

\newcommand{\alphavir}{\alpha_\mathrm{vir}}
\newcommand{\alphavircirc}{\alpha_\mathrm{vir,\circ}}
\newcommand{\sfe}{\mathrm{SFE}}
\newcommand{\sfrff}{\mathrm{SFR}_\mathrm{ff}}

\newcommand{\sigs}{\sigma_s}
\newcommand{\scrit}{s_\mathrm{crit}}
\newcommand{\rhocrit}{\rho_\mathrm{crit}}
\newcommand{\eps}{\epsilon}
\newcommand{\phit}{\phi_t}
\newcommand{\phix}{\phi_x}
\newcommand{\tff}{t_\mathrm{ff}}

\newcommand{\meanrho}{\rho_0}
\newcommand{\means}{s_0}
\newcommand{\deriv}{\,\mathrm{d}}
\newcommand{\mach}{\mathcal{M}}
\newcommand{\macha}{\mathcal{M}_\mathrm{A}}
\newcommand{\cs}{c_\mathrm{s}}
\newcommand{\va}{v_\mathrm{A}}
\newcommand{\ycut}{y_\mathrm{cut}}
\newcommand{\lambdas}{\lambda_\mathrm{s}}
\newcommand{\lambdams}{\lambda_\mathrm{ms}}
\newcommand{\lambdaj}{\lambda_\mathrm{J}}
\newcommand{\lambdajt}{\lambda_\mathrm{J,turb}}
\newcommand{\lambdajmt}{\lambda_\mathrm{J,mag}}
\newcommand{\mc}{M_\mathrm{c}}

\newcommand{\mbe}{M_\mathrm{BE}}
\newcommand{\mcrit}{M_\mathrm{crit}}
\newcommand{\mphi}{M_\Phi}
\newcommand{\sigsfr}{\Sigma_\mathrm{SFR}}
\newcommand{\sigsf}{\Sigma_\mathrm{SF}}
\newcommand{\tsf}{t_\mathrm{SF}}
\newcommand{\siggas}{\Sigma_\mathrm{gas}}
\newcommand{\sigtot}{\Sigma_\mathrm{tot}}
\newcommand{\chisqred}{\chi^2_\mathrm{red}}

\newcommand{\rhosink}{\rho_\mathrm{sink}}
\newcommand{\rsink}{r_\mathrm{sink}}

\newcommand{\msol}{\mbox{$M_{\sun}$}}
\newcommand{\g}{\mathrm{g}}
\newcommand{\cm}{\mathrm{cm}}
\newcommand{\km}{\mathrm{km}}
\newcommand{\pc}{\mathrm{pc}}
\newcommand{\AU}{\mbox{AU}}
\newcommand{\s}{\mathrm{s}}
\newcommand{\yr}{\mathrm{yr}}
\newcommand{\Gauss}{\mathrm{G}}

\newcommand{\bfu}{\mathbf{v}}
\newcommand{\bfB}{\mathbf{B}}
\newcommand{\vect}[1]{{\mathbf{#1}}}

\slugcomment{draft \today}

\shorttitle{The Star Formation Rate}
\shortauthors{Federrath \& Klessen}

\begin{document}

\title{The Star Formation Rate of Turbulent Magnetized Clouds: \\Comparing Theory, Simulations, and Observations}

\author{Christoph~Federrath\altaffilmark{1}, and Ralf~S.~Klessen\altaffilmark{2}}
\email{christoph.federrath@monash.edu}

\altaffiltext{1}{Monash Centre for Astrophysics, School of Mathematical Sciences, Monash University, Vic 3800, Australia}
\altaffiltext{2}{Universit\"at Heidelberg, Zentrum f\"ur Astronomie, Institut f\"ur Theoretische Astrophysik, Albert-Ueberle-Str.~2, 69120 Heidelberg, Germany}

\begin{abstract}
The role of turbulence and magnetic fields is studied for star formation in molecular clouds. We derive and compare six theoretical models for the star formation rate (SFR)---the Krumholz \& McKee (KM), Padoan \& Nordlund (PN), and Hennebelle \& Chabrier (HC) models, and three multi-freefall versions of these, suggested by HC---all based on integrals over the log-normal distribution of turbulent gas. We extend all theories to include magnetic fields, and show that the SFR depends on four basic parameters: (1) virial parameter $\alphavir$; (2) sonic Mach number $\mach$; (3) turbulent forcing parameter $b$, which is a measure for the fraction of energy driven in compressive modes; and (4) plasma $\beta=2\macha^2/\mach^2$ with the Alfv\'en Mach number $\macha$. We compare all six theories with MHD simulations, covering cloud masses of $300$ to $4\times10^6\,\msol$ and Mach numbers \mbox{$\mach=3$--$50$} and \mbox{$\macha=1$--$\infty$}, with solenoidal ($b=1/3$), mixed ($b=0.4$) and compressive turbulent ($b=1$) forcings. We find that the SFR increases by a factor of four between $\mach=5$ and $50$ for compressive turbulent forcing and $\alphavir\sim1$. Comparing forcing parameters, we see that the SFR is more than $10\times$ higher with compressive than solenoidal forcing for $\mach=10$ simulations. The SFR and fragmentation are both reduced by a factor of two in strongly magnetized, trans-Alfv\'enic turbulence compared to hydrodynamic turbulence. All simulations are fit simultaneously by the multi-freefall KM and multi-freefall PN theories within a factor of two over two orders of magnitude in SFR. The simulated SFRs cover the range and correlation of SFR column density with gas column density observed in Galactic clouds, and agree well for star formation efficiencies \mbox{$\sfe=1\%$--$10\%$} and local efficiencies $\eps=0.3$--$0.7$ due to feedback. We conclude that the SFR is primarily controlled by interstellar turbulence, with a secondary effect coming from magnetic fields.
\end{abstract}


\section{Introduction}

Stars form in turbulent, magnetized molecular clouds, as observed in the Milky Way and in other galaxies. Yet, the basic physical processes controlling star formation are still poorly understood. Observations of star-forming clouds show that the star formation rate (SFR) column density $\sigsfr$ varies over four orders of magnitude and exhibits a positive correlation with the gas surface density $\siggas$ \citep{HeidermanEtAl2010}, suggesting that denser gas forms stars at a higher rate. This engenders the central question of how the gas is locally compressed in the interstellar medium, such that dense cores can form and eventually become unstable under their own gravitational attraction to form stars. Gas compression in shocks, induced by large-scale supersonic turbulence might be a key---if not \emph{the} key process---setting the initial conditions for star formation \citep[see, e.g., the reviews by][]{MacLowKlessen2004,McKeeOstriker2007}.

Based on molecular cloud masses in the range \mbox{$M_c=100$ to $10^7\,\msol$} and temperatures $T\lesssim20\,\mathrm{K}$, the clouds should be highly Jeans-unstable and would thus collapse globally. However, molecular clouds do not show systematic, global collapse motions. If they did, the average Galactic SFR in the Milky Way, \mbox{$\textrm{SFR}_\textrm{MW}\approx1$--$2\,\msol\,\yr^{-1}$} \citep{RobitailleWhitney2010,ChomiukPovich2011} would be about two orders of magnitude higher than the observed value \citep{ZuckermanPalmer1974,ZuckermanEvans1974}. However, this stability analysis only takes thermal pressure into account. In reality, clouds are magnetized and subject to strong turbulent motions \citep{ScaloElmegreen2004,ElmegreenScalo2004}.

Originally, it has been thought that primarily magnetic fields would provide stability against fast global collapse, and that only after the neutral species have slowly diffused through the charged particles, star formation would occur in the central regions of magnetized clouds \citep{MestelSpitzer1956,Mouschovias1976,Shu1983}. In this so-called ambipolar-diffusion process, magnetic flux is left behind in the envelope, while the mass increases in the cloud core. Thus, star formation regulated by ambipolar diffusion predicts a higher mass-to-flux ratio in the cores than in the envelopes of the clouds, which is --- however --- typically not observed \citep{CrutcherHakobianTroland2009,MouschoviasTassis2009,LunttilaEtAl2009,SantosLimaEtAl2010,LazarianEsquivelCrutcher2012,BertramEtAl2012}.

An alternative scenario is that the observed supersonic random motions \citep{ZuckermanPalmer1974,ZuckermanEvans1974,Larson1981,SolomonEtAl1987,FalgaronePugetPerault1992,OssenkopfMacLow2002,HeyerBrunt2004,SchneiderEtAl2011,RomanDuvalEtAl2011} regulate star formation. In this picture, turbulent energy stabilizes the clouds on large scales, but at the same time, supersonic turbulence induces local compressions, producing filaments and cores, which are the progenitors of stars. Eventually, both turbulence and magnetic fields play their parts; the only question is: which one is the dominant controlling factor of star formation?

The aim of this paper is to advance our understanding of the relevant physical processes and their parameters controlling the conversion of dense gas into stars, and to explain the observed variations of the SFR column density. We develop and compare six predictive theories --- the original Krumholz \& McKee (KM), Padoan \& Nordlund (PN), and Hennebelle \& Chabrier (HC) theories, and multi-freefall versions of theses three ---, which are all based on integrals over the turbulent density probability distribution function (PDF), explained in detail in the next section. We extend the KM and HC theories, as well as all the multi-freefall theories to include magnetic fields. We evaluate the relative importance of turbulence, its forcing characteristics, and magnetic fields in controlling the SFR and show that the SFR depends on the following four basic parameters:
\begin{enumerate}
\item{virial parameter $\alphavir=2E_\mathrm{kin}/|E_\mathrm{grav}|$,}
\item{sonic Mach number $\mach=\sigma_V/\cs$,}
\item{turbulent forcing parameter $b$, with purely solenoidal (divergence-free) forcing parameterized by $b=1/3$, mixed forcing by $b=0.4$ and purely compressive (curl-free) forcing by $b=1$, and}
\item{the ratio of thermal to magnetic pressure $\beta=2\macha^2/\mach^2$ with the Alfv\'en Mach number $\macha$.}
\end{enumerate}
We test all six theories with numerical simulations of supersonic, magnetized turbulence including self-gravity and sink particles to capture dense, collapsing, star-forming gas. We find that the multi-freefall KM and PN models including magnetic fields provide the best fits to our numerical simulations with typical uncertainties of less than a factor of two. This is an encouraging agreement, given that the SFR varies by two orders of magnitude in the simulations, depending on the four basic cloud parameters listed above.

Comparing our numerical experiments with SFRs measured in Galactic star-forming regions, we find that for typical star formation efficiencies of \mbox{$\sfe=1\%$--$10\%$}, the best-fit local efficiencies due to radiative and mechanical feedback from jets, winds, expanding shells or outflows driven by young stellar objects are \mbox{$\eps=0.3$--$0.7$} with a best-fit value of $\eps\approx0.5$ for $\sfe=3\%$. This suggests that a fraction \mbox{$\eps\approx0.3$--$0.7$} of all the infalling gas onto a typical protostellar core is accreted by the protostar, while a fraction \mbox{$(1-\eps)\approx0.3$--$0.7$} is re-injected into the interstellar medium by jets, winds, and outflows. We find good agreement between the numerical simulations and Galactic observations, suggesting that the observed variations in $\sigsfr$ with $\siggas$ are a result of different combinations of the four basic parameters controlling the SFR: $\alphavir$, $\mach$, $b$, and $\beta$, as listed above. Since molecular clouds are often characterized by virial parameters of order unity, we conclude that the degree of compression induced by the turbulent forcing and sonic Mach number have the strongest influence on the SFR, causing variations by more than an order of magnitude, while magnetic fields can account for reductions of the SFR by a factor of two.

In Section~\ref{sec:theory}, we introduce and discuss the six analytic theories for the SFR, based on the turbulent density PDF, derive and discuss their dependencies, add magnetic fields to the theories that did not include magnetic-field effects in previous derivations, and compare them with each other in detail. We then test the analytic theories with numerical simulations of supersonic, magnetized turbulence by varying the sonic Mach number (\mbox{$\mach=3$--$50$}), the forcing of the turbulence (solenoidal, mixed, compressive), and the magnetic-field strength (yielding Alfv\'en Mach numbers \mbox{$\macha=1.3$--$\infty$}) to cover a comprehensive range of cloud parameters. The simulation methods and setups are explained in Section~\ref{sec:sims}. A detailed time-evolution analysis of column density, magnetic-field morphology, and fragmentation properties is presented in Section~\ref{sec:simresults}. In Section~\ref{sec:model_comparison}, we compare the SFRs measured in the magnetohydrodynamic (MHD) simulations with the six theoretical models, and determine the best-fit theory parameters that are universally applicable and fit all our simulations simultaneously. Section~\ref{sec:obs} presents a comparison of SFR column densities in the simulations with observations of Galactic clouds. We discuss limitations of the theoretical and numerical models, as well as limitations in the comparison with observations in Section~\ref{sec:limitations}. Finally, we list our conclusions and summarize the most important results in Section~\ref{sec:conclusions}. Here, we study the SFR in detail, while in Paper II \citep{FederrathKlessen2012paper2}, we concentrate on the star formation efficiency (SFE).

\section{The SFR from the Statistics of Supersonic Magnetized Turbulence} \label{sec:theory}

\subsection{The Density PDF} \label{sec:pdf}

The probability density function (PDF) of the gas density in a turbulent medium---such as a molecular cloud---is the key ingredient for analytic models of star formation. A log-normal density PDF has been used to explain the mass distribution of cores and stars, the core mass function (CMF) and the stellar initial mass function (IMF) \citep{PadoanNordlund2002,HennebelleChabrier2008,HennebelleChabrier2009,Elmegreen2011,VeltchevKlessenClark2011,DonkovVeltchevKlessen2012,ParravanoSanchezAlfaro2012,Hopkins2012b}, the Kennicutt-Schmidt relation \citep{KrumholzMcKee2005,Tassis2007}, the SFE \citep{Elmegreen2008}, and the SFR \citep{KrumholzMcKee2005,PadoanNordlund2011,HennebelleChabrier2011}. Here we concentrate on the SFR and derive its basic dependencies.

The log-normal PDF of the gas density is defined as,
\begin{equation}
p_s(s)=\frac{1}{\sqrt{2\pi\sigs^2}}\exp\left(-\frac{(s-s_0)^2}{2\sigs^2}\right)\,,
\label{eq:pdf}
\end{equation}
expressed in terms of the logarithmic density,
\begin{equation} \label{eq:s}
s\equiv\ln{(\rho/\meanrho)}\,.
\end{equation}
The PDF is a normal (Gaussian) distribution in $s$, meaning it is a log-normal distribution in $\rho$. The quantities $\meanrho$ and $\means$ denote the mean density and mean logarithmic density, the latter of which is related to the standard deviation $\sigs$ by
\begin{equation}
\means=-\frac{1}{2}\,\sigs^2
\end{equation}
due to the normalization and mass-conservation constraints of the PDF \citep{Vazquez1994,FederrathKlessenSchmidt2008}. The reason to use $s$ instead of $\rho$ in the context of the density PDF, is that $s$ is dimensionless, and that the PDF of $s$ is Gaussian unlike the PDF of $\rho$. This is because the distribution of $\rho$ is generated by a multiplicative process in which shocks are amplified by other shocks as they collide and interact in isothermal supersonic turbulence, with the local Mach number being independent of the local density \citep{Vazquez1994,PassotVazquez1998,KritsukEtAl2007,FederrathDuvalKlessenSchmidtMacLow2010}. Since $s\propto\ln(\rho)$ as defined in Equation~(\ref{eq:s}), this multiplicative process in $\rho$ turns into an additive process in $s$. Following the central limit theorem, a large sum of random variables produces a Gaussian distribution, and thus only $p_s$ is Gaussian, while $p_\rho$ is not. However, $p_s$ can be easily transformed into $p_\rho$ because $p_s \deriv s=p_\rho \deriv\rho$, and thus $p_\rho=p_s/\rho$ \citep{LiKlessenMacLow2003}. We will omit the index $s$ in $p_s$ in the following and simply use $p(s)$ for the PDF given by Equation~(\ref{eq:pdf}).

As soon as significant collapse sets in, the density PDF develops a power-law tail at high densities \citep[e.g.,][]{Klessen2000,KainulainenEtAl2009}, which is discussed in more detail in Section~\ref{sec:pdftails} below, and in Paper II \citep{FederrathKlessen2012paper2}.

\subsection{The Standard Deviation of Density Fluctuations in Supersonic, Magnetized Turbulence} \label{sec:sigs}
The standard deviation $\sigs$ in Equation~(\ref{eq:pdf}), which is a measure of how much the density varies in a turbulent medium, depends on (1) the amount of compression induced by the turbulent forcing mechanism, (2) the Mach number, and (3) the degree of magnetization. First, the turbulent energy injection mechanism determines the amount of compression induced directly by driving turbulence in the interstellar medium (ISM). Various turbulent driving mechanisms have been discussed and compared in \citet{MacLowKlessen2004}. For instance, expanding supernova shells \citep{BalsaraEtAl2004,AvillezBreitschwerdt2005,TamburroEtAl2009} or growing H\textsc{ii} regions around massive stars and clusters of stars \citep{McKee1989,KrumholzMatznerMcKee2006,GritschnederEtAl2009,PetersEtAl2010,GoldbaumEtAl2011} as well as compression of ISM gas in galactic spiral shocks \citep{Elmegreen2009} and gravitational contraction \citep{Hoyle1953,VazquezCantoLizano1998,KlessenHennebelle2010,ElmegreenBurkert2010,FederrathSurSchleicherBanerjeeKlessen2011} are likely exciting a considerable amount of compressible modes that will directly lead to compression, and thus to higher density contrasts on molecular cloud scales in the ISM, while galactic rotation and magnetorotational instabilities \citep[e.g.,][]{PiontekOstriker2004,PiontekOstriker2007} are likely producing more solenoidal modes. Second, higher Mach numbers $\mach$ lead to stronger shocks and thus to higher density contrasts. For instance, the density jump in a non-magnetized, isothermal shock is proportional to $\mach^2$. Finally, higher magnetization dampens density fluctuations as magnetic fields act like a cushion due to the additional magnetic pressure \citep{OstrikerStoneGammie2001,PriceFederrathBrunt2011}.

The actual dependence of turbulent density fluctuations $\sigs$ on the three parameters above (forcing, Mach number, and magnetic field) can be derived from the shock jump conditions of an individual MHD shock, and then averaged over a whole ensemble of such shocks \citep{PadoanNordlund2011}. \citet{MolinaEtAl2012} provide a rigorous derivation of $\sigs$ for different correlations of the magnetic field with density. They distinguish three cases, $B\propto\rho^0$, $B\propto\rho^{1/2}$, and $B\propto\rho^1$. For the intermediate case, \citet{MolinaEtAl2012} derive
\begin{equation}
\sigs^2=\ln\left(1+b^2\mach^2\frac{\beta}{\beta+1}\right)\,,
\label{eq:sigs}
\end{equation}
which is similar to the relation derived in \citet{PadoanNordlund2011}, except for the factor $b^2$, explained below, and except for the definition of $\beta$, for which \citet{PadoanNordlund2011} only take post-shock gas into account (see the more extended discussion on this issue in Section~\ref{sec:PNmodel}). The case $B\propto\rho^1$ is similar to the intermediate case, but is a rather extreme MHD case because magnetic-field lines are assumed to be oriented only perpendicular to the flow direction. So is the other extreme case in which the magnetic field is assumed to be parallel to the flow, yielding $B\propto\rho^0$. In the more realistic case of turbulent flows, field lines become tangled, and the $B$--$\rho$ correlation is a combination of compression of field lines and turbulent dynamo amplification \citep{SchleicherEtAl2010,SurEtAl2010,FederrathSurSchleicherBanerjeeKlessen2011,TurkEtAl2012,SchoberEtAl2012}. In a three-dimensional system with a random distribution of flow velocities and magnetic-field orientations, $B\propto\rho^{1/2}$ provides a reasonable intermediate dependence. We will thus only consider $B\propto\rho^{1/2}$ here, which is favored by simulations \citep{PadoanNordlund1999,CollinsEtAl2011,MolinaEtAl2012}, and also close to what is suggested from observations of magnetic fields in molecular clouds \citep{CrutcherEtAl2010}\footnote{The observationally determined exponent of the \mbox{$B$--$\rho$} correlation is quite uncertain. \citet{Crutcher1999} find $B\propto\rho^{0.47}$, while \citet{CrutcherEtAl2010} find $B\propto\rho^0$ below gas densities of $300\,\cm^{-3}$, and $B\propto\rho^{0.65}$ above. For simplicity, we adopt Equation~(\ref{eq:sigs}), derived for the intermediate case, $B\propto\rho^{1/2}$.}.

In the case of $B\propto\rho^0$, i.e., for no density correlation of the magnetic field, Equation~(\ref{eq:sigs}) reduces to the well-known and frequently used hydrodynamic (HD) expression, $\sigs^2=\ln\left(1+b^2\mach^2\right)$ with $\beta\to\infty$ \citep[e.g.,][]{PadoanNordlundJones1997,PassotVazquez1998,OstrikerStoneGammie2001,LemasterStone2008,FederrathKlessenSchmidt2008,PriceFederrathBrunt2011} as a necessary condition in the purely HD limit. The parameters $b$, $\mach$, and $\beta$ in Equation~(\ref{eq:sigs}) are the turbulent forcing parameter, the rms sonic Mach number, and the ratio of thermal to magnetic pressure, plasma $\beta=P_\mathrm{th}/P_\mathrm{mag}$. Using the definitions of the thermal pressure for an isothermal equation of state $P_\mathrm{th}=\rho\cs^2$, magnetic pressure $P_\mathrm{mag}=B^2/(8\pi)$, Alfv\'en velocity $\va^2=B^2/(4\pi\rho)$, sonic and Alfv\'en Mach numbers, $\mach=\sigma_V/\cs$ and $\macha=\sigma_V/v_\mathrm{A}$, the plasma beta can be expressed as $\beta=2\cs^2/\va^2=2\macha^2/\mach^2$. These are all dimensionless numbers, rendering them particularly useful because they determine the basic properties of turbulent plasmas and can thus be compared directly for any such system. Equation~(\ref{eq:sigs}) can thus also be written as
\begin{equation}
\sigs^2=\ln\left(1+b^2\mach^2\frac{2\macha^2}{\mach^2+2\macha^2}\right)\,.
\label{eq:sigsmacha}
\end{equation}
The forcing parameter $b$ was shown to vary smoothly between $b\approx1/3$ for purely solenoidal (divergence-free) forcing, and $b\approx1$ for purely compressive (curly-free) forcing of the turbulence \citep{FederrathKlessenSchmidt2008,SchmidtEtAl2009,FederrathDuvalKlessenSchmidtMacLow2010,SeifriedSchmidtNiemeyer2011,MicicEtAl2012,KonstandinEtAl2012}. A stochastic mixture of forcing modes in three-dimensional space leads to $b\approx0.4$ \citep[see Figure~8 in][]{FederrathDuvalKlessenSchmidtMacLow2010}.

Using numerical simulations, \citet{MolinaEtAl2012} found that Equations~(\ref{eq:sigs}) and~(\ref{eq:sigsmacha}) work well in the regime $\macha\gtrsim2$, while for $\macha\lesssim2$, the assumption of isotropy entering the analytic derivation of Equations~(\ref{eq:sigs}) and~(\ref{eq:sigsmacha}) breaks down, so we only apply them in the super-Alfv\'enic regime in all the following.

\subsection{Basics of the SFR Derivation} \label{sec:derivation}
Here we present an analytic derivation of the SFR from the statistics of supersonic, isothermal, magnetized turbulence. The main ingredient for this analytic derivation is an integral over the density PDF, Equation~(\ref{eq:pdf}), in order to estimate the gas mass above a given density threshold, contributing to star formation. We will compare different ways of estimating the density threshold, which is the main difference between the three most successful, existing analytic models for the SFR \citep{KrumholzMcKee2005,PadoanNordlund2011,HennebelleChabrier2011}. We will express all quantities in terms of dimensionless numbers, in order to simplify the derivation and to make it more general. We follow the standard terminology and use the \emph{Star Formation Rate per Freefall Time} ($\sfrff$), as coined by \citet{KrumholzMcKee2005}, which is the mass fraction going into stars per time, where the time is expressed in units of the mean freefall time.

The SFR in units of $\msol\,\yr^{-1}$ can be computed by scaling $\sfrff$ with the real cloud mass $\mc$ and the actual freefall time evaluated at the mean density of the cloud, $\tff(\meanrho)$:
\begin{equation} \label{eq:sfrffdef}
\mathrm{SFR} \equiv \frac{\mc}{\tff(\meanrho)}\,\sfrff\,.
\end{equation}
Note that this definition of $\sfrff$ is different from the definition used in \citet{KrumholzTan2007} and \citet{KrumholzDekelMcKee2012}, who use freefall times estimated at different densities and/or use a definition based on column densities, such that the values of $\sfrff$ quoted in those studies and the ones computed here cannot be directly compared. For instance, given an SFR for fixed $M_c$, the dimensionless value of $\sfrff$ would be much smaller, if the freefall time at a high-density tracer was used rather than the freefall time at the mean density of the cloud because $\tff(\rho>\meanrho)$ is shorter than $\tff(\meanrho)$.

The basic idea for an analytic model of $\sfrff$ is to integrate the log-normal density PDF, Equation~(\ref{eq:pdf}), weighted by $\rho/\meanrho$ to get the mass fraction of gas with density above a critical density $\scrit$ (to be determined below in Section~\ref{sec:sixtheories}), and weighted by a freefall-time factor to construct a dimensionless mass rate:
\begin{equation}
\sfrff = \frac{\eps}{\phit} \mathlarger{\int}_{\scrit}^{\infty}{\frac{\tff(\meanrho)}{\tff(\rho)} \frac{\rho}{\meanrho} \, p(s) \deriv s}\,.
\label{eq:sfrffbasic}
\end{equation}
Note that the factor $\tff(\meanrho)/\tff(\rho)$ appears inside the integral because gas with different densities has different freefall times,
\begin{equation} \label{eq:tff}
\tff(\rho)\equiv\left(\frac{3\pi}{32G\rho}\right)^{1/2}\,,
\end{equation}
which should be taken into account in the most general case \citep[see][]{HennebelleChabrier2011}. Previous estimates for $\sfrff$ either used a factor $\tff(\meanrho)/\tff(\meanrho)=1$ \citep{KrumholzMcKee2005}, or a factor $\tff(\meanrho)/\tff(\rhocrit)$ with $\rhocrit=\meanrho\exp{(\scrit)}$ \citep{PadoanNordlund2011}, both of which are independent of density and were thus pulled out of the integral. We will show, however, that it is crucial to take the multi-freefall nature of gas with different densities into account to obtain better models for $\sfrff$.

The constant factor $\eps$ in Equation~(\ref{eq:sfrffbasic}) accounts for the fact that only a certain fraction of the gas above $\scrit$ might actually go into stars. Since individual stars form in accretion disks from which powerful jets, winds, and outflows are launched during the process of stellar birth, it is likely that a certain fraction of the accreted material is re-injected into the ISM, thus leading to $\eps<1$. Theoretical upper limits are in the range \mbox{$\eps\approx0.25$--$0.7$} \citep[e.g.,][]{MatznerMcKee2000}. The observed displacement of the characteristic mass in the IMF \citep[e.g.,][]{Kroupa2001,Chabrier2003} with respect to the CMF \citep[e.g.,][]{JohnstoneEtAl2000} has been taken to argue that $\eps$ might be around 0.3--0.5 \citep{AlvesLombardiLada2007,AndreEtAl2010}; see however \citet{WardEtAl2012}.

The factor $1/\phit$ in Equation~(\ref{eq:sfrffbasic}) is also of order unity and accounts for the uncertainty in the timescale factor $\tff(\meanrho)/\tff(\rho)$, originally introduced in \citet{KrumholzMcKee2005}. We will determine the best-fit values of $\eps$ and $1/\phit$ in Sections~\ref{sec:simresults} and~\ref{sec:obs}, when we compare the theories with simulations and observations.

\subsection{Six Models for the SFR} \label{sec:sixtheories}
In the following, we will solve Equation~(\ref{eq:sfrffbasic}), using different density thresholds $\scrit$, according to the previous analytic studies of the SFR by \citet[][KM]{KrumholzMcKee2005}, \citet[][PN]{PadoanNordlund2011}, and \citet[][HC]{HennebelleChabrier2011}\footnote{Note that the critical densities derived in the following may or may not be related to density or column density thresholds for star formation introduced in observational studies \citep[e.g.,][]{HeidermanEtAl2010,LadaLombardiAlves2010}. Studying such potential relations, however, deserves further consideration in the near future.}. We distinguish six cases, named `KM', `PN', `HC', and `multi-ff KM', `multi-ff PN', `multi-ff HC' as distinguished in \citet{HennebelleChabrier2011}. The first three represent the original analytic derivations by \citet{KrumholzMcKee2005}, \citet{PadoanNordlund2011}, and \citet{HennebelleChabrier2011}, while the last set of three are all based on the multi-freefall expression of the integral~(\ref{eq:sfrffbasic}). The difference for this last set of three is only the model for the critical density, i.e., the lower limit of the integral. We note that the ideas inherent in each of the original theories contributes to our present understanding of the turbulence-regulated SFR. \citet{KrumholzMcKee2005} developed the basic model, \citet{PadoanNordlund2011} extended it to include magnetic fields, and \citet{HennebelleChabrier2011} improved all models by introducing multi-freefall versions of the aforementioned theories, yet without considering magnetic fields. We build on all these approaches and extend the non-magnetic multi-freefall models to include magnetic fields. We then determine the best combination of the aforementioned theoretical ideas to come up with a more universal theoretical model for the SFR. Table~\ref{tab:theories} summarizes all six theoretical models, which are discussed and derived in detail in the following.

\begin{table*}
\caption{Six Analytic Models for the Star Formation Rate per Freefall Time.}
\label{tab:theories}
\def\arraystretch{1.5}
\setlength{\tabcolsep}{0pt}
\begin{tabular}{ p{0.11\linewidth} p{0.13\linewidth} p{0.08\linewidth} p{0.015\linewidth} p{0.24\linewidth} p{0.42\linewidth}}
\hline
\hline
Analytic Model & Freefall-time \mbox{Factor}  & \multicolumn{3}{p{0.34\linewidth}}{\mbox{Critical Density} \mbox{$\rhocrit/\meanrho=\exp(\scrit)$}} & $\sfrff$\\
\hline
KM                      & 1                                                & $(\pi^2/5)\,\phix^2$     &$\times$&  $\alphavir\mach^2\left(1+\beta^{-1}\right)^{-1}$ & $\eps/(2\phit) \left\{1+\mathrm{erf}\left[(\sigs^2-2\scrit)/(8\sigs^2)^{1/2}\right]\right\}$ \\
PN                      & $\tff(\meanrho)/\tff(\rhocrit)$  & $(0.067)\,\theta^{-2}$ &$\times$&  $\alphavir\mach^2 f(\beta)$ & $\eps/(2\phit) \left\{1+\mathrm{erf}\left[(\sigs^2-2\scrit)/(8\sigs^2)^{1/2}\right]\right\} \exp\left[(1/2)\scrit\right]$ \\
HC                      & $\tff(\meanrho)/\tff(\rho)$       & $(\pi^2/5)\,\ycut^{-2}$ &$\times$&  $\alphavir\mach^{-2}\left(1+\beta^{-1}\right) + \tilde{\rho}_\mathrm{crit,turb}$ & $\eps/(2\phit) \left\{1+\mathrm{erf}\left[(\sigs^2-\scrit)/(2\sigs^2)^{1/2}\right]\right\} \exp\left[(3/8)\sigs^2\right]$ \\
multi-ff KM         & $\tff(\meanrho)/\tff(\rho)$       & $(\pi^2/5)\,\phix^2$     &$\times$&  $\alphavir\mach^2\left(1+\beta^{-1}\right)^{-1}$ & $\eps/(2\phit) \left\{1+\mathrm{erf}\left[(\sigs^2-\scrit)/(2\sigs^2)^{1/2}\right]\right\} \exp\left[(3/8)\sigs^2\right]$ \\
multi-ff PN         & $\tff(\meanrho)/\tff(\rho)$       & $(0.067)\,\theta^{-2}$  &$\times$&  $\alphavir\mach^2 f(\beta)$ & $\eps/(2\phit) \left\{1+\mathrm{erf}\left[(\sigs^2-\scrit)/(2\sigs^2)^{1/2}\right]\right\} \exp\left[(3/8)\sigs^2\right]$ \\
multi-ff HC         & $\tff(\meanrho)/\tff(\rho)$       & $(\pi^2/5)\,\ycut^{-2}$ &$\times$&  $\alphavir\mach^{-2}\left(1+\beta^{-1}\right)$ & $\eps/(2\phit) \left\{1+\mathrm{erf}\left[(\sigs^2-\scrit)/(2\sigs^2)^{1/2}\right]\right\} \exp\left[(3/8)\sigs^2\right]$ \\
\hline
\end{tabular}
\\
\textbf{Notes.} The function $f(\beta)$, entering the critical density in the PN and multi-ff PN models is given by Equation~(\ref{eq:fofbeta}). The added turbulent contribution $\tilde{\rho}_\mathrm{crit,turb}$ in the critical density of the HC model is given by Equation~(\ref{eq:rhocrit_turb}).
\end{table*}

\subsubsection{The KM Model}

In the KM model by \citet{KrumholzMcKee2005}, the freefall-time factor $\tff(\meanrho)/\tff(\rho)$ in Equation~(\ref{eq:sfrffbasic}) is simply set to unity. Moreover, \citet{KrumholzMcKee2005} define the critical density $\scrit$ in the lower limit of the $\sfrff$ integral by comparing the \citet{Jeans1902} length
\begin{equation} \label{eq:lambdaj}
\lambdaj(\rho)=\left(\frac{\pi\cs^2}{G\rho}\right)^{1/2}\,,
\end{equation}
evaluated at the mean density with the sonic scale $\lambdas$ (defined in Equation~\ref{eq:ls} below),
\begin{equation} \label{eq:scrit_ls}
\scrit=2\,\ln\left(\phix \frac{\lambdaj(\meanrho)}{\lambdas}\right)\,.
\end{equation}
This choice is motivated by the expectation that the collapse sets in roughly at the sonic scale, where the turbulent fluctuations are of the order of the thermal sound speed, i.e., the local Mach number has dropped to about unity at the sonic scale \citep{VazquezBallesterosKlessen2003,FederrathDuvalKlessenSchmidtMacLow2010}. The global turbulent supersonic support is expected to become insignificant at the sonic scale, such that collapse can proceed below that scale \citep[e.g.,][]{MacLowKlessen2004}. The leading factor 2 in Equation~(\ref{eq:scrit_ls}) stems from the density dependence of the Jeans length, $\lambdaj(\rho)\propto\rho^{-1/2}$, and the numerical factor $\phix$ allows for slight variations in the actual scale on which the collapse sets in. \citet{KrumholzMcKee2005} find $\phix=1.12$ for the simulations by \citet{VazquezBallesterosKlessen2003}. In real molecular clouds, the sonic scale is expected to be of order $0.1\,$pc within factors of a few \citep[e.g.,][]{FalgaronePugetPerault1992,GoodmanEtAl1998,StahlerPalla2004,SchneeEtAl2007,McKeeOstriker2007}.

To make Equation~(\ref{eq:scrit_ls}) more useful, we express all dependent variables for $\scrit$ in terms of dimensionless numbers. This can be achieved by rewriting the Jeans length as
\begin{equation} \label{eq:lj}
\lambdaj(\meanrho)=\left(\frac{\pi\cs^2}{G\rho_0}\right)^{1/2}=\pi\cs\left(\frac{L^3}{6G\mc}\right)^{1/2}\,,
\end{equation}
where we have assumed a spherical cloud with diameter $L$, mass $\mc$, and isothermal sound speed $\cs$. Since the velocity fluctuations in a turbulent medium depend on the length scale $\ell$ as
\begin{equation} \label{eq:sigmav}
\sigma_v(\ell)=\sigma_V\,(\ell/L)^p \,,
\end{equation}
where $\sigma_V\approx 1\,\km\,\s^{-1}$ is the three-dimensional, non-thermal velocity dispersion on the scale $L\approx1\,\pc$, and $p\approx0.5$ from observations in Galactic clouds \citep{Larson1981,SolomonEtAl1987,OssenkopfMacLow2002,HeyerBrunt2004,HeyerEtAl2009,RomanDuvalEtAl2011}, the Galactic Central Molecular Zone \citep{JonesEtAl2012,ShettyEtAl2012}, and from numerical simulations \citep{KritsukEtAl2007,SchmidtEtAl2009,FederrathDuvalKlessenSchmidtMacLow2010}, the sonic scale can be written as
\begin{equation} \label{eq:ls}
\lambda_\mathrm{s}=L\left(\cs/\sigma_V\right)^{1/p}\,.
\end{equation}
Substituting Equations~(\ref{eq:lj}) and~(\ref{eq:ls}) into Equation~(\ref{eq:scrit_ls}), we find
\begin{eqnarray} \label{eq:scritvir} \label{eq:scrit_km_nob}
\scrit{_\mathrm{,KM}} \,& = &\, \ln\left[\frac{\phix^2\pi^2}{5}\,\frac{5\sigma_V^2 L}{6G\mc}\,\left(\frac{\sigma_V}{\cs}\right)^{2(1-p)/p}\right] \nonumber \\
                                       \,& = &\, \ln\left[(\pi^2/5)\phix^2\,\alphavir\,\mach^{2}\right]\,,
\end{eqnarray}
where we have identified the virial parameter for a spherical, uniform-density cloud with velocity dispersion $\sigma_V$ on the diameter scale $L$,
\begin{equation} \label{eq:alphavircirc}
\alphavircirc=5\sigma_V^2 L/(6G\mc)\,,
\end{equation}
and the rms Mach number, $\mach=\sigma_V/\cs$, and used $p=0.5$ in the second step. This derivation is essentially identical to the one presented in \citet{KrumholzMcKee2005}, with the exception that we use the more general expression for the virial parameter here,
\begin{equation} \label{eq:alphavir}
\alphavir = 2E_\mathrm{kin}/\left|E_\mathrm{grav}\right|\,,
\end{equation}
the ratio of twice the kinetic energy to the gravitational energy. This general form reduces to $\alphavircirc$ given by Equation~(\ref{eq:alphavircirc}) with $E_\mathrm{kin}=\mc\sigma_V^2/2$ and $E_\mathrm{grav}=-3G\mc^2/(5R)$ for a spherical, homogeneous cloud with radius $R=L/2$. We emphasize that the definition of $\alphavircirc$ is based on global parameters, assuming a spherical cloud with uniform density. This is far from realistic, given that clouds are in fact highly inhomogeneous and non-spherical. In the analytic derivations, however, this simplification given by Equation~(\ref{eq:alphavircirc}) is necessary to enable a mathematical analysis of the problem. In the simulations discussed in Section~\ref{sec:sims} below, however, we will directly compute the virial parameter from the gravitational potential of the actual, three-dimensional, inhomogeneous spatial gas distribution, providing a more general and accurate measure of the virial parameter given by the general form, Equation~(\ref{eq:alphavir}). This is discussed further below when we compare the theories to numerical simulations and in Section~\ref{sec:alphavir}.

The original model by \citet{KrumholzMcKee2005} neglects magnetic fields. Here, magnetic-field effects are partially added automatically by using Equation~(\ref{eq:sigs}) for $\sigs$, such that $\sigs$ decreases with increasing magnetic energy, as derived in \citet{MolinaEtAl2012}. This however only changes $\sigs$, while a modification of $\scrit$ is also necessary to fully account for magnetic-pressure effects on $\sfrff$.

Here we provide and apply a simple rule to include magnetic-field effects in the expression for the critical density $\scrit$. The key idea is to replace the thermal pressure by the sum of the thermal and magnetic pressures:
\begin{eqnarray}
P_\mathrm{th} & \rightarrow & P_\mathrm{th} + P_\mathrm{mag} \nonumber \\
\iff \rho\cs^2 & \to & \rho\cs^2 + (1/2)\rho\va^2\,, \label{eq:magextension}
\end{eqnarray}
where the second line implies isothermal gas. Using $\va^2=2\cs^2\beta^{-1}$ with the definition of plasma $\beta=P_\mathrm{th}/P_\mathrm{mag}$ in Section~\ref{sec:sigs}, we can thus simply replace the sound speed by an effective sound speed,
\begin{equation} \label{eq:csmag}
\cs \to \cs\left(1+\beta^{-1}\right)^{1/2}\,.
\end{equation}
Since $\mach=\sigma_V/\cs$, we can also replace the sonic Mach number by an effective Mach number to take magnetic pressure into account:
\begin{equation} \label{eq:machmag}
\mach \to \mach\left(1+\beta^{-1}\right)^{-1/2}\,.
\end{equation}
Doing this for $\scrit{_\mathrm{,KM}}$ in Equation~(\ref{eq:scrit_km_nob}) yields the magnetic version of the critical density,
\begin{equation} \label{eq:scrit_km}
\scrit{_\mathrm{,KM}} = \ln\left[(\pi^2/5)\phix^2\,\alphavir\,\mach^{2}\left(1+\beta^{-1}\right)^{-1}\right]\,.
\end{equation}
Even though we simply replaced the thermal sound speed by an effective, magnetic sound speed to derive this expression, it has a deeper physical meaning. What we physically do in the derivation of $\scrit$ is to replace the thermal Jeans length in the numerator of Equation~(\ref{eq:scrit_ls}) with the magnetothermal Jeans length,
\begin{equation} \label{eq:ljmt}
\lambdajmt = \left(\frac{\pi\cs^2\left(1+\beta^{-1}\right)}{G\rho}\right)^{1/2}\,,
\end{equation}
and the sonic scale in the denominator with the magnetosonic scale,
\begin{equation} \label{eq:lsms}
\lambdams = L\left[\cs\left(1+\beta^{-1}\right)^{1/2}/\sigma_V\right]^{1/p}\,.
\end{equation}
We note that the magnetic modifications given by Equations~(\ref{eq:magextension}) only account for magnetic pressure, i.e., isotropic pressure induced by the small-scale magnetic field. It does not account for mean magnetic-field effects, and as such will only be a valid extension to MHD as long as the turbulence remains trans- to super-Alfv\'enic because sub-Alfv\'enic turbulence with a strong mean magnetic field component is anisotropic, which is discussed at more detail below.

Finally, solving the general $\sfrff$-integral (Equation~\ref{eq:sfrffbasic}) with $\scrit=\scrit{_\mathrm{,KM}}$ from Equation~(\ref{eq:scrit_km}) and unity for the freefall-time factor (see Table~\ref{tab:theories} for a summary), the SFR per freefall time in the KM model is
\begin{eqnarray} \label{sfrff_km}
\sfrff{_\mathrm{,KM}} \,& = &\, \frac{\eps}{\phit} \mathlarger{\int}_{\scrit{_\mathrm{,KM}}}^{\infty} {\exp(s)\,p(s) \deriv s} \nonumber \\
                                      \,& = &\, \frac{\eps}{2\phit} \left[1+\mathrm{erf}\left(\frac{\sigs^2-2\scrit{_\mathrm{,KM}}}{\sqrt{8\sigs^2}}\right)\right]\,.
\end{eqnarray}
This derivation is identical to the one in \citet{KrumholzMcKee2005}, except for the extension to include magnetic fields in the theory based on the plasma $\beta$ terms in $\sigs$, Equation~(\ref{eq:sigs}), and in the critical density, Equation~(\ref{eq:scrit_km}).

\subsubsection{The PN Model} \label{sec:PNmodel}
\citet{PadoanNordlund2011} use $\tff(\meanrho)/\tff(\rhocrit)$ as the freefall-time factor $\tff(\meanrho)/\tff(\rho)$ in Equation~(\ref{eq:sfrffbasic}), such that the freefall time of the critical density is used for all densities above the critical density to estimate $\sfrff$. Unlike \citet{KrumholzMcKee2005} who relate the critical density $\scrit$ to the Jeans length and the sonic scale, \citet{PadoanNordlund2011} related the critical density to the magnetic shock jump conditions and to the magnetic critical mass for collapse. Starting with their assumed balance of thermal plus magnetic pressure by turbulent ram pressure on the cloud scale,
\begin{equation}
\rho_\mathrm{MHD}\left(\cs^2+\frac{1}{2}\va^2\right) = \meanrho\left(\frac{\sigma_V}{2}\right)^2\,,
\end{equation}
and using the definitions for $\mach$ and $\beta$ from Section~\ref{sec:sigs}, \citet{PadoanNordlund2011} arrive at an expression for the density jump
\begin{equation}
\rho_\mathrm{MHD} = \meanrho \frac{\mach^2}{4}\frac{\beta}{\beta+1}\,.
\end{equation}
This leads to the post-shock thickness
\begin{equation} \label{eq:shockthickness}
\lambda_\mathrm{MHD} = \theta L \, \frac{4}{\mach^2}\frac{\beta+1}{\beta}\,,
\end{equation}
since $\rho_\mathrm{MHD}/\meanrho=\theta L/\lambda_\mathrm{MHD}$ with the numerical parameter $\theta\lesssim1$, the fraction of the cloud size forming the largest shocks. Thus, $\theta L$ can be interpreted as the turbulent injection or forcing scale. In numerical simulations, most of the kinetic, turbulent energy is usually injected at a wavenumber $k=2$ in units of $2\pi/L$, corresponding to half of the total cloud size \citep[e.g.,][]{KritsukEtAl2007,SchmidtEtAl2009,FederrathDuvalKlessenSchmidtMacLow2010}, as in the simulations discussed below in Section~\ref{sec:sims}. Thus, $\theta\approx1/2$, but there might be some corrections to that particular scale \citep{WangGeorge2002}. \citet{PadoanNordlund2011} take $\theta\approx0.35$. Here, we will simply interpret $\theta$ as a numerical factor of order unity, accounting for any uncertainty in the post-shock thickness with respect to the total cloud scale $L$ in Equation~(\ref{eq:shockthickness}).

In order to derive a critical density for star formation, \citet{PadoanNordlund2011} compare the mass of a sphere with radius $\lambda_\mathrm{MHD}/2$ to the critical mass for collapse. \citet{McKee1989} define the critical mass for collapse of a magnetized gas sphere as
\begin{equation} \label{eq:mcrit}
\mcrit \approx \mbe + \mphi\,,
\end{equation}
where
\begin{equation} \label{eq:mbe}
\mbe = 1.182 \cs^3 G^{-3/2} \rho^{-1/2}
\end{equation}
is the Bonnor-Ebert mass \citep{Ebert1955,Bonnor1956} and
\begin{equation} \label{eq:mphi}
\mphi = m_\Phi\frac{\pi R^2\,B}{G^{1/2}} = m_\Phi^3\frac{9\pi^{5/2}}{2G^{3/2}}\rho^{-1/2}\va^3
\end{equation}
is the magnetic critical mass for a sphere with radius $R$, threaded by a magnetic field $B$, where we have used the Alfv\'en velocity $\va=B/(4\pi\rho)^{1/2}$ in the second step. The numerical factor $m_\Phi$ in Equation~(\ref{eq:mphi}) can vary depending on the geometry and model taken, e.g., \citet{PadoanNordlund2011} take $m_\Phi=0.17$ with a reference to \citet{TomisakaIkeuchiNakamura1988}, while \citet{McKee1989} use $m_\Phi=0.12$, and \citet{Strittmatter1966} derive $m_\Phi=(12\pi^2/5)^{-1/2}\approx0.21$ for a non-rotating cloud and $m_\Phi=(9\pi^4/10)^{-1/2}\approx0.11$ for an oblate spheroidal cloud with eccentricity approaching unity \citep[see][]{NakanoNakamura1978}.

Finally, inserting Equations~(\ref{eq:mbe}) and~(\ref{eq:mphi}) into Equation~(\ref{eq:mcrit}) and setting the critical mass $\mcrit(\rhocrit)=(4\pi/3)(\lambda_\mathrm{MHD}/2)^3\rhocrit$ with the post-shock thickness given by Equation~(\ref{eq:shockthickness}), yields the critical density,
\begin{equation} \label{eq:scrit_pn}
\scrit{_\mathrm{,PN}} = \ln{\left[0.067\theta^{-2} \alphavir \mach^2 f(\beta) \right]}
\end{equation}
with
\begin{equation} \label{eq:fofbeta}
f(\beta) \equiv \frac{\left(1+0.925 \beta^{-3/2}\right)^{2/3}}{\left(1+\beta^{-1}\right)^2}\,.
\end{equation}
Note that $\scrit{_\mathrm{,PN}}$ has the same dependence on $\alphavir$ and $\mach$ as $\scrit{_\mathrm{,KM}}$ in Equation~(\ref{eq:scrit_km}).

\citet{PadoanNordlund2011} use a rather special definition of $\beta$, which is the average post-shock $\beta$. From a semi-analytical comparison of the mean magnetic field with the rms magnetic field, they derive a criterion for $\beta$ based on the average Alfv\'en Mach number, which \citet{PadoanNordlund2011} simply use as a switch between MHD and purely HD turbulence. However, it is not straightforward to derive a post-shock value of $\beta$ because it involves a density-threshold dependence \citep[see discussion in][]{PadoanNordlund2011}. Moreover, the switch discussed by \citet{PadoanNordlund2011} is a semi-analytical criterion derived from their simulations. We therefore decide to ignore this special definition of $\beta$ for simplicity and apply Equation~(\ref{eq:scrit_pn}) with our definition of $\beta$ (see Section~\ref{sec:sigs}), which includes all, and not just the post-shock gas. This is consistent with the definition of all other dynamical quantities of interest, e.g., $\alphavir$, $\mach$, $\macha$, $\meanrho$, etc.

Using $\tff(\meanrho)/\tff(\rhocrit{_\mathrm{,PN}})$ and inserting $\scrit{_\mathrm{,PN}}$ into the general Equation~(\ref{eq:sfrffbasic}) for $\sfrff$ yields
\begin{eqnarray} \label{eq:sfrff_pn}
\sfrff{_\mathrm{,PN}} = \frac{\eps}{\phit} \exp\left(\frac{1}{2}\scrit{_\mathrm{,PN}}\right) \mathlarger{\int}_{\scrit{_\mathrm{,PN}}}^{\infty} {\exp(s)\,p(s) \deriv s} \nonumber \\
                                     = \frac{\eps}{2\phit} \exp\left(\frac{1}{2}\scrit{_\mathrm{,PN}}\right) \left[1+\mathrm{erf}\left(\frac{\sigs^2-2\scrit{_\mathrm{,PN}}}{\sqrt{8\sigs^2}}\right)\right] \nonumber \\
\end{eqnarray}
for the PN model.

\subsubsection{The HC Model}
\citet{HennebelleChabrier2011} were the first to argue that the freefall-time factor $\tff(\meanrho)/\tff(\rho)$ must be used in Equation~(\ref{eq:sfrffbasic}), such that different densities contribute to $\sfrff$ with their individual freefall time (see Equation~\ref{eq:tff}). The full HC model for $\sfrff$ is based on the mass spectrum of gravitationally bound structures, as derived in \citet{HennebelleChabrier2008,HennebelleChabrier2009}:
\begin{equation}
\mathcal{N}(M) = \frac{\deriv(N/V)}{\deriv M} \propto \,-\frac{1}{M}\frac{\deriv s}{\deriv M}\,\exp(s)\,p(s)\,,
\end{equation}
which is essentially Equation~(6) in \citet{HennebelleChabrier2011}, except for the freefall time factor. The SFR in the HC model is then given by the integral over the mass spectrum, weighted by the mass and the freefall time factor:
\begin{eqnarray} \label{eq:sfrff_hc}
\sfrff \,& = &\, -\frac{\eps}{\phit} \mathlarger{\int}_{0}^{M_\mathrm{cut}} \frac{M\deriv M}{M}\frac{\deriv s}{\deriv M} \frac{\tff(\meanrho)}{\tff(\rho)}\,\exp(s)\,p(s) \nonumber \\
        \,& = &\, \frac{\eps}{\phit} \mathlarger{\int}_{\scrit}^{\infty}{\frac{\tff(\meanrho)}{\tff(\rho)} \frac{\rho}{\meanrho} \, p(s) \deriv s}\,.
\end{eqnarray}
Note that the first equality is the same as Equation~(7) in \citet{HennebelleChabrier2011}\footnote{Equation~(7) in \citet{HennebelleChabrier2011} contains an error in that the factor $\deriv M/M$ in their integral must instead read $\deriv M$ (P.~Hennebelle \& G.~Chabrier 2012, private communication), which simplifies the equation significantly because the mass and radius dependencies drop entirely and the integral can be completely rewritten in terms of $s$ and solved analytically (see our Equation~\ref{eq:sfrff_hc}).}. It can be simplified to the second line in Equation~(\ref{eq:sfrff_hc}), by transforming the mass variable into the logarithmic density variable $s$ and changing the limits of the integral accordingly. We emphasize that the second equality in Equation~(\ref{eq:sfrff_hc}) is exactly the same as the general model for $\sfrff$ given by Equation~(\ref{eq:sfrffbasic}) above.

In the HC model, the critical density $\scrit{_\mathrm{,HC}}$ is defined by requiring that the turbulent Jeans length $\lambdajt$ at the critical density is a fraction $\ycut$ of the cloud scale $L$. \citet{HennebelleChabrier2011} do not provide an explicit physical interpretation of this choice, but a follow-up study is in preparation (P.~Hennebelle \& G.~Chabrier 2012, private communication). The turbulent Jeans length is obtained by adding an effective turbulent pressure \citep[see][]{Chandrasekhar1951a,Chandrasekhar1951b,BonazzolaEtAl1987}\footnote{The concept of turbulent pressure is also used to derive accretion rates and luminosities during high-mass star formation in massive turbulent cores \citep{McKeeTan2002,McKeeTan2003}.} to the sound speed in the purely thermal Jeans length, Equation~(\ref{eq:lambdaj}):
\begin{eqnarray} \label{eq:lambdajt_def}
\lambdajt \,& \equiv &\, \left(\frac{\pi\cs^2+(\pi/3)\sigma_v^2(\lambdajt)}{G\rho}\right)^{1/2} \nonumber \\
                 \,& = &\, \left(\frac{\pi\cs^2+\pi\lambdajt\sigma_V^2/(3L)}{G\rho}\right)^{1/2} \,,
\end{eqnarray}
in which the turbulent velocity dispersion, Equation~(\ref{eq:sigmav}), must be evaluated on the scale $\ell=\lambdajt$, such that the turbulent Jeans length is implicitly defined by Equation~(\ref{eq:lambdajt_def}). Rewriting yields a quadratic equation with two solutions:
\begin{equation}
\lambdajt(\rho)=\frac{\pi\sigma_V^2 \pm \sqrt{36\pi\cs^2GL^2\rho+\pi^2\sigma_V^4}}{6GL\rho}\,
\end{equation}
for which only the positive root is physical because the Jeans length must become larger when adding a stabilizing pressure---in this case a turbulent pressure. Naturally, this expression reduces to the thermal Jeans length for $\sigma_V\to0$. Now, setting the turbulent Jeans length equal to $\ycut L$ as defined in \citet{HennebelleChabrier2011}, and identifying the virial parameter, Equation~(\ref{eq:alphavircirc}), and the Mach number $\mach=\sigma_V/\cs$, finally yields the critical density threshold in the HC model:
\begin{equation} \label{eq:scrit_hc}
\scrit{_\mathrm{,HC}} = \ln{\left[ \tilde{\rho}_\mathrm{crit,th} + \tilde{\rho}_\mathrm{crit,turb} \right]} \,,
\end{equation}
where the (magneto)thermal contribution is
\begin{equation} \label{eq:rhocrit_th}
\tilde{\rho}_\mathrm{crit,th} \equiv (\pi^2/5)\ycut^{-2} \alphavir \mach^{-2} (1+\beta^{-1}) \,,
\end{equation}
and the turbulent contribution is
\begin{equation} \label{eq:rhocrit_turb}
\tilde{\rho}_\mathrm{crit,turb} \equiv (\pi^2/15)\,\ycut^{-1}\,\alphavir \,.
\end{equation}
Note that the dependence of the thermal contribution to $\scrit{_\mathrm{,HC}}$ on $\alphavir$ is the same as in the KM and PN models, while the dependence on the Mach number is $\mach^{-2}$, which is the opposite of the dependence in the KM and PN models, for both of which $\rhocrit\propto\mach^{+2}$ (see Table~\ref{tab:theories} for a summary of all analytic models).

The original HC model does not take magnetic fields into account, but we have extended the HC theory to MHD here by replacing the sonic Mach number in Equation~(\ref{eq:rhocrit_th}) with the magnetic version in the same way as done for the KM model via Equation~(\ref{eq:machmag}). The magnetic correction factor $(1+\beta^{-1})$ in Equation~(\ref{eq:rhocrit_th}) simply becomes unity in the hydrodynamical limit ($\beta\to\infty$).

The SFR in the HC model is thus given by integrating Equation~(\ref{eq:sfrff_hc}) or equivalently Equation~(\ref{eq:sfrffbasic}) with $\scrit{_\mathrm{,HC}}$, which yields
\begin{eqnarray} \label{eq:sfrff_hc_solution}
\sfrff{_\mathrm{,HC}} & = & \frac{\eps}{\phit} \mathlarger{\int}_{\scrit{_\mathrm{,HC}}}^{\infty} {\exp\left(\frac{3}{2}s\right) \, p(s) \deriv s} \nonumber \\
                                     & = & \frac{\eps}{2\phit} \exp\left(\frac{3}{8}\sigs^2\right) \left[1+\mathrm{erf}\left(\frac{\sigs^2-\scrit{_\mathrm{,HC}}}{\sqrt{2\sigs^2}}\right)\right] . \nonumber \\
\end{eqnarray}

\subsubsection{The Multi-freefall KM Model}
Following \citet{HennebelleChabrier2011}, we define all three multi-freefall versions of the KM, PN, and HC models by solving the generalized, multi-freefall integral, Equation~(\ref{eq:sfrffbasic}). The analytic solution of that equation for an arbitrary threshold $\scrit$ is
\begin{equation} \label{eq:sfrff_basicsolution}
\sfrff = \frac{\eps}{2\phit} \exp\left(\frac{3}{8}\sigs^2\right) \left[1+\mathrm{erf}\left(\frac{\sigs^2-\scrit}{\sqrt{2\sigs^2}}\right)\right],
\end{equation}
which is identical to Equation~(8) in \citet{HennebelleChabrier2011}, and identical to the HC model, Equation~(\ref{eq:sfrff_hc_solution}), except that the critical density is defined according to either the KM, PN, or HC models.
Thus, the multi-ff KM model is defined by using the threshold density $\scrit=\scrit{_\mathrm{,KM}}$ from Equation~(\ref{eq:scrit_km}) in the generalized solution of the multi-freefall $\sfrff$, Equation~(\ref{eq:sfrff_basicsolution}).

\subsubsection{The Multi-freefall PN Model}
The multi-ff PN model is defined by using the threshold density $\scrit=\scrit{_\mathrm{,PN}}$ from Equation~(\ref{eq:scrit_pn}) in the generalized solution of the multi-freefall $\sfrff$, Equation~(\ref{eq:sfrff_basicsolution}).

\subsubsection{The Multi-freefall HC Model}
The multi-ff HC model is defined by taking the threshold density $\scrit=\scrit{_\mathrm{,HC}}$ from Equation~(\ref{eq:scrit_hc}), but only with the thermal contribution $\tilde{\rho}_\mathrm{crit,th}$ from Equation~(\ref{eq:rhocrit_th}), while setting the turbulent contribution $\tilde{\rho}_\mathrm{crit,turb}=0$, and using that threshold density in the generalized solution of the multi-freefall $\sfrff$, Equation~(\ref{eq:sfrff_basicsolution}). We do this to be consistent with the definition in \citet{HennebelleChabrier2011}. Note that the thermal density threshold is derived by requiring that the thermal Jeans length at that density, $\lambdaj$ is equal to $\ycut L$, while the full HC model includes the turbulent contribution, which is obtained by setting $\lambdajt=\ycut L$ (see the derivation of the HC model above).

\begin{figure*}[t]
\centerline{
\includegraphics[width=0.95\linewidth]{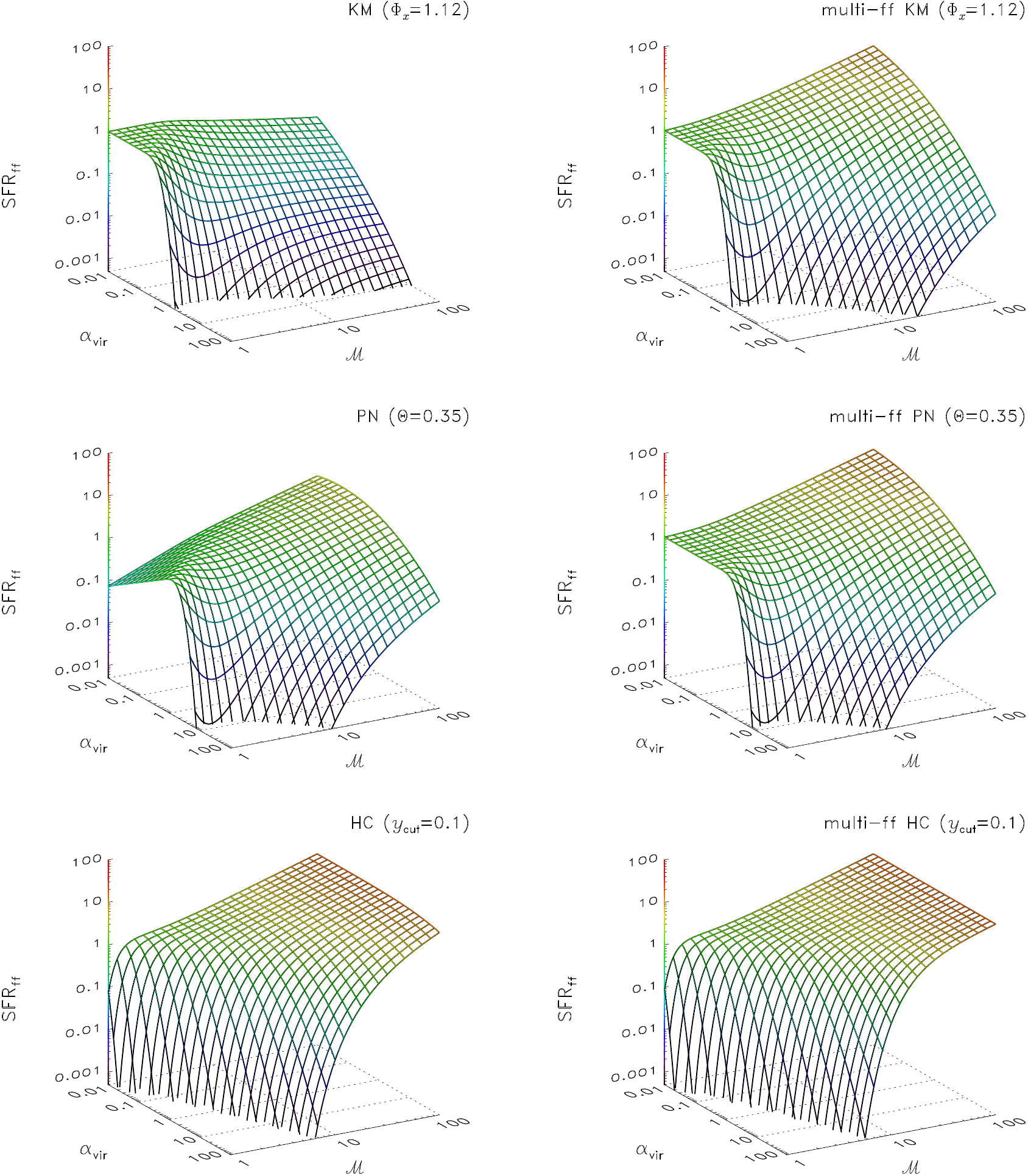}
}
\caption{Comparison of the six analytic models for the star formation rate per freefall time, $\sfrff$: KM, PN, HC (left panels) and multi-freefall KM, PN, HC (right panels). See Table~\ref{tab:theories} and the derivations in Section~\ref{sec:derivation} for details of the different analytic models and functions plotted ($\eps/\phit=1$ in each panel). The dependence of $\sfrff$ on the virial parameter $\alphavir$, and the sonic Mach number $\mach$ are shown in each panel for a turbulent forcing parameter $b=0.4$, corresponding to a statistical mixture of solenoidal and compressive modes in the turbulent forcing. All models are plotted without taking magnetic fields into account, i.e., plasma $\beta\to\infty$.}
\label{fig:sfrffmodel_mix}
\end{figure*}

\subsection{Dependencies of the Analytically Derived $\sfrff$} \label{sec:theoryresults}
After the detailed derivation of the six different $\sfrff$ models, we can now start to compare them. Figure~\ref{fig:sfrffmodel_mix} shows all six $\sfrff$ models: KM, PN, HC (left panels), and multi-ff KM, PN, HC (right panels) for a turbulent forcing parameter $b=0.4$, corresponding to a statistical mixture of solenoidal and compressive modes in the turbulent forcing \citep[][Figure~8]{FederrathDuvalKlessenSchmidtMacLow2010}. When looking at the derivations above, it becomes clear that $\sfrff$ is a function of $\alphavir$, $\mach$, $b$, and $\beta$. The dependencies enter through the definition of the critical densities in the different models, and through the variance of turbulent density fluctuations, Equation~(\ref{eq:sigs}). We plot the analytically derived $\sfrff$ as a function of the virial parameter $\alphavir$ and the sonic Mach number $\mach$ in each panel. Note that all these models are plotted for $\beta\to\infty$, i.e., without taking magnetic fields into account yet. As shown in Table~\ref{tab:theories}, each model has two fudge factors of order unity. The first one is $1/\phit$ for all models (where the local efficiency was set to $\eps=1$ for simplicity in all models, to facilitate the comparison), while the second one is $\phix$, $\theta$, and $\ycut$ for the (multi-freefall) KM, PN, and HC models, respectively. We plot all curves for $\eps/\phit=1$ to enable direct comparisons, and used the favored values of the fudge factors by the different authors, $\phix=1.12$ \citep{KrumholzMcKee2005}, $\theta=0.35$ \citep{PadoanNordlund2011}, and $\ycut=0.1$ \citep{HennebelleChabrier2011}.

\paragraph{Dependence on $\alphavir$}
Let us first concentrate on the dependence of $\sfrff$ on the virial parameter. Since the virial parameter, Equations~(\ref{eq:alphavircirc}) and~(\ref{eq:alphavir}), is defined here as the ratio of twice the kinetic to the gravitational energy, it essentially measures how strongly the system is bound, and whether it is contracting ($\alphavir\lesssim1$) or expanding ($\alphavir\gtrsim1$). Thus, we generally expect that the SFR should decrease with increasing $\alphavir$ because the cloud then becomes less bound and less likely to form stars. Indeed, the analytic SFR generally decreases with increasing $\alphavir$ in all models with the exception of the original PN model, for which $\sfrff$ first increases for $\alphavir\lesssim1$ and then decreases for $\alphavir\gtrsim1$. The increase comes from the freefall-time factor $\tff(\meanrho)/\tff(\rhocrit)$ in the PN model, which leads to the factor $\exp(\scrit{_\mathrm{,PN}}/2)$ in Equation~(\ref{eq:sfrff_pn}), and with the critical density from Equation~(\ref{eq:scrit_pn}) to $\sfrff\propto\alphavir$ for small $\alphavir$. As expected though, this direct proportionality disappears in the multi-freefall PN model, as in the other two multi-freefall models (multi-ff KM and multi-ff HC).

\paragraph{Dependence on $\mach$}
The expected dependence of $\sfrff$ on the sonic Mach number is that $\sfrff$ should increase with increasing $\mach$ because higher Mach number means stronger and denser local compression, leading to higher SFRs. Indeed, the Mach number dependence is generally similar in all models, i.e., $\sfrff$ increases with $\mach$, with the exception of the original KM model, which has the weakest dependence on $\mach$. For large $\alphavir$, $\sfrff{_\mathrm{,KM}}$ increases, but only slowly, while for small $\alphavir$, it stays constant or even decreases with increasing $\mach$. Both the HC and multi-freefall HC models have the strongest positive correlation with the Mach number, such that for $\mach\gtrsim10$, $\sfrff{_\mathrm{,HC}}$ hardly depends on $\alphavir$ anymore \citep[see also,][Figure~1]{HennebelleChabrier2011}\footnote{Note that the three different sonic Mach numbers shown in Figure~1 of \citet{HennebelleChabrier2011} are actually $\mach=4.5$, 9, and 18, and not 4, 9, and 16 as indicated in their figure caption (P.~Hennebelle \& G.~Chabrier 2012, private communication).}. The strong increase of $\sfrff$ with $\mach$ in the two HC cases comes from the Mach number dependence of the thermal contribution to $\scrit{_\mathrm{,HC}}$, which is $\tilde{\rho}_\mathrm{crit,th} \propto \mach^{-2}$, leading to a decreasing threshold density in the HC models, and thus to a higher SFR. This is the opposite compared to the KM and PN models, for both of which the critical density increases with the square of the Mach number (see Equations~\ref{eq:scrit_km} and~\ref{eq:scrit_pn}, respectively, or Table~\ref{tab:theories}).

We also note the local minima of $\sfrff$ around $\mach\approx2$ in all models, except the HC and multi-freefall HC models. Those minima are spurious because they occur close to $\mach=1$, for which the basic approach of shock-induced star formation must eventually break down as the system becomes transonic. Shocks require $\mach>1$, by definition, but for rms Mach 1--2, a significant fraction of the system is transonic to subsonic. We thus conclude that all six models break down for the low Mach number regime, $\mach\lesssim2$. The rms sonic Mach numbers in real molecular clouds usually exceed unity by far \citep{Larson1981,FalgaronePugetPerault1992,RomanDuvalEtAl2011,SchneiderEtAl2012}, such that our analytic models are generally applicable to typical molecular clouds with $\mach>2$.

\begin{figure}[t]
\centerline{
\includegraphics[width=0.9\linewidth]{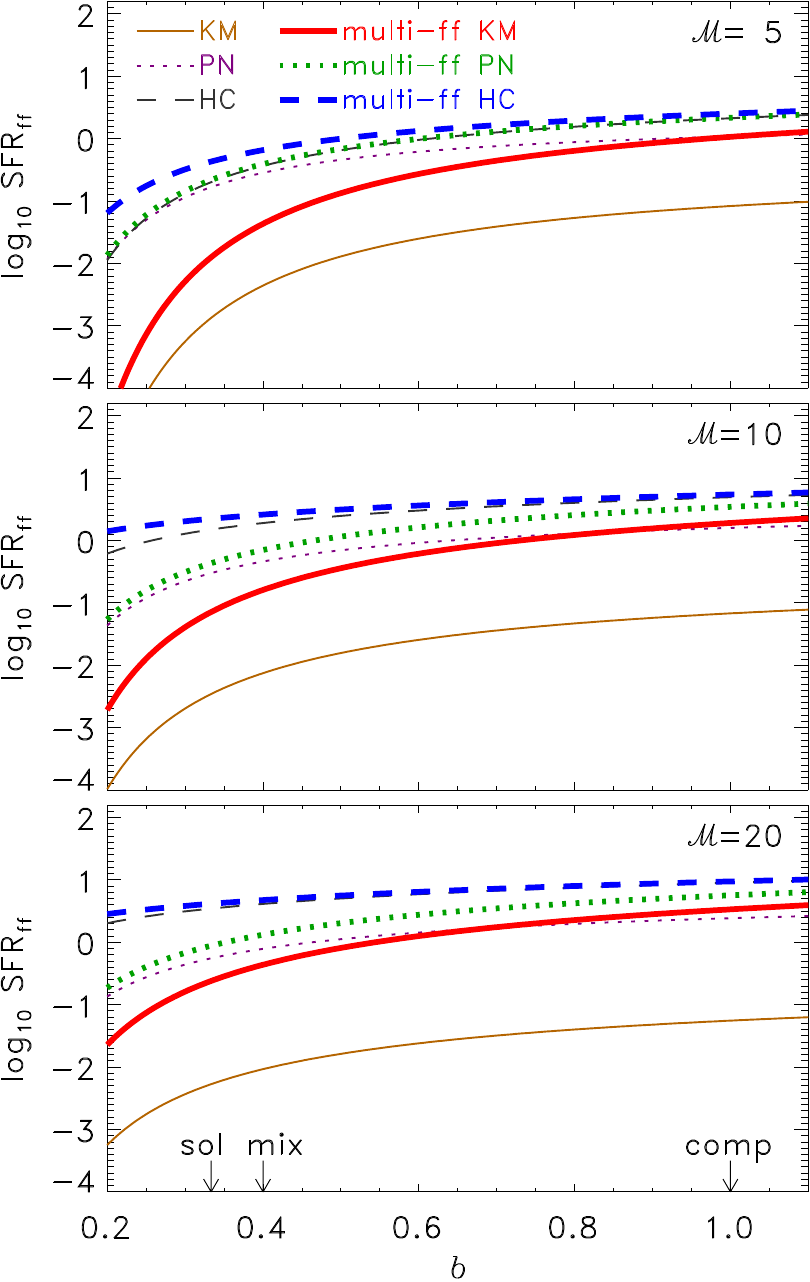}
}
\caption{$\sfrff$ as a function of the forcing parameter $b$ in Equation~(\ref{eq:sigs}) for sonic Mach numbers $\mach=5$ (top), $\mach=10$ (middle), and $\mach=20$ (bottom). All curves are plotted for $\alphavir=1$, $\eps/\phit=1$, and the favored fudge factors by \citet{KrumholzMcKee2005}, \citet{PadoanNordlund2011}, and \citet{HennebelleChabrier2011}: $\phix=1.12$, $\theta=0.35$, and $\ycut=0.1$, respectively. Only purely hydrodynamic cases are shown ($\beta\to\infty$). The star formation rate increases monotonically from $b=1/3$ (solenoidal turbulent forcing), over $b=0.4$ (mixed forcing), to $b=1$ (compressive forcing).}
\label{fig:sfrff_on_b}
\end{figure}

\paragraph{Dependence on $b$}
While the dependence on Mach number enters $\sfrff$ both through $\scrit$ and $\sigs^2$, the forcing dependence only enters through the forcing parameter $b$ in $\sigs^2$, Equation~(\ref{eq:sigs}). Figure~\ref{fig:sfrff_on_b} shows $\sfrff$ as a function of the forcing parameter $b$ for all models and three different Mach numbers ($\mach=5$, 10, and 20). All curves are plotted for $\alphavir=1$, $\beta\to\infty$, $\eps/\phit=1$, and the standard fudge factors $\phix=1.12$, $\theta=0.35$, and $\ycut=0.1$, respectively. We see that $\sfrff$ increases monotonically with $b$, from $b=1/3$ (solenoidal forcing), over $b=0.4$ (mixed forcing), to $b=1$ (compressive forcing) in all models. This is expected because the density variance becomes larger for more compressive forcing, pushing a significant fraction of the gas to higher densities \citep{FederrathKlessenSchmidt2008,FederrathDuvalKlessenSchmidtMacLow2010,KonstandinEtAl2012}. Similar to the behavior with increasing Mach number, increasing the amount of direct compression induced by the turbulent forcing leads to higher local densities, and thus to higher SFRs with a typical increase of about an order of magnitude for compressive forcing compared to solenoidal forcing.

\begin{figure}[t]
\centerline{
\includegraphics[width=0.9\linewidth]{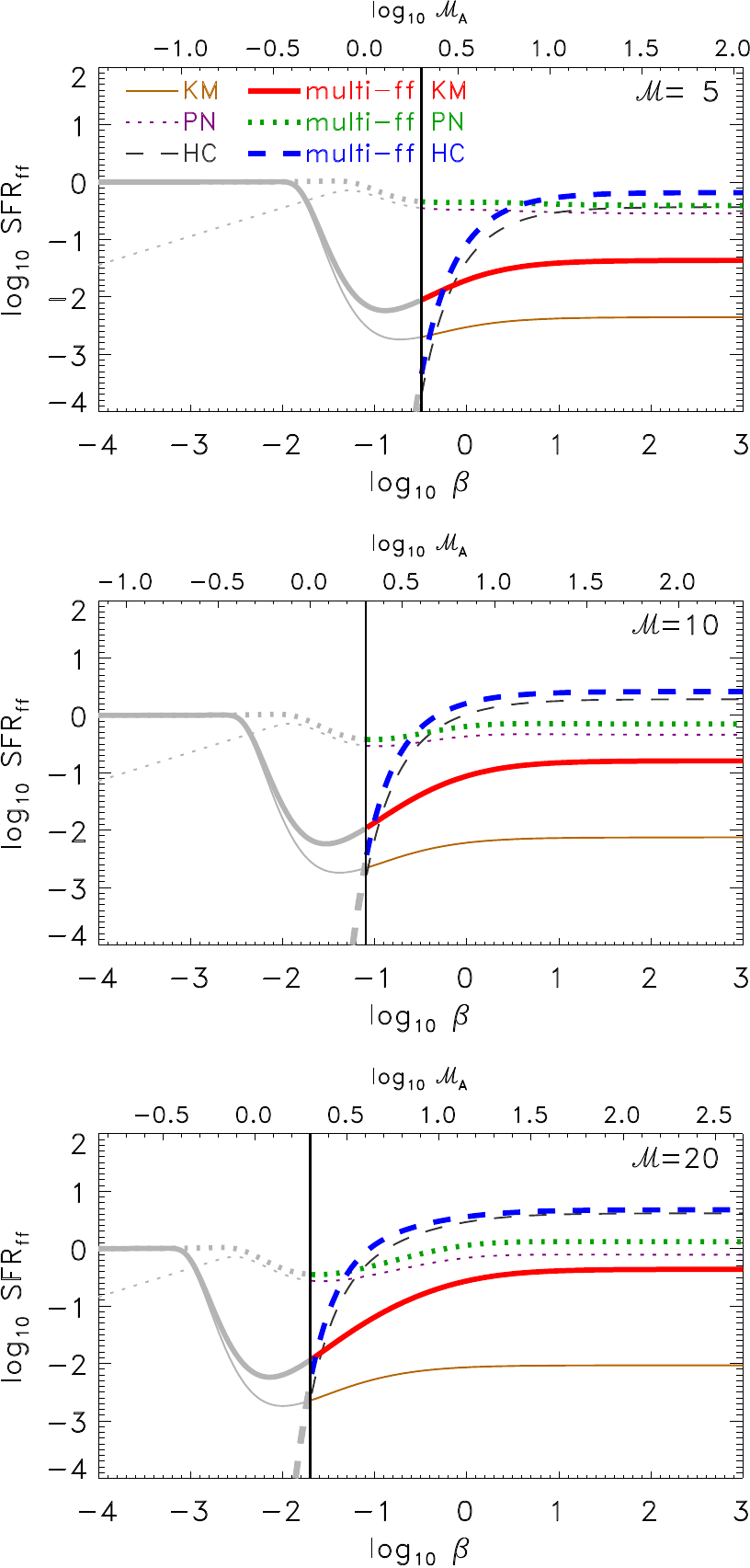}
}
\caption{Same as Figure~\ref{fig:sfrff_on_b}, but $\sfrff$ is shown as a function of plasma $\beta$, the ratio of thermal to magnetic pressure (bottom abscissa) or as a function of the Alfv\'en Mach number, $\macha=\mach\sqrt{\beta/2}$ (top abscissa) for mixed forcing ($b=0.4$). Since the sonic Mach number is $\mach=5$, 10, and 20 (top to bottom panels), the $\macha$-axis varies between the three panels. The solid, vertical line separates $\macha\!<2\!$ from $\macha\!>\!2$. Analytic predictions below $\macha\lesssim2$ are inaccurate \citep{MolinaEtAl2012} and only shown in gray.}
\label{fig:sfrff_on_beta}
\end{figure}

\paragraph{Dependence on $\beta$}
We expect that by adding magnetic energy to the system, the SFR should decrease because magnetic energy adds a stabilizing pressure to the system, counteracting gravitational collapse. Figure~\ref{fig:sfrff_on_beta} shows the dependence of $\sfrff$ on plasma $\beta$ in the six analytic models. We emphasize that only the original PN model had a magnetic-field dependence, coming from the dependence of $\scrit{_\mathrm{,PN}}$ on $\beta$ in Equation~(\ref{eq:scrit_pn}), and from the dependence of $\sigs$ on $\beta$ in Equation~(\ref{eq:sigs}). However, we have extended all other analytic models (KM, HC, and multi-ff KM, PN, HC) to MHD, simply by applying the MHD version of $\sigs$, Equation~(\ref{eq:sigs}) in all models, and replacing the sonic Mach number in the expressions for the critical density by the magnetic version $\mach\to\mach/\sqrt{1+1/\beta}$, introduced in Equation~(\ref{eq:machmag}).

As found in a detailed comparison of the analytically-derived $\sigs$ with numerical simulations of MHD turbulence in \citet{MolinaEtAl2012}, the standard deviation--Mach number relation, Equation~(\ref{eq:sigs}), breaks down for $\macha\lesssim2$ because strongly sub-Alfv\'enic flows become highly anisotropic \citep[e.g.,][]{MacLow1999,ChoVishniac2000,ChoLazarian2003,BeresnyakEtAl2005,BruntFederrathPrice2010a,EsquivelLazarian2011}. Since the magnetic-field dependence of $\sfrff$ was introduced as an isotropic magnetic-pressure extension, the behavior of the analytic models for $\macha\lesssim2$ is likely invalid. Thus, we only consider the trans- to super-Alfv\'enic regime with $\macha\gtrsim2$. In this regime, $\sfrff$ decreases with increasing magnetic energy, i.e., decreasing $\beta$ or $\macha$ in all models, as expected when adding a stabilizing magnetic pressure.

\section{Testing the Analytic Theories for the SFR with Numerical Simulations} \label{sec:sims}
In order to test the analytic predictions of the star formation rate (SFR) models in Section~\ref{sec:theory}, we perform a series of numerical simulations of driven, supersonic turbulence, including magnetic fields, gravity, and a model for collapse and accretion of star-forming regions to measure the SFR. Ideally, we would like to sample as much of the parameter space as possible with the simulations. Since the analytic SFR depends on $\alphavir$, $\mach$, $b$, and $\beta$ (see Section~\ref{sec:theoryresults}), we have to restrict ourselves to testing only a subset of those because the simulations are computationally too expensive to scan through the entire parameter range. We thus concentrate here on the Mach number and forcing dependence, as well as the dependence on the magnetic field, but only consider models with an initial virial parameter of around unity. However, as the turbulence produces strong spatial density variations, the virial parameter can change by an order of magnitude from its initial value given by Equation~(\ref{eq:alphavircirc}) when the turbulence is fully established because the mass is rearranged into complex filamentary and sheet-like structures. To take this into account, we always compute instantaneous values of $\alphavir$, based on the spatial distribution of the gas (Equation~\ref{eq:alphavir}), as for all other parameters, and then average them over space and time. The time interval for averaging is chosen such that it covers the whole star formation sequence in the simulations, from the time when the turbulence is fully established, as explained in more detail in Section~\ref{sec:ics} below. First however, we explain our numerical scheme in Section~\ref{sec:methods}, the forcing of the turbulence in Section~\ref{sec:forcing}, and the sink particles introduced to model core and star formation in Section~\ref{sec:sinks}.

\subsection{Numerical Methods} \label{sec:methods}
We use the adaptive mesh refinement \citep[AMR,][]{BergerColella1989} code FLASH\footnote{\url{http://flash.uchicago.edu/site/flashcode/}} \citep{FryxellEtAl2000,DubeyEtAl2008} in version 2.5 to integrate the ideal, three-dimensional, MHD equations, including self-gravity,
\begin{eqnarray} \label{eq:mhd}
& & \frac{\partial \rho}{\partial t} + \nabla\cdot\left(\rho \bfu\right)=0\,, \nonumber\\
& & \rho\left(\frac{\partial}{\partial t} + \bfu\cdot\nabla\right)\bfu = \frac{(\bfB\cdot\nabla)\bfB}{4\pi} - \nabla P_\star + \rho\left({\bf g} + {\bf F_\mathrm{stir}}\right)\,, \nonumber\\
& & \frac{\partial E}{\partial t} + \nabla\cdot\left[\left(E+P_\star\right)\bfu - \frac{\left(\bfB\cdot\bfu\right)\bfB}{4\pi}\right] = \rho\bfu\cdot\left({\bf g}+{\bf F_\mathrm{stir}}\right)\,, \nonumber\\
& & \frac{\partial \bfB}{\partial t} = \nabla\times\left(\bfu\times\bfB\right)\,,\quad\nabla\cdot\bfB = 0\,,
\end{eqnarray}
where the gravitational acceleration of the gas ${\bf g}$, is the sum of the self-gravity of the gas and the contribution of sink particles (a subgrid model for collapse and accretion of star-forming regions in the simulations, explained in Section~\ref{sec:sinks} below):
\begin{eqnarray} \label{eq:grav}
& & {\bf g} = -\nabla\Phi_\mathrm{gas} + {\bf g}_\mathrm{sinks}\,, \nonumber\\
& & \nabla^2\Phi_\mathrm{gas} = 4\pi G\rho\,.
\end{eqnarray}
In the ideal MHD Equations~(\ref{eq:mhd}), $\rho$, $\bfu$, $P_\star=P_\mathrm{th}+ 1/(8\pi)\left|\bfB\right|^2$, $\bfB$, and $E=\rho \epsilon_\mathrm{int} + (\rho/2)\left|\bfu\right|^2 + 1/(8\pi)\left|\bfB\right|^2$ denote gas density, velocity, pressure (thermal plus magnetic), magnetic field, and total energy density (internal plus kinetic, plus magnetic), respectively. The MHD equations are closed with a polytropic equation of state, $P_\mathrm{th}=\cs^2\rho^\Gamma$ with $\Gamma=1$, such that the gas remains isothermal with a constant sound speed $\cs=0.2\,\km\,\s^{-1}$, corresponding to a temperature of $T\approx11\,\mathrm{K}$ for gas with a mean molecular weight of 2.3. This is a reasonable approximation for dense, molecular gas of solar metallicity, over a wide range of densities \citep{WolfireEtAl1995,OmukaiEtAl2005,PavlovskiSmithMacLow2006,GloverMacLow2007a,GloverMacLow2007b,GloverFederrathMacLowKlessen2010,HillEtAl2011,HennemannEtAl2012}. Moreover, \citet{GloverClark2012} find that the SFR is almost insensitive to the metallicity. Reducing the metallicity of the gas by two orders of magnitude reduces the time-averaged SFR by less than a factor of two. Thus, our conclusions remain intact, even though we neglect the detailed chemistry, cooling and heating processes in molecular clouds in this study.

We solve the MHD Equations~(\ref{eq:mhd}) on three-dimensional, periodic grids with maximum resolutions of $N_\mathrm{res}^3=128^3$--$1024^3$ grid points. These are all uniform-grid simulations, except for the $N_\mathrm{res}=1024$ simulation, where we use a root grid with $512^3$ cells and one level of AMR with a refinement criterion to ensure that the local Jeans lengths is covered with at least 32 grid cells, in order to resolve turbulent vorticity and magnetic-field amplification on the Jeans scale \citep{SurEtAl2010,FederrathSurSchleicherBanerjeeKlessen2011,TurkEtAl2012}. We use a positive-definite MHD Riemann solver \citep{BouchutKlingenbergWaagan2007,BouchutKlingenbergWaagan2010,Waagan2009}, which has been tested for efficiency, robustness, and accuracy in \citet{WaaganFederrathKlingenberg2011}. This study shows that the MHD scheme keeps $\nabla\cdot\bfB$ errors at a negligible level, and allows us to model extremely high-Mach turbulence without producing unphysical states. This is particularly important for this study because we model supersonic turbulence on the largest scales of molecular clouds with rms Mach numbers as high as $\mach\approx50$ and compressive forcing, which produces density contrasts by several orders of magnitude, sometimes between two adjacent grid cells because of multiple interactions of shocks and strong rarefaction waves, even before gravitational collapse sets in. Grid-based HD solvers often produce negative densities in such situations because of numerical post-shock oscillations. Such unphysical states are avoided by construction in the HLL3R Riemann scheme \citep{WaaganFederrathKlingenberg2011} used here. The self-gravity of the gas, i.e., the gas--gas gravitational interaction (Equation~\ref{eq:grav}) is computed using a multi-grid Poisson solver \citep[the FLASH2.5 version discussed in][]{Ricker2008}, while the sink particle interactions are computed by direct $N$-body summation, as explained in Section~\ref{sec:sinks} below. We note that the gravitational potential $\Phi_\mathrm{gas}$ is computed with respect to the periodic boundary conditions specified in the simulations.

The ideal MHD Equations~(\ref{eq:mhd}) do not contain any explicit kinematic viscosity and magnetic resistivity terms. However, any numerical scheme has an effective numerical viscosity $\nu$ and magnetic resistivity $\eta$ due to the necessary discretization of the MHD equations. Even though the numerical viscosity depends on the specifications of the algorithm, it can be used to mimic the effects of explicit viscosity and resistivity \citep{BenziEtAl2008}. It is important to realize, though, that the kinematic and magnetic Reynolds numbers that can be achieved with ideal MHD depend on the grid resolution. As shown in \citet{FederrathEtAl2011PRL}, compressible, ideal MHD turbulence resolved with $128^3$ grid cells reaches kinematic Reynolds numbers $\mathrm{Re}=L\sigma_V/\nu\approx1500$ and magnetic Reynolds numbers $\mathrm{Rm}=L\sigma_V/\eta\approx3000$. For \citet{Burgers1948} scaling of the turbulence $\sigma_v(\ell)\propto\ell^{1/2}$ (Equation~\ref{eq:sigmav} with $p=1/2$), the Reynolds numbers scale $\propto N_\mathrm{res}^{3/2}$ as opposed to \citet{Kolmogorov1941c} scaling of the turbulence, $\sigma_v(\ell)\propto\ell^{1/3}$ (Equation~\ref{eq:sigmav} with $p=1/3$), leading to a Reynolds-number scaling $\propto N_\mathrm{res}^{4/3}$. Thus, even in our highest resolution simulation with $N_\mathrm{res}=1024$, we only achieve Reynolds numbers, $\mathrm{Re}\approx(2.4$--$3.4)\!\times\!10^4$ and $\mathrm{Rm}\approx(4.8$--$6.8)\!\times\!10^4$, depending on the scaling of the turbulence. In summary, although the flows we model exhibit fully developed turbulence \citep{Frisch1995}, their Reynolds number are still considerably smaller than the ones inferred for real molecular clouds \citep[see, e.g.,][]{SchoberEtAl2012}. We will thus study the resolution dependence of our results for the SFR below.

\subsection{Turbulent Forcing} \label{sec:forcing}
Previous numerical studies of non-driven turbulence have shown that supersonic turbulence decays in about a crossing time, irrespective of whether magnetic fields are included or not \citep{ScaloPumphrey1982,MacLowEtAl1998,StoneOstrikerGammie1998,MacLow1999}. The observed presence of turbulence has thus lead to the conclusion that interstellar turbulence should be driven by some physical stirring mechanisms. Those mechanisms include supernova explosions and expanding, ionizing shells from previous cycles of star formation \citep{McKee1989,KrumholzMatznerMcKee2006,BalsaraEtAl2004,BreitschwerdtEtAl2009,PetersEtAl2011,GoldbaumEtAl2011,LeeMurrayRahman2012}, gravitational collapse and accretion of material \citep{VazquezCantoLizano1998,KlessenHennebelle2010,ElmegreenBurkert2010,VazquezSemadeniEtAl2010,FederrathSurSchleicherBanerjeeKlessen2011,RobertsonGoldreich2012}, and galactic spiral-arm compression of H\textsc{I} clouds \citep{DobbsBonnell2008,DobbsEtAl2008} and magnetorotational instability (MRI) \citep{PiontekOstriker2007,TamburroEtAl2009}. On smaller scales, jets and outflows from young stellar objects have been suggested to drive turbulence \citep{NormanSilk1980,BanerjeeKlessenFendt2007,NakamuraLi2008,CunninghamEtAl2009,CarrollFrankBlackman2010,WangEtAl2010}. Turbulence in high-redshift galaxies is also likely driven by feedback from previous cycles of star formation \citep{GreenEtAl2010}. A summary and comparison of driving mechanisms for interstellar turbulence is provided in \citet{MacLowKlessen2004} and \citet{Elmegreen2009}. \citet{MacLowKlessen2004} conclude that expanding shells are likely the dominant driver of interstellar turbulence in the star-forming parts of the Galaxy. More recently, \citet{LeeMurrayRahman2012} also noted that the kinetic energy injected per unit time by star-forming complexes via expansion of bubbles is about 2/3 of the luminosity required to maintain the observed velocity dispersions, supporting the view that expanding bubbles driven by massive star clusters from previous star formation are a major driver of turbulence in the Milky Way \citep[see e.g., the Cygnus X giant molecular cloud studied in][]{SchneiderEtAl2011}.

It is important to realize that all these potential drivers (maybe with the exception of the MRI) are expected to primarily drive compressible modes in the velocity field, but do not directly excite solenoidal modes. However, even though the turbulence in molecular clouds might be driven compressively, solenoidal modes are indirectly excited by nonlinear interactions of multiple colliding shock fronts \citep{Vishniac1994,SunTakayama2003,KritsukEtAl2007,FederrathDuvalKlessenSchmidtMacLow2010}, by baroclinity, rotation and shear \citep{DelSordoBrandenburg2011}, and by viscosity \citep{MeeBrandenburg2006,FederrathEtAl2011PRL}, such that supersonic turbulence driven by even \emph{purely} compressive forcing contains about half of its kinetic power in solenoidal modes and the other half in compressible modes in the inertial range \citep[][Figure~14]{FederrathDuvalKlessenSchmidtMacLow2010}.

Modeling physical turbulent stirring mechanisms in numerical simulations requires assumptions about the spatial and temporal correlation of the turbulent forcing events. It is also still a matter of debate which of the physical mechanisms dominates the injection of turbulent energy on different cloud scales. Given these uncertainties, instead of trying to mimic one or more of the potential physical drivers of turbulence, we here use simulations of the so-called `driven turbulence in a box'. From these simplified and idealized simulations, we can draw statistical conclusions about the role of turbulence for star formation, given average properties of a cloud ($\alphavir$, $\mach$, $b$, and $\beta$). In particular, our turbulent forcing approach allows us to evaluate the role of the mixture of velocity modes excited by a physical driver.

In practice, the stochastic forcing term ${\bf F_\mathrm{stir}}$ is applied as a source term in Equations~(\ref{eq:mhd}) to drive turbulence in the simulations. ${\bf F_\mathrm{stir}}$ only contains large-scale modes, $1<k<3$, where most of the power is injected at the $k=2$ mode in Fourier space, which corresponds to half of the box size $L$ in physical space. We thus model turbulent forcing on large scales, as favored by molecular cloud observations \citep[e.g.,][]{OssenkopfMacLow2002,HeyerWilliamsBrunt2006,BruntHeyerMacLow2009,GaenslerEtAl2011,RomanDuvalEtAl2011}. Smaller scales, $k>3$ are not affected directly by the forcing, such that turbulence can develop self-consistently on these scales. We use the Ornstein-Uhlenbeck (OU) process to model ${\bf F_\mathrm{stir}}$, which is a well-defined stochastic process with a finite autocorrelation timescale \citep{EswaranPope1988,SchmidtHillebrandtNiemeyer2006}, leading to a smoothly varying stochastic force field in space and time. Details about the OU process and the forcing applied in this study can be found in \citet{SchmidtEtAl2009}, \citet{FederrathDuvalKlessenSchmidtMacLow2010}, and \citet{KonstandinEtAl2012}. However, the essential point of our forcing approach is that we can adjust the mixture of solenoidal and compressive modes of ${\bf F_\mathrm{stir}}$. This is achieved by decomposing a given vector field with random mixtures into its solenoidal and compressive parts, by applying the projection tensor $\mathcal{\underline{P}}^{\,\zeta}(\vect{k})$ in Fourier space. In index notation, this tensor reads
\begin{equation} \label{eq:projectionoperator}
\mathcal{P}_{ij}^\zeta = \zeta\,\mathcal{P}_{ij}^\perp+(1-\zeta)\,\mathcal{P}_{ij}^\parallel = \zeta\,\delta_{ij}+(1-2\zeta)\,\frac{k_i k_j}{|k|^2}\;,
\end{equation}
where $\delta_{ij}$ is the Kronecker symbol, and $\mathcal{P}_{ij}^\perp = \delta_{ij} - k_i k_j / k^2$ and $\mathcal{P}_{ij}^\parallel = k_i k_j / k^2$ are the solenoidal and compressive projection operators, respectively. The ratio of compressive power to total power in ${\bf F_\mathrm{stir}}$ can be derived from Equation~(\ref{eq:projectionoperator}) by evaluating the norm of the compressive component of the projection tensor and dividing it by the total injected power, resulting in
\begin{equation} \label{eq:forcing_ratio}
\frac{F_\mathrm{comp}}{F_\mathrm{tot}} = \frac{(1-\zeta)^2}{1-2\zeta+3\zeta^2}\,,
\end{equation}
for three-dimensional space \citep{SchmidtEtAl2009,FederrathDuvalKlessenSchmidtMacLow2010}. The projection operator serves to construct a purely solenoidal force field by setting $\zeta=1$, while for $\zeta=0$, a purely compressive force field is obtained. Any combination of solenoidal and compressive modes can be constructed by choosing $\zeta\in[0,1]$. Here we compare simulations with $\zeta=1$ (sol), $\zeta=1/2$ (mix), and $\zeta=0$ (comp). A detailed study of the forcing dependence of the $b$-parameter entering the expression for the variance of the density PDF, Equations~(\ref{eq:sigs}) and~(\ref{eq:sigsmacha}), is provided in \citet[][Figure~8]{FederrathDuvalKlessenSchmidtMacLow2010}, where they measure $b$ as a function of the forcing parameter $\zeta$.

\subsection{Sink Particles and Resolution Criteria} \label{sec:sinks}
In order to model collapse and accretion of star-forming gas in the simulations, we use a subgrid model called `sink particles', which is a method originally invented by \citet{BateBonnellPrice1995} for Smoothed Particle Hydrodynamics, and first adopted for Eulerian, AMR simulations by \citet{KrumholzMcKeeKlein2004}. In \citet{KrumholzMcKeeKlein2004}, a Lagrangian sink particle is introduced, if the gas reaches a given density. However, sink particles are supposed to represent bound objects that are going into collapse, and thus, a density threshold as the only criterion for sink particle creation is insufficient \citep{FederrathBanerjeeClarkKlessen2010}. Based on the ideas of \citet{BateBonnellPrice1995} and \citet{KrumholzMcKeeKlein2004}, we use an advanced AMR-based approach for sink particles, in which only bound and collapsing gas is accreted, thus avoiding the creation of spurious sink particles \citep[for a detailed analysis, see][]{FederrathBanerjeeClarkKlessen2010}. The key feature of this approach is to define a control volume around cells that exceed the density threshold set by the resolution criterion to avoid artificial fragmentation. \citet{TrueloveEtAl1997} found that the Jeans length must be resolved with at least 4 grid cells to avoid artificial fragmentation, leading to a resolution-dependent density threshold criterion for the creation of sink particles:
\begin{equation} \label{eq:rhosink}
\rhosink = \frac{\pi\cs^2}{4G\,\rsink^2}\,,
\end{equation}
where the sink particle accretion radius $\rsink$ is set to 2.5 grid-cell lengths at the maximum level of refinement, corresponding to half a Jeans length at $\rhosink$, such that the Jeans length is still resolved with 5 grid cells prior to potential sink particle creation to avoid artificial fragmentation. Grid cells exceeding the density threshold given by Equation~(\ref{eq:rhosink}), however, do not form sink particles right away. First, a spherical control volume with radius $\rsink$ is defined around the cell exceeding $\rhosink$ within which additional checks for gravitational instability and collapse are performed. We check whether the gas
\begin{itemize}
\item is on the highest level of refinement,
\item is converging from all directions in the rest frame of the central cell (negative radial velocity),
\item is at a local gravitational potential minimum,
\item is bound $\left(\left|E_\mathrm{grav}\right|>E_\mathrm{kin}+E_\mathrm{th}+E_\mathrm{mag}\right)$,
\item is Jeans-unstable, and
\item is not within $\rsink$ of an existing sink particle.
\end{itemize}
If all these checks are passed, a sink particle is created in the center of the control volume \citep[see][]{FederrathBanerjeeClarkKlessen2010}. This procedure avoids spurious sink particle formation, and allows us to trace only truly collapsing and star-forming gas. Given the checks above, it is clear that in some cases, a sink particle is not necessarily formed even though the density threshold is exceeded. This does not mean, however, that such gas would be subject to artificial gravitational fragmentation. Since the checks did not allow sink particle creation, the gas in the control volume was not collapsing and/or not bound, so there is no need to worry about artificial fragmentation at this stage, even though the density threshold was exceeded. This can happen quite frequently in supersonic turbulence because shocks can push the gas density above the threshold, even though this gas is not necessarily gravitationally bound after the shock passage.

Once a sink particle is created, it can gain mass by accreting gas from the AMR grid, but only if this gas exceeds the threshold density, is inside the sink particle accretion radius, is bound to the particle, and is collapsing toward it. If all these criteria are fulfilled, the excess mass above the density threshold defined by Equation~(\ref{eq:rhosink}) is removed from the MHD system and added to the sink particle, such that mass, momentum and angular momentum are conserved by construction \citep[see][for details]{FederrathBanerjeeClarkKlessen2010,FederrathBanerjeeSeifriedClarkKlessen2011}.

All contributions to the gravitational interactions between the gas on the grid and the sink particles are computed by direct $N$-body summation over all grid cells and sink particles (gas--sink, sink--gas, and sink--sink), using gravitational spline softening inside the sink particle radius to avoid singularities during close encounters. The softening only affects scales that are anyway below the grid-resolution cutoff set by the sink particle accretion radius. A second-order accurate Leapfrog integrator is used to advance the sink particles on a time step that allows us to resolve close and highly eccentric orbits of sink particles without introducing significant errors on super-resolution grid scales.

\begin{table*}
\caption{Basic Parameters of the Numerical Models of Forced, Supersonic, Self-gravitating, (M)HD Turbulence.}
\label{tab:sims}
\def\arraystretch{1.2}
\setlength{\tabcolsep}{3.5pt}
\begin{tabular}{l|rcccccccc|ccccc}
\hline
\hline
Model & $N_\mathrm{res}$ & Forcing & $\meanrho$ & $L$ & $M_c$ & $\sigma_V$ & $B_0$ & $\beta_0$ & $\alphavircirc$ & $\alphavir$ & $\mach$ & $b$ & $\beta$ & $\macha$\\
& & & $[\g\,\cm^{-3}]$ & $[\pc]$ & $[\msol]$ & $[\km\,\s^{-1}]$ & $[\mu\Gauss]$ & & & & & & \\
(1) & (2) & (3) & (4) & (5) & (6) & (7) & (8) & (9) & (10) & (11) & (12) & (13) & (14) & (15) \\
\hline
01) GT256sM3                       & 256  & sol  & $  5.8\!\times\!10^{-19}$ & $  3.3\!\times\!10^{-1}$ & $  3.1\!\times\!10^{2}$ & $0.59$ & $ 0$ & $\infty$ & $0.07$ & $ 1.4$ & $ 2.9$ & $1/3$ & $\infty$ & $\infty$ \\
02) GT512sM3                       & 512  & sol  & $  5.8\!\times\!10^{-19}$ & $  3.3\!\times\!10^{-1}$ & $  3.1\!\times\!10^{2}$ & $0.59$ & $ 0$ & $\infty$ & $0.07$ & $ 1.4$ & $ 3.0$ & $1/3$ & $\infty$ & $\infty$ \\
03) GT256mM3                       & 256  & mix  & $  5.8\!\times\!10^{-19}$ & $  3.3\!\times\!10^{-1}$ & $  3.1\!\times\!10^{2}$ & $0.61$ & $ 0$ & $\infty$ & $0.08$ & $ 1.1$ & $ 3.1$ & $0.4$ & $\infty$ & $\infty$ \\
04) GT256cM3                       & 256  & comp & $  5.8\!\times\!10^{-19}$ & $  3.3\!\times\!10^{-1}$ & $  3.1\!\times\!10^{2}$ & $0.58$ & $ 0$ & $\infty$ & $0.07$ & $0.46$ & $ 2.9$ & $1$   & $\infty$ & $\infty$ \\
05) GT512cM3                       & 512  & comp & $  5.8\!\times\!10^{-19}$ & $  3.3\!\times\!10^{-1}$ & $  3.1\!\times\!10^{2}$ & $0.58$ & $ 0$ & $\infty$ & $0.07$ & $0.48$ & $ 2.9$ & $1$   & $\infty$ & $\infty$ \\
\hline
06) GT256sM5                       & 256  & sol  & $  3.3\!\times\!10^{-21}$ & $  2.0\!\times\!10^{0 }$ & $  3.9\!\times\!10^{2}$ & $ 1.0$ & $ 0$ & $\infty$ & $ 1.0$ & $ 8.0$ & $ 5.0$ & $1/3$ & $\infty$ & $\infty$ \\
07) GT256mM5                       & 256  & mix  & $  3.3\!\times\!10^{-21}$ & $  2.0\!\times\!10^{0 }$ & $  3.9\!\times\!10^{2}$ & $0.99$ & $ 0$ & $\infty$ & $0.98$ & $ 5.4$ & $ 5.0$ & $0.4$ & $\infty$ & $\infty$ \\
08) GT256cM5                       & 256  & comp & $  3.3\!\times\!10^{-21}$ & $  2.0\!\times\!10^{0 }$ & $  3.9\!\times\!10^{2}$ & $0.91$ & $ 0$ & $\infty$ & $0.82$ & $ 1.5$ & $ 4.5$ & $1$   & $\infty$ & $\infty$ \\
\hline
09) GT128sM10                      & 128  & sol  & $  8.2\!\times\!10^{-22}$ & $  8.0\!\times\!10^{0 }$ & $  6.2\!\times\!10^{3}$ & $ 2.1$ & $ 0$ & $\infty$ & $ 1.1$ & $ 11.$ & $ 10.$ & $1/3$ & $\infty$ & $\infty$ \\
10) GT256sM10                      & 256  & sol  & $  8.2\!\times\!10^{-22}$ & $  8.0\!\times\!10^{0 }$ & $  6.2\!\times\!10^{3}$ & $ 2.1$ & $ 0$ & $\infty$ & $ 1.1$ & $ 12.$ & $ 10.$ & $1/3$ & $\infty$ & $\infty$ \\
11) GT512sM10                      & 512  & sol  & $  8.2\!\times\!10^{-22}$ & $  8.0\!\times\!10^{0 }$ & $  6.2\!\times\!10^{3}$ & $ 2.1$ & $ 0$ & $\infty$ & $ 1.1$ & $ 12.$ & $ 10.$ & $1/3$ & $\infty$ & $\infty$ \\
12) GT512mM10\phantom{B10} (s1)    & 512  & mix  & $  8.2\!\times\!10^{-22}$ & $  8.0\!\times\!10^{0 }$ & $  6.2\!\times\!10^{3}$ & $ 2.1$ & $ 0$ & $\infty$ & $ 1.1$ & $ 4.5$ & $ 11.$ & $0.4$ & $\infty$ & $\infty$ \\
13) GT512mM10B1\phantom{0} (s1)    & 512  & mix  & $  8.2\!\times\!10^{-22}$ & $  8.0\!\times\!10^{0 }$ & $  6.2\!\times\!10^{3}$ & $ 2.1$ & $ 1$ & $ 8.2$ & $ 1.1$ & $ 5.4$ & $ 10.$ & $0.4$ & $ 2.8$ & $ 12.$ \\
14) GT512mM10\phantom{B10} (s2)    & 512  & mix  & $  8.2\!\times\!10^{-22}$ & $  8.0\!\times\!10^{0 }$ & $  6.2\!\times\!10^{3}$ & $ 2.2$ & $ 0$ & $\infty$ & $ 1.2$ & $ 8.4$ & $ 11.$ & $0.4$ & $\infty$ & $\infty$ \\
15) GT512mM10B1\phantom{0} (s2)    & 512  & mix  & $  8.2\!\times\!10^{-22}$ & $  8.0\!\times\!10^{0 }$ & $  6.2\!\times\!10^{3}$ & $ 2.2$ & $ 1$ & $ 8.2$ & $ 1.2$ & $ 9.5$ & $ 11.$ & $0.4$ & $ 1.8$ & $ 10.$ \\
16) GT256mM10\phantom{B10} (s3)    & 256  & mix  & $  8.2\!\times\!10^{-22}$ & $  8.0\!\times\!10^{0 }$ & $  6.2\!\times\!10^{3}$ & $ 2.0$ & $ 0$ & $\infty$ & $ 1.0$ & $ 5.9$ & $ 10.$ & $0.4$ & $\infty$ & $\infty$ \\
17) GT512mM10\phantom{B10} (s3)    & 512  & mix  & $  8.2\!\times\!10^{-22}$ & $  8.0\!\times\!10^{0 }$ & $  6.2\!\times\!10^{3}$ & $ 2.0$ & $ 0$ & $\infty$ & $ 1.0$ & $ 5.9$ & $ 10.$ & $0.4$ & $\infty$ & $\infty$ \\
18) GT512mM10B1\phantom{0} (s3)    & 512  & mix  & $  8.2\!\times\!10^{-22}$ & $  8.0\!\times\!10^{0 }$ & $  6.2\!\times\!10^{3}$ & $ 2.0$ & $ 1$ & $ 8.2$ & $0.97$ & $ 6.4$ & $ 9.9$ & $0.4$ & $ 3.6$ & $ 13.$ \\
19) GT256mM10B3\phantom{0} (s3)    & 256  & mix  & $  8.2\!\times\!10^{-22}$ & $  8.0\!\times\!10^{0 }$ & $  6.2\!\times\!10^{3}$ & $ 1.8$ & $ 3$ & $0.92$ & $0.81$ & $ 8.4$ & $ 9.0$ & $0.4$ & $0.20$ & $ 2.9$ \\
20) GT512mM10B3\phantom{0} (s3)    & 512  & mix  & $  8.2\!\times\!10^{-22}$ & $  8.0\!\times\!10^{0 }$ & $  6.2\!\times\!10^{3}$ & $ 1.8$ & $ 3$ & $0.92$ & $0.83$ & $ 8.7$ & $ 9.1$ & $0.4$ & $0.18$ & $ 2.7$ \\
21) GT256mM10B10 (s3)              & 256  & mix  & $  8.2\!\times\!10^{-22}$ & $  8.0\!\times\!10^{0 }$ & $  6.2\!\times\!10^{3}$ & $ 1.8$ & $10$ & $0.08$ & $0.79$ & $ 6.6$ & $ 8.9$ & $0.4$ & $0.04$ & $ 1.3$ \\
22) GT128cM10                      & 128  & comp & $  8.2\!\times\!10^{-22}$ & $  8.0\!\times\!10^{0 }$ & $  6.2\!\times\!10^{3}$ & $ 1.8$ & $ 0$ & $\infty$ & $0.81$ & $ 1.2$ & $ 9.0$ & $1$   & $\infty$ & $\infty$ \\
23) GT256cM10                      & 256  & comp & $  8.2\!\times\!10^{-22}$ & $  8.0\!\times\!10^{0 }$ & $  6.2\!\times\!10^{3}$ & $ 1.8$ & $ 0$ & $\infty$ & $0.85$ & $ 1.1$ & $ 9.2$ & $1$   & $\infty$ & $\infty$ \\
24) GT512cM10                      & 512  & comp & $  8.2\!\times\!10^{-22}$ & $  8.0\!\times\!10^{0 }$ & $  6.2\!\times\!10^{3}$ & $ 1.9$ & $ 0$ & $\infty$ & $0.87$ & $ 1.1$ & $ 9.4$ & $1$   & $\infty$ & $\infty$ \\
\hline
25) GT256sM20                      & 256  & sol  & $  2.1\!\times\!10^{-22}$ & $  3.2\!\times\!10^{1 }$ & $  9.9\!\times\!10^{4}$ & $ 4.1$ & $ 0$ & $\infty$ & $ 1.0$ & $ 11.$ & $ 20.$ & $1/3$ & $\infty$ & $\infty$ \\
26) GT256mM20                      & 256  & mix  & $  2.1\!\times\!10^{-22}$ & $  3.2\!\times\!10^{1 }$ & $  9.9\!\times\!10^{4}$ & $ 4.2$ & $ 0$ & $\infty$ & $ 1.1$ & $ 4.5$ & $ 21.$ & $0.4$ & $\infty$ & $\infty$ \\
27) GT256cM20                      & 256  & comp & $  2.1\!\times\!10^{-22}$ & $  3.2\!\times\!10^{1 }$ & $  9.9\!\times\!10^{4}$ & $ 4.0$ & $ 0$ & $\infty$ & $ 1.0$ & $0.60$ & $ 20.$ & $1$   & $\infty$ & $\infty$ \\
\hline
28) GT256sM50                      & 256  & sol  & $  3.3\!\times\!10^{-23}$ & $  2.0\!\times\!10^{2 }$ & $  3.9\!\times\!10^{6}$ & $ 10.$ & $ 0$ & $\infty$ & $ 1.1$ & $ 12.$ & $ 52.$ & $1/3$ & $\infty$ & $\infty$ \\
29) GT512sM50                      & 512  & sol  & $  3.3\!\times\!10^{-23}$ & $  2.0\!\times\!10^{2 }$ & $  3.9\!\times\!10^{6}$ & $ 10.$ & $ 0$ & $\infty$ & $ 1.1$ & $ 13.$ & $ 52.$ & $1/3$ & $\infty$ & $\infty$ \\
30) GT256mM50                      & 256  & mix  & $  3.3\!\times\!10^{-23}$ & $  2.0\!\times\!10^{2 }$ & $  3.9\!\times\!10^{6}$ & $ 10.$ & $ 0$ & $\infty$ & $ 1.0$ & $ 7.0$ & $ 51.$ & $0.4$ & $\infty$ & $\infty$ \\
31) GT512mM50                      & 512  & mix  & $  3.3\!\times\!10^{-23}$ & $  2.0\!\times\!10^{2 }$ & $  3.9\!\times\!10^{6}$ & $ 10.$ & $ 0$ & $\infty$ & $ 1.1$ & $ 7.4$ & $ 51.$ & $0.4$ & $\infty$ & $\infty$ \\
32) GT256cM50                      & 256  & comp & $  3.3\!\times\!10^{-23}$ & $  2.0\!\times\!10^{2 }$ & $  3.9\!\times\!10^{6}$ & $ 9.8$ & $ 0$ & $\infty$ & $0.95$ & $0.54$ & $ 49.$ & $1$   & $\infty$ & $\infty$ \\
33) GT512cM50                      & 512  & comp & $  3.3\!\times\!10^{-23}$ & $  2.0\!\times\!10^{2 }$ & $  3.9\!\times\!10^{6}$ & $ 9.9$ & $ 0$ & $\infty$ & $0.99$ & $0.56$ & $ 50.$ & $1$   & $\infty$ & $\infty$ \\
34) GT1024cM50                     & 1024 & comp & $  3.3\!\times\!10^{-23}$ & $  2.0\!\times\!10^{2 }$ & $  3.9\!\times\!10^{6}$ & $ 10.$ & $ 0$ & $\infty$ & $1.00$ & $0.55$ & $ 50.$ & $1$   & $\infty$ & $\infty$ \\
\hline
\end{tabular}
\\
\textbf{Notes.} Column (1): simulation name. Columns (2--10): maximum grid resolution in one direction of the three-dimensional, cubic domain, mode of forcing (solenoidal, mixed, compressive), mean density, linear box size, total mass, velocity dispersion on the box scale, mean magnetic-field strength (in the $z$-direction of the domain), initial plasma $\beta_0$, and virial parameter based on Equation~(\ref{eq:alphavircirc}). Columns (11--15): time-averaged virial parameter based on Equation~(\ref{eq:alphavir}), computed directly from the three-dimensional gas distribution, the sonic Mach number, forcing parameter, ratio of thermal to magnetic pressure (plasma $\beta$), and Alfv\'en Mach number. To guide the eye, horizontal lines separate models with different sonic Mach number.
\end{table*}

\subsection{Initial Conditions, Procedures, and List of Models} \label{sec:ics}
Starting from a uniform density distribution and zero velocities, the forcing term ${\bf F_\mathrm{stir}}$ in Equations~(\ref{eq:mhd}) excites turbulent motions. First, we evolve the MHD equations for two turbulent crossing times, $2T=L/(\mach\cs)$ without self-gravity, in order to establish fully developed, compressible turbulence \citep[e.g.,][]{KlessenHeitschMacLow2000,Klessen2001,HeitschMacLowKlessen2001,FederrathKlessenSchmidt2009,FederrathDuvalKlessenSchmidtMacLow2010,PriceFederrath2010,MicicEtAl2012}. We do not include the gravity terms until $t=2\,T$, in order to avoid that our measurements of the SFR are contaminated by this rather artificial initial transient phase, during which the system is building up a turbulent cascade \citep{SchmidtEtAl2009}. After that, we solve the full system of MHD Equations~(\ref{eq:mhd}) and~(\ref{eq:grav}) including self-gravity and formation of sink particles. For practical purposes, we reset the time $t=2\,T$ to $t=0\,\tff(\meanrho)$, which is the time when turbulence is fully established and star formation is allowed to proceed. We note that this procedure is slightly different from setting up a simulation with power-law velocity scaling drawn from Gaussian random seeds as an initial condition, commonly applied in numerical star formation studies \citep[e.g.,][]{BateBonnellBromm2003,ClarkEtAl2005,KrumholzKleinMcKee2007,PriceBate2008,PriceBate2009,SmithClarkBonnell2008,FederrathBanerjeeClarkKlessen2010,WalchEtAl2010,GirichidisEtAl2011}. In those cases, the initial random velocity field is imposed on top of a given density profile (often constant density or radial power-law distributions), such that density and velocity fields have no causal connection. Here, the initial density and velocity fields at $t=0$ are consistently coupled via the equations of (magneto)hydrodynamics. We also keep driving the turbulence instead of imposing only an initial Gaussian perturbation as in the studies mentioned above.

All our numerical simulations and their basic parameters are listed in Table~\ref{tab:sims}. Each model has a unique name, starting with `GT' (for `GravTurb'), followed by the maximum grid resolution (`128', `256', `512', and `1024'), the forcing type (`s':solenoidal, `m':mixed, and `c':compressive), and the Mach number (`M3', `M5', `M10', `M20', and `M50'). Models with an initially uniform magnetic field in the $z$-direction through the simulation box are additionally denoted with `B1', `B3', and `B10', corresponding to $B_0=1$, 3, and $10\,\mu\Gauss$, respectively. Different random sequences with the same statistical properties for the turbulent forcing are indicated by `(s1)', `(s2)', and `(s3)' at the end of the model name, indicating that random `(seed1)', `(seed2)', or `(seed3)' was used. Columns 2--10 in Table~\ref{tab:sims} list the maximum numerical resolution, type of forcing, mean density $\meanrho$, box size $L$, the total mass $M_c$, large-scale velocity dispersion $\sigma_V$, initial magnetic-field strength $B_0$, initial plasma $\beta_0$, and virial parameter $\alphavircirc$ computed with Equation~(\ref{eq:alphavircirc}).

Columns 11--15 are derived quantities, measured as space and time averages after turbulence is fully established, $t\geq0$, until 20\% of the original cloud mass is accreted onto sink particles, i.e., the star formation efficiency has reached $\sfe=20\%$. We list the average virial parameter $\alphavir$, the sonic Mach number $\mach$, forcing parameter $b$, plasma $\beta$, and Alfv\'en Mach number $\macha$. The instantaneous virial parameter, Equation~(\ref{eq:alphavir}), in column 11 of Table~\ref{tab:sims} is computed as $\alphavir = 2 E_\mathrm{kin} / |E_\mathrm{grav}| = \sum M_i v_i^2 / |\sum M_i\Phi_{\mathrm{gas},i}|$ from the gravitational potential $\Phi_\mathrm{gas}$ returned by the Poisson solver (see Section~\ref{sec:methods}), as a sum over all grid cells $i$ with mass $M_i$ and velocity $v_i$. We note that this is different from the value $\alphavircirc$ obtained from Equation~(\ref{eq:alphavircirc}) and listed in column 10, which assumes a homogenous, spherical density distribution. In contrast, we obtain highly inhomogeneous density distributions in our compressible, turbulent clouds. We thus prefer to compute $\alphavir$ based on the three-dimensional density field as explained above\footnote{Note that a similar approach is used in \emph{Herschel} observations by \citet{AndreEtAl2010} to estimate the stability of interstellar filaments. That is based on column density instead of volume density, but takes the spatial (projected) distribution of matter into account, rather than estimating the dynamical state of the cloud based on the spherical, uniform-density approximation in Equation~(\ref{eq:alphavircirc}).}. In analogy, the sonic and Alfv\'en Mach numbers, as well as $\beta$ are computed as spatial root-mean-squared averages over all cells in the simulation box as a function of time, followed by averaging over time. We will show in the next section that this approach is justified because we find that all those parameters do not vary significantly with time during star formation. The value of the forcing parameter $b$ was not determined by averaging because it was already measured in \citet[][Figure~8]{FederrathDuvalKlessenSchmidtMacLow2010}, giving best-fit values $b=1/3$, 0.4, and 1 for solenoidal, naturally-mixed, and compressive forcing of the turbulence, respectively.

We do not include any data or discussion of the state of the clouds after $\sfe=20\%$ is reached because at that point in time, local feedback processes would have likely altered the subsequent evolution of the clouds so drastically that we cannot trust our results for higher $\sfe$. Even before that, inclusion of feedback processes might change the results, at least locally. For example, we expect the amount of accreted gas to be reduced, if feedback were included \citep[e.g.,][]{WangEtAl2010,PetersEtAl2011}. This fact can be accounted for by adjusting the local efficiency parameter $\eps$ introduced in Equation~(\ref{eq:sfrffbasic}) to values $\eps<1$ for all the models discussed here. We get back to this issue when we compare our simulations with the observational data in Section~\ref{sec:obs}.

The basic model parameters in Table~\ref{tab:sims} were chosen to roughly follow observed properties of molecular clouds, covering a range of cloud sizes $L\approx0.3$--$200\,\pc$, masses $M_c\approx300$ to $4\times10^6\,\msol$, and velocity dispersions $\sigma_V\approx0.6$--$10\,\km\,\s^{-1}$ \citep[e.g.,][]{Larson1981,SolomonEtAl1987,FalgaronePugetPerault1992}, with typical cloud scalings summarized and discussed in \citet{MacLowKlessen2004} and \citet{McKeeOstriker2007}. However, even though most real clouds may roughly follow such an average scaling, the scatter around that average is typically about an order of magnitude or more in terms of mass, density, and velocity dispersion for a given cloud size \citep[e.g.,][]{HeyerEtAl2009,RomanDuvalEtAl2010}. The procedure used here to determine the initial cloud parameters in the simulations is as follows. First, for a given target Mach number, we determine the appropriate size of the cloud by inverting the observed velocity dispersion--size relation given by Equation~(\ref{eq:sigmav}). Having the size and velocity dispersion, we then set the virial parameter given by Equation~(\ref{eq:alphavircirc}) to a value close to unity. The only exceptions are the $\mach\sim3$ models, where we set it to $\alphavircirc\approx0.07$ because this turned out to give actual virial parameters $\alphavir$ closer to unity after the turbulence had been fully established (compare columns 10 and 11 in Table~\ref{tab:sims}). Using the initial guess of $\alphavircirc$, we then solve for the mass of the cloud, by inverting Equation~(\ref{eq:alphavircirc}). From the mass and size, we compute the mean density of the model cloud.

It is important to note that the actual virial parameter obtained after two turbulent crossing times can be up to an order of magnitude different from the initial guess provided by Equation~(\ref{eq:alphavircirc}), depending on the Mach number and forcing of the model (see Table~\ref{tab:sims}). This is because the density distribution in the state of fully developed supersonic turbulence is highly inhomogeneous and is not well described by Equation~(\ref{eq:alphavircirc}). Thus, we do not know the virial parameter that arises in the regime of fully developed turbulence a priori. The $\alphavir$ in the turbulent phase is typically higher (except for the compressive forcing cases at high Mach numbers, $\mach\sim20$ and 50) than the one computed from Equation~(\ref{eq:alphavircirc}), also because we use periodic boundary conditions. This reduces the gravitational binding energy of the system compared to an isolated system (as assumed in Equation~\ref{eq:alphavircirc}). Real clouds are neither periodic nor isolated, but using periodic boundaries, we mimic the effects of the surrounding medium on the region studied in our computational boxes (discussed further in Section~\ref{sec:limitations}). We emphasize that the virial parameters obtained here are consistent with observations, given that observational estimates of $\alphavir$ are usually obtained based on Equation~(\ref{eq:alphavircirc}) or column-density versions of it.

Magnetic-field strengths for the MHD simulations were chosen to be consistent with the range observed in clouds \citep[e.g.,][]{Crutcher1999,CrutcherEtAl2010}. We vary the magnetic field for simulations with mixed forcing and fixed sonic Mach number of $\mach\sim10$, which gives us a good indication of the role of magnetic fields for typical molecular cloud properties. \citet{HeilesTroland2005} and \citet{CrutcherEtAl2010} show that most clouds with number densities in the range \mbox{$10$--$10^4\,\cm^{-3}$} have magnetic-field strengths in the range \mbox{$B_z\approx1$--$10\,\mu\Gauss$}, with an apparent peak of the distribution at around $B_z\approx3\,\mu\Gauss$. Our MHD simulations have mean densities of about $200\,\cm^{-3}$, so we decided to compare models with $B_z = 1$, 3, and $10\,\mu\Gauss$, in order to cover the observed range of line-of-sight magnetic-field strengths.

\section{Simulation Results} \label{sec:simresults}
After the initial turbulent state has been established by driving for two crossing times (see Section~\ref{sec:ics}) in each simulation, we study the subsequent evolution under the influence of self-gravity by looking at column density projections of the simulated clouds and their magnetic-field morphology (Section~\ref{sec:morphology}). We then discuss the time evolution of $\alphavir$, $\mach$, $\macha$, and $\sfe$ and measure the SFR in Section~\ref{sec:timeevol}.
%A detailed analysis of density and column density PDFs, and Fourier spectra is provided in a companion paper (Federrath \& Klessen 2012, Paper II, in preparation).

\begin{figure*}[t]
\centerline{
\includegraphics[width=0.99\linewidth]{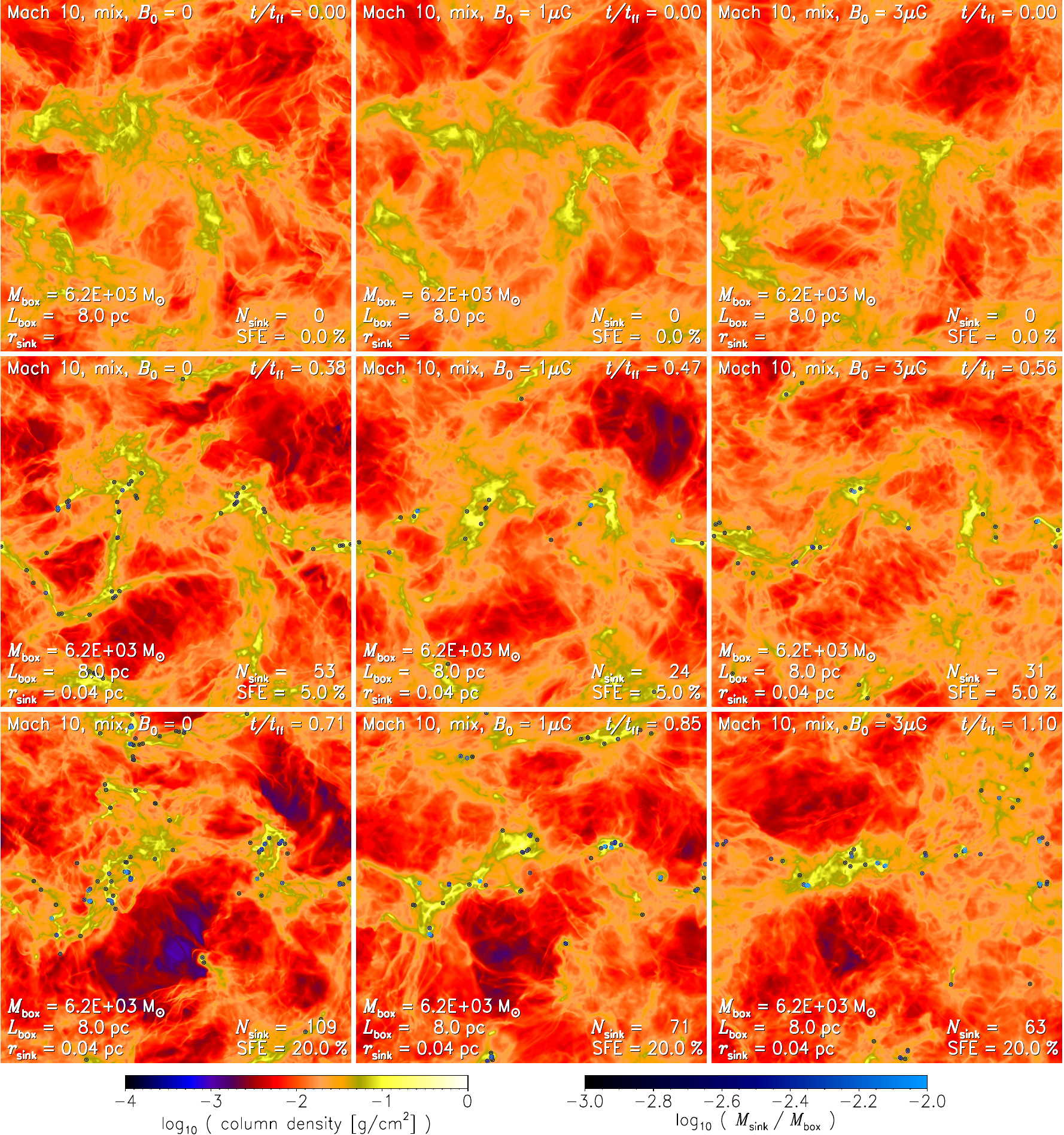}
}
\caption{Time evolution of column density projections of the simulations with mixed forcing at $\mach=10$ with initial magnetic-field strengths $B_0=0\,\mu\Gauss$ (left panels), $B_0=1\,\mu\Gauss$ (middle panels), and $B_0=3\,\mu\Gauss$ (right panels). The times shown correspond to the initial, fully developed turbulent state, $t=0$ (top panels), and when the star formation efficiency reached $\sfe=5\%$ and $20\%$ (middle and bottom panels, respectively), i.e., 5\% and 20\% of the gas was accreted by sink particles (shown as circles with the sink particle radius). The higher the magnetic field, the slower the star formation (see time in the top right corner), and the fewer sink particles form (see bottom right corner of each image) due to the increasing magnetic pressure.}
\label{fig:mhdimages}
\end{figure*}

\subsection{Cloud and Magnetic-field Morphology} \label{sec:morphology}

\subsubsection{Effects of the Magnetic Field}

Figure~\ref{fig:mhdimages} shows the time evolution of column density snapshots (from top to bottom) for models with mixed forcing at $\mach=10$ and $512^3$ resolution for initial magnetic fields $B_0=0$, 1, and $3\,\mu\Gauss$ (left, middle, and right panels). Key initial parameters (box size, total mass, etc.), the time in units of $\tff(\meanrho)$, the $\sfe$, and the number of sink particles formed are given in each panel. The top row shows the gas at $t=0$, i.e., when turbulence is fully developed and self-gravity is switched on. We see shocks and large-scale structure induced by the large-scale turbulence with column density contrasts ranging over more than four orders of magnitude. Comparing the purely HD run (left) with the two magnetized runs (middle and right), we see that shocks become smoother and density contrasts slightly decrease as the magnetic-field strength increases. This is because magnetic fields act like a cushion, reducing density fluctuations, due to the additional magnetic pressure parameterized either by plasma $\beta$ or the Alfv\'en Mach number $\macha$ \citep[see Equation~\ref{eq:sigs} or~\ref{eq:sigsmacha}, and][]{MolinaEtAl2012}, the time-averaged values of which are given in Table~\ref{tab:sims}. At later times, the gas starts collapsing locally at sites previously compressed by the supersonic turbulence, at which point local filaments become more and more massive as they accrete gas from the surrounding and eventually become so dense that these cores have to be replaced with sink particles, allowing us to advance the simulations to later times (see Section~\ref{sec:sinks}). The radius $\rsink$ of the sink particles is determined by the numerical resolution constraint, and is given in each panel, as soon as sink particles have formed. Our resolution is insufficient to resolve individual stars, but the sink particles can be regarded as dense, bound cores in our simulations.

\begin{figure}[t]
\centerline{
\includegraphics[width=0.8\linewidth]{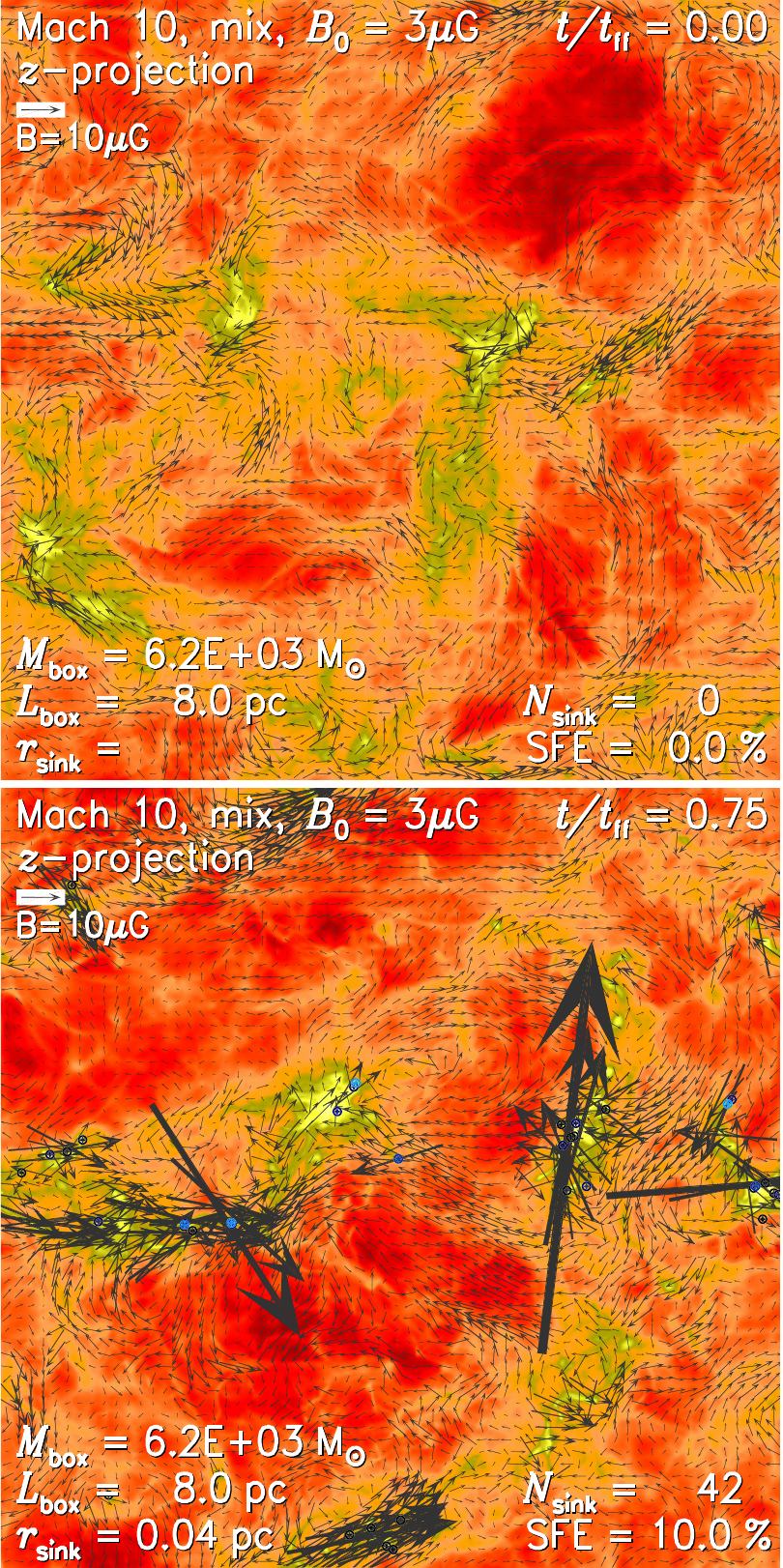}
}
\caption{Column density and magnetic-field vectors in simulation model \#20 (GT512mM10B3; see Table~\ref{tab:sims}) with $B_0=3\,\mu\Gauss$ for $t/\tff=0$ (top) and $\sfe=10\%$ (bottom). The magnetic field is amplified in the core and cluster regions, where compression and turbulent dynamo action both contribute to increasing the field strength locally \citep[][]{SurEtAl2010,FederrathSurSchleicherBanerjeeKlessen2011}. The magnetic field frequently changes direction and strength in the cores in this model of super-Alfv\'enic turbulence with $\macha\approx2.7$ (see Table~\ref{tab:sims}), best seen in the movies (see additional online material). The colors are identical to Figure~\ref{fig:mhdimages}.}
\label{fig:mhdimagesB}
\end{figure}

Comparing the runs with different magnetic-field strengths in Figure~\ref{fig:mhdimages}, we see two important effects with increasing magnetic field: (1) a reduction of fragmentation, i.e., fewer sink particles have formed by the end of the simulations at $\sfe=20\%$ and (2) reaching a given $\sfe$ takes longer, i.e., the core formation rate and hence the SFR are reduced. For instance, when the SFE has reached 20\%, the runs with $B_0=0$, 1, and $3\,\mu\Gauss$ have formed 109, 71, and 63 sink particles in $0.71$, $0.85$, and $1.1\,\tff(\meanrho)$, respectively.

The higher the magnetic field, the larger the topologically-connected structures, compared to the more fragmented and dispersed filaments in the purely hydrodynamical run. Comparing numerical simulations and observations with filament-tracking tools \citep[e.g.,][]{AndreEtAl2010,MenshchikovEtAl2010,ArzoumanianEtAl2011,HillEtAl2011,SchneiderEtAl2012} or polarization analyses \citep[e.g.,][]{BurkhartLazarianGaensler2012} may eventually help to reveal the role of magnetic fields. In particular, the orientation of magnetic fields might tell us about its dynamical influence \citep{SchneiderEtAl2010,LiHenning2011,PerettoEtAl2012}. In Figure~\ref{fig:mhdimagesB}, we show the time evolution of column density snapshots with local magnetic-field vectors computed by a mass-weighted average along the line of sight superimposed, for the run with $B_0=3\,\mu\Gauss$ for $t/\tff=0$ (top) and $\sfe=10\%$ (bottom). The magnetic field grows due to compression of the field lines and due to dynamo action \citep{SurEtAl2010,FederrathSurSchleicherBanerjeeKlessen2011,BertramEtAl2012}, particularly in regions where dense cores accumulate and form clusters. The magnetic field is very intermittent and shows no particularly preferred direction in the cluster centers because the gas motions are so chaotic that the magnetic-field direction changes frequently. The magnetic field is of moderate strength compared to the turbulence in this case, shown by the average super-Alfv\'enic Mach number in this simulation, $\macha\approx2.7$ (see Table~\ref{tab:sims}). The field strengths are consistent with observations in typical molecular clouds. On scales larger than molecular clouds and on Galactic scales though, the turbulence might be trans-Alfv\'enic rather than super-Alfv\'enic, which would naturally lead to more aligned magnetic field structures there \citep[e.g.,][]{BeckEtAl1996,HeilesTroland2005,LiHenning2011}.

\begin{figure*}[t]
\centerline{
\includegraphics[width=0.95\linewidth]{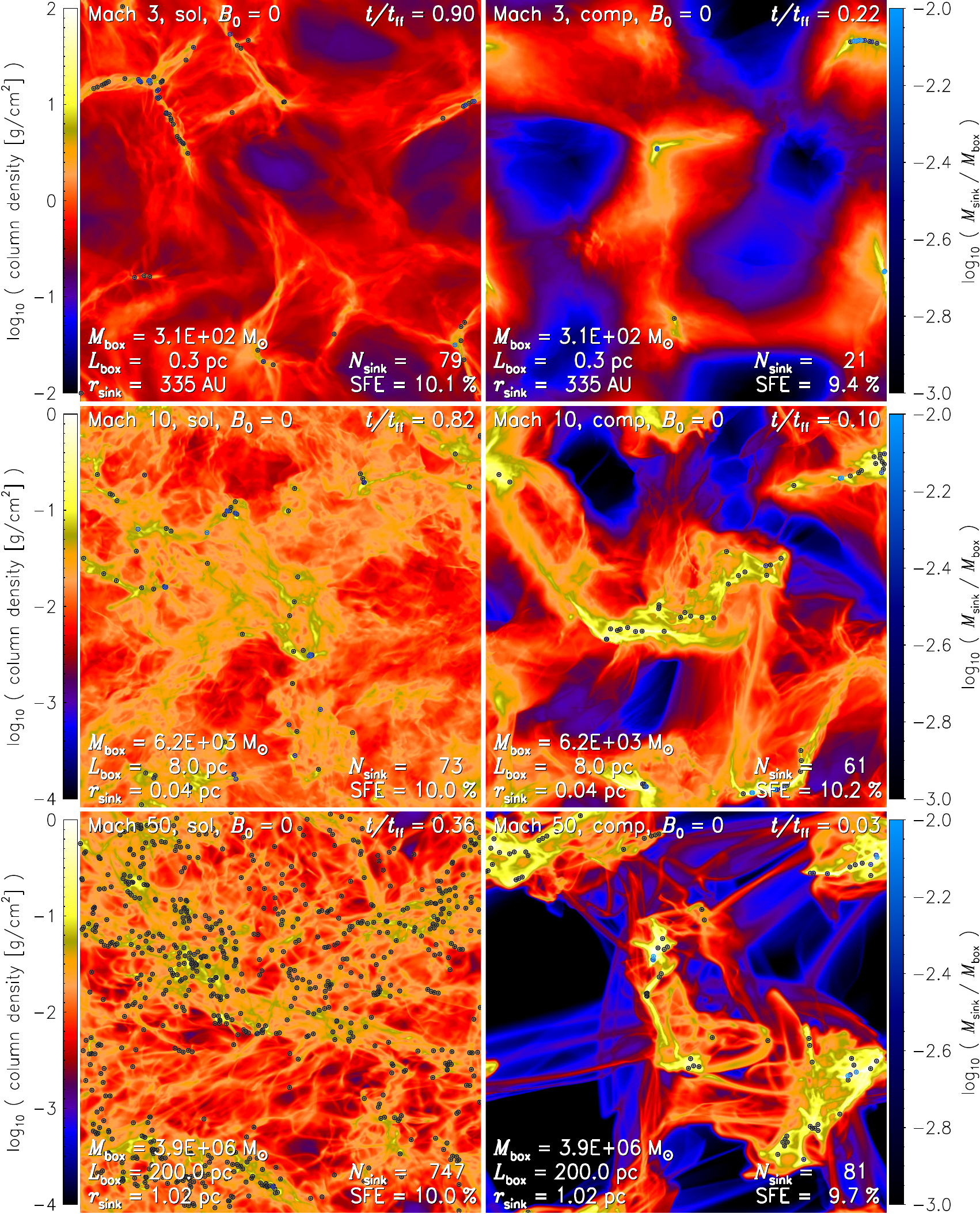}
}
\caption{Column density projections of the simulations with solenoidal forcing (left panels) and compressive forcing (right panels) for Mach numbers $\mach\sim3$ (top), $\mach\sim10$ (middle), and $\mach\sim50$ (bottom), when 10\% of the initial gas mass is accreted by sink particles (shown as circles with the sink particle radius). The mass and size of the three-dimensional domains, and the number of sink particles formed, are given in each panel. In addition to the morphological differences between the forcings for a given Mach number, the elapsed time in units of the freefall time at the mean density (see label in the top right corner of each panel) is significantly different between the two extreme cases of turbulent forcing, suggesting extremely different star formation rates for solenoidal and compressive forcing.}
\label{fig:hdimages}
\end{figure*}

\subsubsection{Effects of Turbulent Forcing and Sonic Mach Number}

After having looked at the time evolution of column density snapshots in mixed, $\mach\sim10$ simulations, we now focus on the gas morphology when $\sfe=10\%$, representing a typical molecular cloud value, but comparing different forcing and sonic Mach numbers. Figure~\ref{fig:hdimages} shows column density projections of the $512^3$ runs with solenoidal forcing (left panels) and compressive forcing (right panels) at $\mach\sim3$ (top), $\mach\sim10$ (middle), and $\mach\sim50$ (bottom). Note the different length and mass scales probed in these images, with box sizes of $L=0.3$, 8 and $200\,\pc$, and masses of $M_c=310$, $6.2\times10^3$ and $3.9\times10^6\,\msol$, respectively. Since the resolution is fixed, the sink particle radii vary from $\rsink=335\,\AU$ over $\rsink=0.04\,\pc$, up to $1\,\pc$. Thus, neither of those represents stars, but rather star clusters in the largest-scale runs and potentially protostellar accretion envelopes in the smallest-scale runs. The scale and mass sequence from the bottom to the top panels in Figure~\ref{fig:hdimages} can be interpreted as zooms into patches of larger-scale runs and re-simulating these patches with higher resolution in successively smaller boxes. Clearly, these images emphasize how artificial this kind of numerical experiment is, yet real molecular clouds exhibit similar hierarchical structures \citep{FalgaronePugetPerault1992,OssenkopfMacLow2002}, often characterized as fractals \citep{Scalo1990,ElmegreenFalgarone1996,StutzkiEtAl1998,SanchezEtAl2005,RomanDuvalEtAl2010}. The fractal dimension $D$ inferred from different techniques ($\Delta$-variance, box counting, \mbox{mass--size} relation, and perimeter-area method) was shown to vary between $D\approx2.6$ for purely solenoidal and $D\approx2.3$ for purely compressive forcing, in the range of observational determinations \citep{FederrathKlessenSchmidt2009}, and is consistent with theoretical ideas to explain the slope of the stellar IMF \citep{ChabrierHennebelle2011}. As can be seen in Figure~\ref{fig:hdimages}, compressive forcing produces more sheet-like structures (planar shocks), while solenoidal forcing produces more volume-filling structures, providing a visual explanation for the dependence of $D$ on the forcing.

Besides the morphological distinctions, the most striking difference between solenoidal and compressive forcings is the timescale of core and star formation (compare $t/\tff$ in the upper right corner of each panel in Figure~\ref{fig:hdimages}). For fixed Mach number, cloud size, and mass, compressive forcing accelerates the conversion of gas into stars compared to solenoidal forcing by factors of 4, 8, and 12 for the $\mach\sim3$, 10, and 50 runs, respectively, when $\sfe\sim10\%$. This result emphasizes the important role of the turbulent forcing for setting the SFR.

\subsection{Time Evolution of $\alphavir$, $\mach$, $\macha$, and $\sfe$} \label{sec:timeevol}

\subsubsection{Effects of the Forcing, Random Seed, and Resolution}

\begin{figure}[t]
\centerline{
\includegraphics[width=0.99\linewidth]{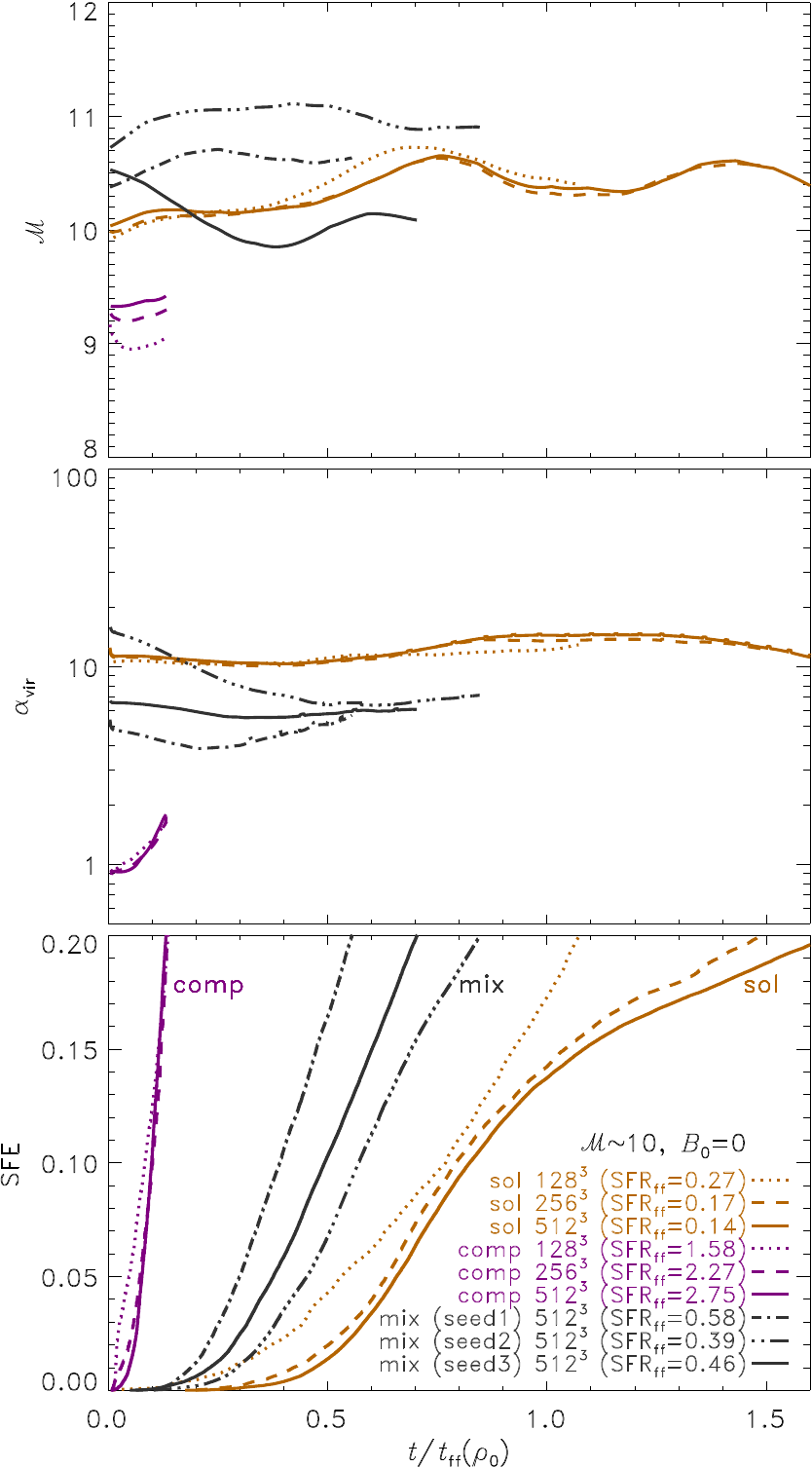}
}
\caption{Time evolution of the rms Mach number $\mach$ (top), the virial parameter $\alphavir$ (middle), and the star formation efficiency $\sfe$ (bottom) for models with fixed Mach number $\mach\sim10$, and without magnetic field, but with solenoidal forcing (gold) and compressive forcing (violet) at numerical resolutions of $128^3$ (dotted), $256^3$ (dashed), and $512^3$ (solid), as well as for three mixed-forcing models (black), each forced with a different random number sequence at fixed $512^3$ resolution: seed1 (dash-dotted), seed2 (triple-dot-dashed), and seed3 (solid). The time axis is scaled in units of the freefall time at the mean density of the respective simulation (see Table~\ref{tab:sims}). Values of $\sfrff$, measured from linear fits to the $\sfe$--time curves (bottom panel) in the range $\sfe=[4\%,\,20\%]$ are given for each model in the legend.}
\label{fig:accr_res}
\end{figure}

We now turn to the detailed time evolution and determination of the SFR in the simulations. Figure~\ref{fig:accr_res} shows the time evolution of dynamical quantities $\mach$, $\alphavir$, and $\sfe$ for models with $\mach\sim10$ and $B_0=0$ for solenoidal and compressive forcing at numerical resolutions of $128^3$, $256^3$, and $512^3$ grid cells, and for mixed forcing with three different random seeds of the turbulent forcing. The Mach number (top panel) shows variations of order 10\% around the target Mach number of $\mach\sim10$ for each simulation, and some systematic variations with different forcing and random seeds. The differences between solenoidal, mixed, and compressive forcings are caused by stronger dissipation with more compressive forcing, requiring a higher forcing amplitude to reach the same Mach number than in solenoidal forcing. We adjust the amplitude of the forcing such that the gas reaches a given Mach number in the fully developed turbulent phase. Since the value of $\mach$ depends on nonlinear dissipation properties, i.e., strengths of shocks and amount of vorticity generated, and thus on the Mach number of the turbulence \citep{FederrathEtAl2011PRL}, the time-averaged rms Mach number for a given forcing amplitude cannot be predicted a priori and must be adjusted iteratively by running test simulations with different forcing amplitude and measuring the time-averaged rms Mach number, resulting in some deviation of the actual Mach number from the target Mach number (see the time-averaged $\mach$ for each model in Table~\ref{tab:sims}). The temporal fluctuations and the differences between random seeds, however, are purely statistical. In order to compare our simulation data with the analytic theories, we thus always use the volume- and time-averaged quantities entering the theoretical models from Section~\ref{sec:theoryresults}.

The middle panel of Figure~\ref{fig:accr_res} shows $\alphavir(t)$. As for $\mach(t)$, the resolution dependence is only marginal, and significantly less than the statistical fluctuations \citep[see also][showing that one-point statistics are typically well converged with grid resolutions of $256^3$ cells]{KitsionasEtAl2009,PriceFederrath2010,KritsukEtAl2011Codes}. This demonstrates that the length scales of the dominant gravitational structures are resolved well enough in our present numerical experiments and that our definition of $\alphavir$ is robust with respect to changes in the numerical resolution.

The difference of $\alphavir\equiv2\,E_\mathrm{kin}/|E_\mathrm{grav}|$ between the forcings deserves some attention. While the $\mach\sim10$ runs with solenoidal forcing have $\alphavir\approx12$, the compressive ones have $\alphavir\approx1.1$ (see Table~\ref{tab:sims}), even though the Mach number is similar and the mass of the clouds is identical. In Figure~\ref{fig:hdimages} we saw that compressive forcing produces more locally compressed structures than solenoidal forcing, resulting in an overall higher gravitational binding energy $|E_\mathrm{grav}|$ compared to solenoidal forcing. The total kinetic energy $E_\mathrm{kin}$ on the other hand is the same within a factor of $\sim2$, which means that the factor of $\sim10$ difference in $\alphavir$ is primarily due to the difference in $|E_\mathrm{grav}|$. This shows that comparing simple theoretical estimates of the virial parameter, solely based on the total mass as a measure for $|E_\mathrm{grav}|$ (as, e.g., assumed in Equation~\ref{eq:alphavircirc}), should be considered with great caution because such an estimate ignores the internal structure of the clouds. Thus, we prefer to estimate $\alphavir$ based on the actual spatial distribution, as we have done in Figure~\ref{fig:accr_res} and in Table~\ref{tab:sims} for all models. We emphasize that this direct comparison of $\alphavircirc$ with $\alphavir$ performed here means that observational estimates of the virial parameter based on global measures such as described by Equation~(\ref{eq:alphavircirc}) or alike are only accurate within an order of magnitude. Measurements of gravitational (in)stability based on the actual column density distribution of filaments in \emph{Herschel} observations of the Gould Belt (GB) by \citet{AndreEtAl2010}, for example, are thus likely more accurate and meaningful than estimates based on uniform-density, spherical approximations such as Equation~(\ref{eq:alphavircirc}).

The bottom panel of Figure~\ref{fig:accr_res} shows the time evolution of the total mass accreted by sink particles, divided by the total cloud mass, i.e., the $\sfe$. We measure the slope of these curves by fitting a linear function in the interval $\sfe=[4\%\,,20\%]$, which gives the $\sfrff$ for each model quoted in the legend. We choose to set the lower limit of the fit range to $\sfe=4\%$ because the initial accretion phase is highly nonlinear with a fast increase in slope, after which the accretion becomes roughly linear in time, such that the slope is reasonably well defined for most of the models.

First, we study the dependence of $\sfrff$ on the random seed. The three models with mixed forcing and different random seeds (seed1, seed2, seed3) exhibit variations in $\sfrff$ by a factor of 1.5. However, other seeds might deviate further from this, such that the factor 1.5 in $\sfrff$ is a lower limit for the uncertainty introduced by the random seed. When we compare mixed-forcing models at $\mach\sim10$ with different magnetic-field strengths later, we always compare runs with seed3 because the $\sfrff$ for seed3 is in between the ones measured for seed1 and seed2, thus giving the best average behavior for the data at hand.

Finally, we investigate the resolution dependence of our simulations with solenoidal and compressive forcings in Figure~\ref{fig:accr_res}. For resolutions of $128^3$, $256^3$, and $512^3$ grid cells, we find that $\sfrff=0.27$, 0.17, and 0.14 for solenoidal forcing, and $\sfrff=1.58$, 2.27, and 2.75 for compressive forcing, respectively. Thus, with increasing resolution, $\sfrff$ is decreasing for solenoidal forcing, but increasing for compressive forcing. The difference in $\sfrff$ between $128^3$ and $256^3$ is a factor of 0.63 for solenoidal forcing, and a factor of 1.44 for compressive forcing. These factors become smaller when we compare the $256^3$ with the $512^3$ simulations, giving factors of 0.82 for solenoidal forcing and 1.21 for compressive forcing. Thus our results converge with increasing resolution. Moreover, we can estimate $\sfrff$ in the limit of infinite resolution from extrapolating the convergence behavior. Doing this, we see that our measurements of $\sfrff$ at $128^3$ resolution are converged only within a factor of about 2.5, so we discard the two $128^3$ simulations (GT128sM10 and GT128cM10) in all the following. In contrast, the $256^3$ data are converged to within a factor of 1.5 for both solenoidal and compressive forcings, which is similar to the uncertainty introduced by varying the random seed as discussed in the previous paragraph. Thus, differences larger than a factor of 1.5 in $\sfrff$ between models with different physical parameters are likely of physical rather than numerical or statistical origin. For instance, the $\sfrff$ for compressive forcing with $\mach\sim10$ is more than an order of magnitude larger than the $\sfrff$ of the respective solenoidal simulation, demonstrating the physical importance of the turbulent forcing for controlling the SFR.

\subsubsection{Effects of Increasing the Sonic Mach Number}

\begin{figure}[t]
\centerline{
\includegraphics[width=0.99\linewidth]{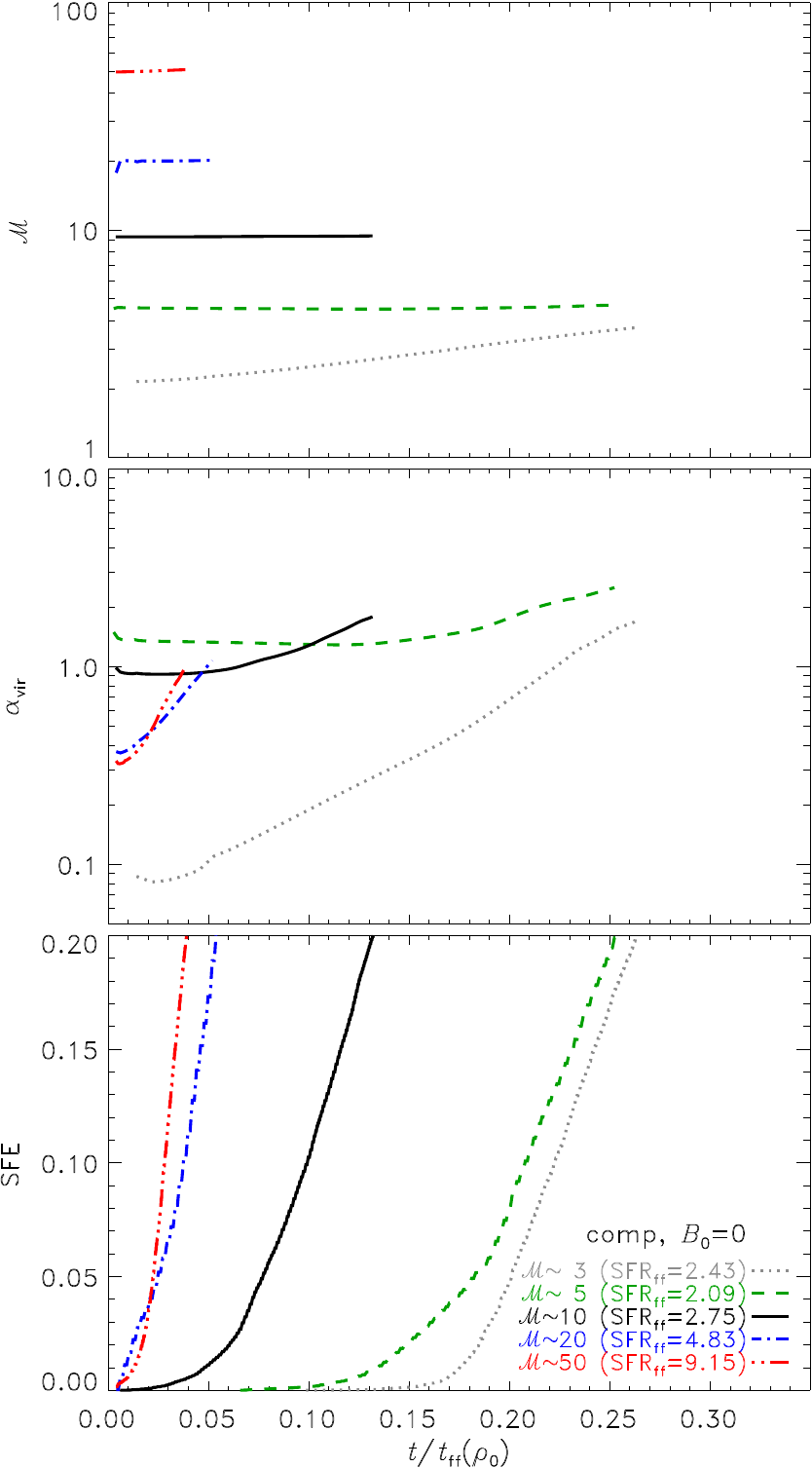}
}
\caption{Same as Figure~\ref{fig:accr_res}, but for compressive-forcing models with sonic Mach number, $\mach\sim3$, 5, 10, 20, and 50.}
\label{fig:accr_mach}
\end{figure}

Having looked at models with different forcing, random seed, and resolution, we now study models with varying Mach number. Figure~\ref{fig:accr_mach} shows the same as Figure~\ref{fig:accr_res}, but for compressive-forcing models with Mach $\mach\sim3$, 5, 10, 20, and 50. The Mach number increases slightly with time in all models, which is caused by local accelerations during collapse. This only accounts for a few percent at most. The exception is the $\mach\sim3$ model, for which $\mach$ increases by almost a factor of two (top panel). This is because the $\mach\sim3$ model is more gravitationally unstable initially, indicated by the virial parameter (middle panel), which increases to around unity, similar to the other models. The $\sfrff$ generally increases with Mach number due to the stronger local compressions created at higher $\mach$ (bottom panel). The only exception is again the $\mach\sim3$ model, which has a slightly higher $\sfrff$ than the $\mach\sim5$ model because the $\mach\sim3$ model is more unstable and starts collapsing globally, while this is not the case in the other models. The difference of $\sfrff$ between $\mach\sim5$ and $50$ is about a factor of 4.4.

\subsubsection{Effects of Increasing the Magnetic-field Strength}

\begin{figure}[t]
\centerline{
\includegraphics[width=0.99\linewidth]{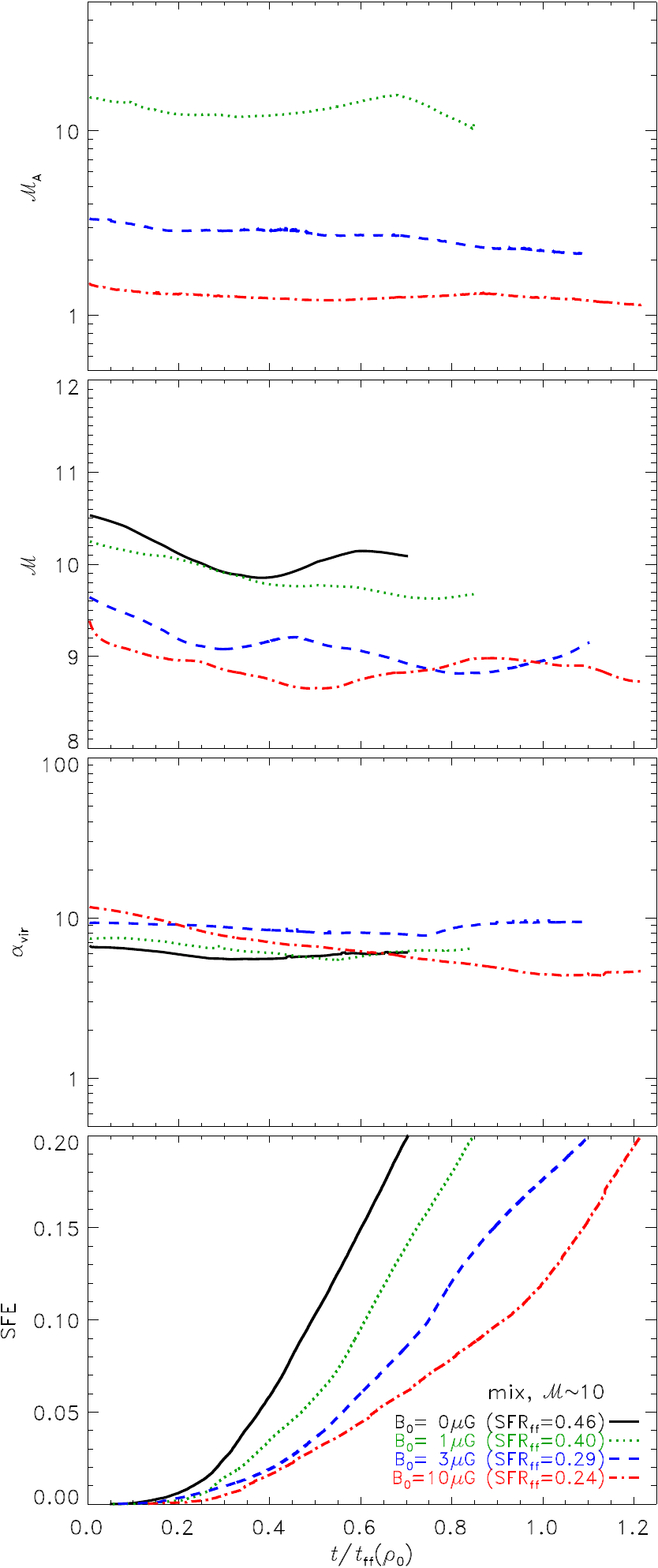}
}
\caption{Same as Figure~\ref{fig:accr_res}, but for mixed-forcing models ($b=0.4$) at $\mach\sim10$ with different initial magnetic-field strengths $B_0=0$, 1, 3, and $10\,\mu\Gauss$. The additional panel on the top shows the time evolution of the Alfv\'en Mach number in the MHD simulations.}
\label{fig:accr_mag}
\end{figure}

Finally, in Figure~\ref{fig:accr_mag} we investigate the time evolution of models with different initial magnetic-field strengths, $B_0=0$, 1, 3, and $10\,\mu\Gauss$ (initial plasma $\beta_0=8.2$, 0.92, and 0.082; see Table~\ref{tab:sims}). The panels are the same as in Figure~\ref{fig:accr_res}, except for an additional panel on the top, showing the Alfv\'en Mach number $\macha$. Apart from some temporal fluctuations, $\macha$, $\mach$, and $\alphavir$ are fairly constant over time. Both $\macha$ and $\mach$ show some minor systematic decrease, which is caused by dynamo action, amplifying the magnetic field by converting turbulent energy into magnetic energy \citep{BrandenburgSubramanian2005}. Most of the dynamo action, however, took place already during the first two turbulent crossing times, $t<0\,\tff$, during which the turbulence becomes fully established (compare columns 9 and 14 of Table~\ref{tab:sims}). The dynamo is nearly saturated at $t=0$ with only very slow linear amplification happening afterward. In addition, field lines are compressed during local collapse, amplifying the field further in dense cores and clusters (see Figure~\ref{fig:mhdimagesB}). 

Most importantly, the last panel of Figure~\ref{fig:accr_mag} shows that the $\sfrff$ decreases monotonically with increasing magnetic field because of the stabilizing effect of the magnetic pressure. The strongest magnetic field case studied here ($B_0=10\,\mu\Gauss$, $\macha\approx1.3$) has an $\sfrff\approx0.24$, which is almost a factor of two smaller than in the respective purely hydrodynamical run ($B_0=0$, $\sfrff\approx0.46$). A similar reduction of the SFR with strong magnetic fields compared to purely hydrodynamical or weakly magnetized models is reported in \citet{PadoanNordlund2011} and \citet{PadoanHaugboelleNordlund2012}, who find a maximum reduction by a factor of $\sim3$. This is a significant, but relatively small effect compared to the influence of different forcing on the $\sfrff$ (see above). Magnetic fields reduce $\sfrff$, but are unlikely the major player in controlling the SFR, provided that molecular cloud turbulence is super-Alfv\'enic or at most trans-Alfv\'enic. This seems to be the case in most clouds. However, as pointed out earlier, on larger scales than molecular clouds, i.e., in the warmer, mainly atomic part of the ISM, turbulence may be trans-Alfv\'enic or even sub-Alfv\'enic \citep{HeilesTroland2005,LiHenning2011,HeyerBrunt2012}, rendering magnetic fields potentially more important in the process of molecular cloud formation. Still, even inside molecular clouds, magnetic fields seem to reduce fragmentation significantly (see Figure~\ref{fig:mhdimages}), thus potentially having a strong impact on the mass distribution of cores and stars \citep[see also][]{PriceBate2007,HennebelleTeyssier2008,BuerzleEtAl2011,PetersEtAl2011,HennebelleEtAl2011}.

\section{Comparing SFRs in the MHD Simulations with Theoretical Predictions} \label{sec:model_comparison}

\begin{figure*}[t]
\centerline{
\includegraphics[width=0.99\linewidth]{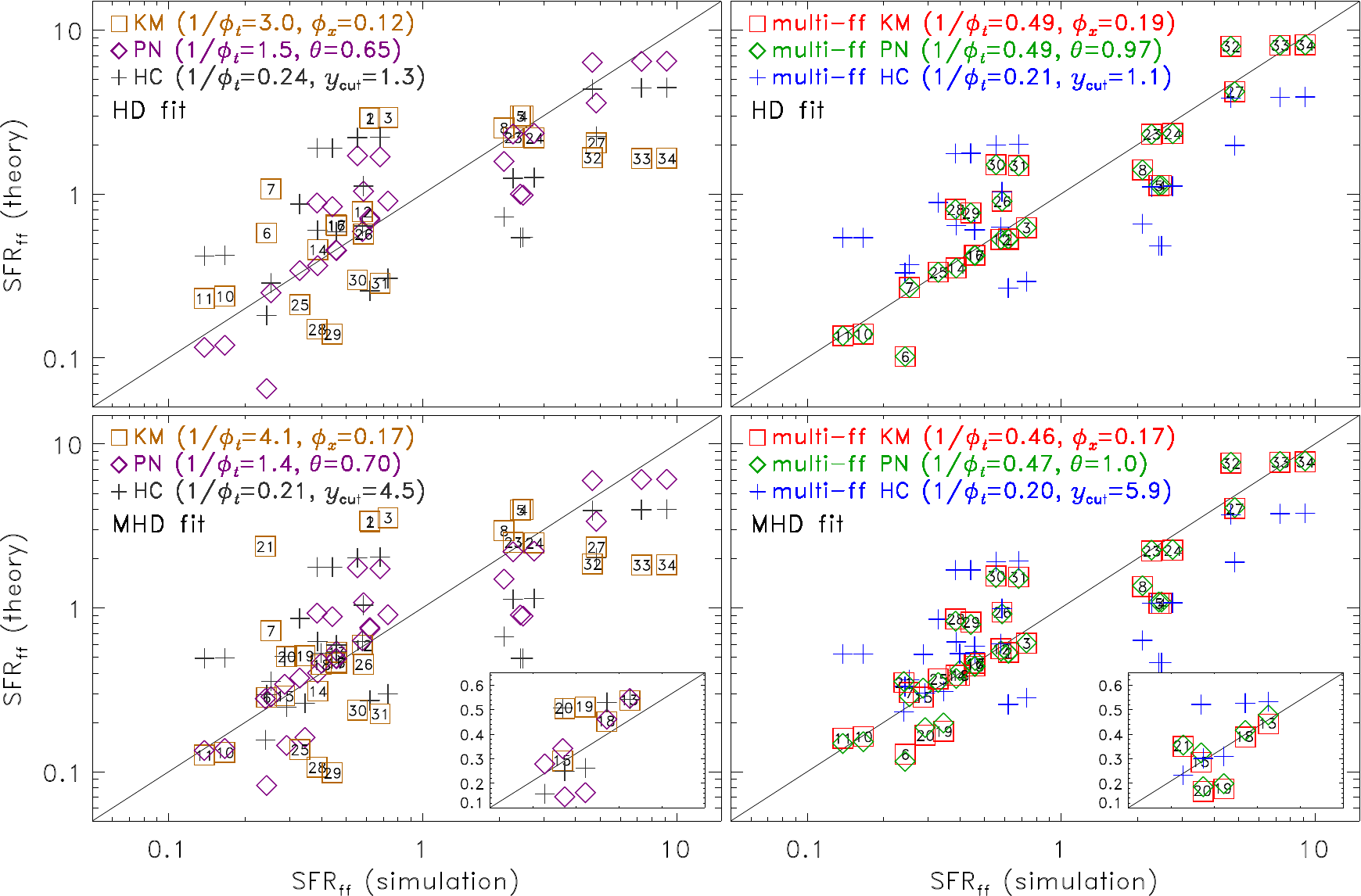}
}
\caption{$\sfrff$ (theory) for the six theories listed in Table~\ref{tab:theories}: KM (boxes), PN (diamonds), and HC (crosses) in the left panels, and the corresponding multi-freefall versions of the theories in the right panels, computed based on the numerical simulation parameters $\alphavir$, $\mach$, $b$, and $\beta$ listed in Table~\ref{tab:sims} and compared with the $\sfrff$ (simulation). The simulation number is given in each of the KM boxes. The analytic model predictions, $\sfrff$ (theory), were fitted to $\sfrff$ (simulation) with the fit parameters $\eps/\phit$ (where $\eps=1$ by definition in the simulations) and the fudge factors $\phix$ (KM), $\theta$ (PN), and $\ycut$ (HC). The best-fit parameters are given in the legend. The fits in the top panels only used the hydrodynamic models for which $B_0=0$, while the fits in the bottom panels include all MHD models listed in Table~\ref{tab:sims} (except for the low-resolution $128^3$-models). A zoom of the region containing the MHD models is shown in the inset plots in the bottom panels, where only the six MHD simulations are included. The diagonal solid line in each plot represents perfect agreement between $\sfrff$ (theory) and $\sfrff$ (simulation). The best-fit parameters with uncertainties and $\chi^2$-values are listed in Table~\ref{tab:fits}. Each simulation--theory data pair is listed in Table~\ref{tab:sfrffs}.}
\label{fig:model_comparison}
\end{figure*}

Using the dimensionless parameters $\alphavir$, $\mach$, $b$, and $\beta$ (or $\macha$) measured for each numerical simulation and listed in the last five columns of Table~\ref{tab:sims}, we can now compute the $\sfrff$ predicted by each of the six theories: KM, PN, HC, and multi-freefall KM, PN, HC, introduced in Section~\ref{sec:theory} (summarized in Table~\ref{tab:theories}), and compare it to the simulated $\sfrff$. The comparison between $\sfrff$ (theory) and $\sfrff$ (simulation) is shown in Figure~\ref{fig:model_comparison} (left panels: KM, PN, HC; right panels: multi-freefall KM, PN, HC). The $\sfrff$ in each of the six theoretical models is fully determined by $\alphavir$, $\mach$, $b$, and $\beta$, except for the parameters $\eps/\phit$ and the fudge factors $\phix$ (KM), $\theta$ (PN), or $\ycut$ (HC). In the simulations, the local efficiency $\eps=1$ because we did not include any form of feedback, but $1/\phit$ and the theory fudge factors are free parameters. In order to constrain them for each theory, we perform two-parameter fits of $\sfrff$ (theory) to $\sfrff$ (simulation). The best-fit parameters are listed in the legend of Figure~\ref{fig:model_comparison}. Table~\ref{tab:fits} additionally lists uncertainty estimates for the parameters, together with $\chi^2$-values, the number of degrees of freedom (DOF) in the fits, and the reduced $\chisqred=\chi^2/\mathrm{DOF}$. The $\chisqred$ is a quantitative indicator for the goodness of fit, with better fits having smaller $\chisqred$. To separate the effects of the magnetic field, we only use purely HD models ($B_0=0$) in the top panels of Figure~\ref{fig:model_comparison} (HD fit), while we include all MHD models in the bottom panels (MHD fit). This distinction is also made in Table~\ref{tab:fits}. Inset plots in the bottom panels show a zoom-in on the MHD models only. The solid diagonal line in each panel represents $\sfrff\mathrm{(theory)}=\sfrff\mathrm{(simulation)}$, i.e., perfect agreement between theory and simulation.

\begin{table}
\caption{$\sfrff$(Theory)--$\sfrff$(Simulation) Fit Parameters (Fig.~\ref{fig:model_comparison}).}
\label{tab:fits}
\def\arraystretch{1.3}
\setlength{\tabcolsep}{2.5pt}
\begin{tabular}{lllrcr}
\hline
\hline
(1) & (2) & (3) & (4) & (5) & (6) \\
Theory (\textit{HD fit}) & $1/\phit$ & Fudge Factor & $\chi^2$ & DOF & $\chisqred$ \\
\hline
             KM & $ 3.00\pm\textrm{n/a}$ & $\phix=0.12\pm\textrm{n/a}$ & $127$ & $24$ & $ 5.3$ \\
             PN & $ 1.50\pm0.16$ & $\theta=0.65\pm0.05$ & $ 46$ & $24$ & $ 1.9$ \\
             HC & $0.24\pm\textrm{n/a}$ & $\ycut= 1.3\pm\textrm{n/a}$ & $135$ & $24$ & $ 5.6$ \\
    multi-ff KM & $0.49\pm0.06$ & $\phix=0.19\pm0.02$ & $ 32$ & $24$ & $ 1.3$ \\
    multi-ff PN & $0.49\pm0.06$ & $\theta=0.97\pm0.10$ & $ 32$ & $24$ & $ 1.3$ \\
    multi-ff HC & $0.21\pm\textrm{n/a}$ & $\ycut= 1.1\pm\textrm{n/a}$ & $149$ & $24$ & $ 6.2$ \\
\hline
Theory (\textit{MHD fit}) \hspace{-7pt} & & & & & \\
\hline
             KM & $ 4.10\pm\textrm{n/a}$ & $\phix=0.17\pm\textrm{n/a}$ & $172$ & $30$ & $ 5.7$ \\
             PN & $ 1.40\pm0.14$ & $\theta=0.70\pm0.04$ & $ 54$ & $30$ & $ 1.8$ \\
             HC & $0.21\pm\textrm{n/a}$ & $\ycut= 4.5\pm\textrm{n/a}$ & $147$ & $30$ & $ 4.9$ \\
    multi-ff KM & $0.46\pm0.06$ & $\phix=0.17\pm0.02$ & $ 39$ & $30$ & $ 1.3$ \\
    multi-ff PN & $0.47\pm0.06$ & $\theta= 1.0\pm0.1$ & $ 37$ & $30$ & $ 1.2$ \\
    multi-ff HC & $0.20\pm\textrm{n/a}$ & $\ycut= 5.9\pm\textrm{n/a}$ & $152$ & $30$ & $ 5.1$ \\
\hline
\end{tabular}
\\
\textbf{Notes.} Column 1: Theoretical model according to Table~\ref{tab:theories}. Columns 2 and 3: Fit parameters for the HD fit set (top) and MHD fit set (bottom), corresponding to the top and bottom panels in Figure~\ref{fig:model_comparison}. Column 4: $\chi^2$ of the fit. Column 5: Number of degrees of freedom (DOF), i.e., the number of numerical models used for fitting (see Table~\ref{tab:sims}) minus 2 (the number of fit parameters). The last column (6) shows the reduced $\chisqred=\chi^2/\mathrm{DOF}$, enabling a direct comparison of the fit quality between the HD and MHD fit sets. Smaller $\chisqred$ indicate better fits. Uncertainty estimates for the fit parameters in columns 2 and 3 are only shown for models with $\chisqred\!<\!2$.
\end{table}

\begin{table*}
\caption{$\sfrff$ in the Simulations Listed in Table~\ref{tab:sims} and Theoretical Predictions for the Best-fit MHD Parameters in Table~\ref{tab:fits}.}
\label{tab:sfrffs}
\def\arraystretch{1.2}
\setlength{\tabcolsep}{5.6pt}
\begin{tabular}{lrc|ccc|ccc}
\hline
\hline
Model & $\sfrff$: & Simulation & KM & PN & HC & multi-ff KM & multi-ff PN & multi-ff HC \\
(1) & & (2) & (3) & (4) & (5) & (6) & (7) & (8) \\
\hline
01) GT256sM3                       & & $6.2\!\times\!10^{-1}$ & $3.4\!\times\!10^{+0}$ & $7.6\!\times\!10^{-1}$ & $2.7\!\times\!10^{-1}$ & $5.3\!\times\!10^{-1}$ & $5.3\!\times\!10^{-1}$ & $2.6\!\times\!10^{-1}$ \\
02) GT512sM3                       & & $6.2\!\times\!10^{-1}$ & $3.3\!\times\!10^{+0}$ & $7.4\!\times\!10^{-1}$ & $2.7\!\times\!10^{-1}$ & $5.3\!\times\!10^{-1}$ & $5.3\!\times\!10^{-1}$ & $2.6\!\times\!10^{-1}$ \\
03) GT256mM3                       & & $7.3\!\times\!10^{-1}$ & $3.5\!\times\!10^{+0}$ & $9.1\!\times\!10^{-1}$ & $3.0\!\times\!10^{-1}$ & $6.1\!\times\!10^{-1}$ & $6.1\!\times\!10^{-1}$ & $2.8\!\times\!10^{-1}$ \\
04) GT256cM3                       & & $2.5\!\times\!10^{+0}$ & $4.0\!\times\!10^{+0}$ & $8.9\!\times\!10^{-1}$ & $4.9\!\times\!10^{-1}$ & $1.1\!\times\!10^{+0}$ & $1.1\!\times\!10^{+0}$ & $4.6\!\times\!10^{-1}$ \\
05) GT512cM3                       & & $2.4\!\times\!10^{+0}$ & $4.0\!\times\!10^{+0}$ & $9.1\!\times\!10^{-1}$ & $4.9\!\times\!10^{-1}$ & $1.1\!\times\!10^{+0}$ & $1.1\!\times\!10^{+0}$ & $4.6\!\times\!10^{-1}$ \\
\hline
06) GT256sM5                       & & $2.4\!\times\!10^{-1}$ & $2.8\!\times\!10^{-1}$ & $8.2\!\times\!10^{-2}$ & $3.0\!\times\!10^{-1}$ & $1.3\!\times\!10^{-1}$ & $1.2\!\times\!10^{-1}$ & $3.3\!\times\!10^{-1}$ \\
07) GT256mM5                       & & $2.5\!\times\!10^{-1}$ & $7.2\!\times\!10^{-1}$ & $2.9\!\times\!10^{-1}$ & $3.6\!\times\!10^{-1}$ & $3.0\!\times\!10^{-1}$ & $2.9\!\times\!10^{-1}$ & $3.6\!\times\!10^{-1}$ \\
08) GT256cM5                       & & $2.1\!\times\!10^{+0}$ & $3.0\!\times\!10^{+0}$ & $1.5\!\times\!10^{+0}$ & $6.7\!\times\!10^{-1}$ & $1.4\!\times\!10^{+0}$ & $1.4\!\times\!10^{+0}$ & $6.3\!\times\!10^{-1}$ \\
\hline
09) GT128sM10                      & & $2.7\!\times\!10^{-1}$ & n/a & n/a & n/a & n/a & n/a & n/a \\
10) GT256sM10                      & & $1.7\!\times\!10^{-1}$ & $1.3\!\times\!10^{-1}$ & $1.4\!\times\!10^{-1}$ & $4.9\!\times\!10^{-1}$ & $1.6\!\times\!10^{-1}$ & $1.5\!\times\!10^{-1}$ & $5.2\!\times\!10^{-1}$ \\
11) GT512sM10                      & & $1.4\!\times\!10^{-1}$ & $1.3\!\times\!10^{-1}$ & $1.3\!\times\!10^{-1}$ & $4.9\!\times\!10^{-1}$ & $1.6\!\times\!10^{-1}$ & $1.5\!\times\!10^{-1}$ & $5.2\!\times\!10^{-1}$ \\
12) GT512mM10\phantom{B10} (seed1) & & $5.8\!\times\!10^{-1}$ & $5.9\!\times\!10^{-1}$ & $6.3\!\times\!10^{-1}$ & $6.2\!\times\!10^{-1}$ & $5.6\!\times\!10^{-1}$ & $5.5\!\times\!10^{-1}$ & $6.0\!\times\!10^{-1}$ \\
13) GT512mM10B1\phantom{0} (seed1) & & $4.6\!\times\!10^{-1}$ & $5.4\!\times\!10^{-1}$ & $5.5\!\times\!10^{-1}$ & $5.4\!\times\!10^{-1}$ & $4.4\!\times\!10^{-1}$ & $4.8\!\times\!10^{-1}$ & $5.3\!\times\!10^{-1}$ \\
14) GT512mM10\phantom{B10} (seed2) & & $3.9\!\times\!10^{-1}$ & $3.1\!\times\!10^{-1}$ & $4.0\!\times\!10^{-1}$ & $6.2\!\times\!10^{-1}$ & $3.9\!\times\!10^{-1}$ & $3.7\!\times\!10^{-1}$ & $6.2\!\times\!10^{-1}$ \\
15) GT512mM10B1\phantom{0} (seed2) & & $2.9\!\times\!10^{-1}$ & $2.9\!\times\!10^{-1}$ & $3.4\!\times\!10^{-1}$ & $5.1\!\times\!10^{-1}$ & $2.8\!\times\!10^{-1}$ & $3.2\!\times\!10^{-1}$ & $5.2\!\times\!10^{-1}$ \\
16) GT256mM10\phantom{B10} (seed3) & & $4.6\!\times\!10^{-1}$ & $4.6\!\times\!10^{-1}$ & $4.9\!\times\!10^{-1}$ & $5.9\!\times\!10^{-1}$ & $4.6\!\times\!10^{-1}$ & $4.4\!\times\!10^{-1}$ & $5.8\!\times\!10^{-1}$ \\
17) GT512mM10\phantom{B10} (seed3) & & $4.6\!\times\!10^{-1}$ & $4.6\!\times\!10^{-1}$ & $4.9\!\times\!10^{-1}$ & $5.9\!\times\!10^{-1}$ & $4.6\!\times\!10^{-1}$ & $4.4\!\times\!10^{-1}$ & $5.8\!\times\!10^{-1}$ \\
18) GT512mM10B1\phantom{0} (seed3) & & $4.0\!\times\!10^{-1}$ & $4.5\!\times\!10^{-1}$ & $4.6\!\times\!10^{-1}$ & $5.3\!\times\!10^{-1}$ & $3.9\!\times\!10^{-1}$ & $4.2\!\times\!10^{-1}$ & $5.3\!\times\!10^{-1}$ \\
19) GT256mM10B3\phantom{0} (seed3) & & $3.4\!\times\!10^{-1}$ & $5.1\!\times\!10^{-1}$ & $1.6\!\times\!10^{-1}$ & $2.6\!\times\!10^{-1}$ & $1.8\!\times\!10^{-1}$ & $2.0\!\times\!10^{-1}$ & $3.1\!\times\!10^{-1}$ \\
20) GT512mM10B3\phantom{0} (seed3) & & $2.9\!\times\!10^{-1}$ & $5.0\!\times\!10^{-1}$ & $1.4\!\times\!10^{-1}$ & $2.5\!\times\!10^{-1}$ & $1.7\!\times\!10^{-1}$ & $1.9\!\times\!10^{-1}$ & $3.0\!\times\!10^{-1}$ \\
21) GT256mM10B10 (seed3)           & & $2.4\!\times\!10^{-1}$ & $2.4\!\times\!10^{+0}$ & $2.8\!\times\!10^{-1}$ & $1.6\!\times\!10^{-1}$ & $3.5\!\times\!10^{-1}$ & $3.6\!\times\!10^{-1}$ & $2.3\!\times\!10^{-1}$ \\
22) GT128cM10                      & & $1.6\!\times\!10^{+0}$ & n/a & n/a & n/a & n/a & n/a & n/a \\
23) GT256cM10                      & & $2.3\!\times\!10^{+0}$ & $2.5\!\times\!10^{+0}$ & $2.2\!\times\!10^{+0}$ & $1.1\!\times\!10^{+0}$ & $2.2\!\times\!10^{+0}$ & $2.3\!\times\!10^{+0}$ & $1.1\!\times\!10^{+0}$ \\
24) GT512cM10                      & & $2.8\!\times\!10^{+0}$ & $2.5\!\times\!10^{+0}$ & $2.2\!\times\!10^{+0}$ & $1.1\!\times\!10^{+0}$ & $2.2\!\times\!10^{+0}$ & $2.3\!\times\!10^{+0}$ & $1.1\!\times\!10^{+0}$ \\
\hline
25) GT256sM20                      & & $3.3\!\times\!10^{-1}$ & $1.4\!\times\!10^{-1}$ & $3.7\!\times\!10^{-1}$ & $8.6\!\times\!10^{-1}$ & $3.7\!\times\!10^{-1}$ & $3.5\!\times\!10^{-1}$ & $8.5\!\times\!10^{-1}$ \\
26) GT256mM20                      & & $5.9\!\times\!10^{-1}$ & $4.5\!\times\!10^{-1}$ & $1.1\!\times\!10^{+0}$ & $1.0\!\times\!10^{+0}$ & $9.4\!\times\!10^{-1}$ & $9.2\!\times\!10^{-1}$ & $9.9\!\times\!10^{-1}$ \\
27) GT256cM20                      & & $4.8\!\times\!10^{+0}$ & $2.3\!\times\!10^{+0}$ & $3.4\!\times\!10^{+0}$ & $2.0\!\times\!10^{+0}$ & $4.0\!\times\!10^{+0}$ & $4.1\!\times\!10^{+0}$ & $1.9\!\times\!10^{+0}$ \\
\hline
28) GT256sM50                      & & $3.8\!\times\!10^{-1}$ & $1.1\!\times\!10^{-1}$ & $9.3\!\times\!10^{-1}$ & $1.8\!\times\!10^{+0}$ & $8.6\!\times\!10^{-1}$ & $8.3\!\times\!10^{-1}$ & $1.7\!\times\!10^{+0}$ \\
29) GT512sM50                      & & $4.4\!\times\!10^{-1}$ & $9.9\!\times\!10^{-2}$ & $8.8\!\times\!10^{-1}$ & $1.8\!\times\!10^{+0}$ & $8.2\!\times\!10^{-1}$ & $7.9\!\times\!10^{-1}$ & $1.7\!\times\!10^{+0}$ \\
30) GT256mM50                      & & $5.5\!\times\!10^{-1}$ & $2.4\!\times\!10^{-1}$ & $1.8\!\times\!10^{+0}$ & $2.0\!\times\!10^{+0}$ & $1.6\!\times\!10^{+0}$ & $1.5\!\times\!10^{+0}$ & $1.9\!\times\!10^{+0}$ \\
31) GT512mM50                      & & $6.8\!\times\!10^{-1}$ & $2.3\!\times\!10^{-1}$ & $1.7\!\times\!10^{+0}$ & $2.0\!\times\!10^{+0}$ & $1.5\!\times\!10^{+0}$ & $1.5\!\times\!10^{+0}$ & $1.9\!\times\!10^{+0}$ \\
32) GT256cM50                      & & $4.7\!\times\!10^{+0}$ & $1.9\!\times\!10^{+0}$ & $6.0\!\times\!10^{+0}$ & $3.9\!\times\!10^{+0}$ & $7.6\!\times\!10^{+0}$ & $7.7\!\times\!10^{+0}$ & $3.7\!\times\!10^{+0}$ \\
33) GT512cM50                      & & $7.3\!\times\!10^{+0}$ & $1.8\!\times\!10^{+0}$ & $6.1\!\times\!10^{+0}$ & $4.0\!\times\!10^{+0}$ & $7.7\!\times\!10^{+0}$ & $7.8\!\times\!10^{+0}$ & $3.7\!\times\!10^{+0}$ \\
34) GT1024cM50                     & & $9.1\!\times\!10^{+0}$ & $1.8\!\times\!10^{+0}$ & $6.1\!\times\!10^{+0}$ & $4.0\!\times\!10^{+0}$ & $7.8\!\times\!10^{+0}$ & $7.9\!\times\!10^{+0}$ & $3.8\!\times\!10^{+0}$ \\
\hline
\end{tabular}
\\
\textbf{Notes.} Column (1): simulation model. Column (2): $\sfrff$ measured in the simulations. Columns (3--8): Theoretical $\sfrff$ computed for the simulation parameters $\alphavir$, $\mach$, $b$, and $\beta$ (or equivalently $\macha$) listed in Table~\ref{tab:sims} in the KM (3), PN (4), and HC (5) theories, as well as, in the multi-freefall KM (6), multi-freefall PN (7), and multi-freefall HC (8) theories, using the best-fit MHD parameters from Table~\ref{tab:fits}. No theoretical values were computed for the GT128sM10 and GT128cM10 simulations because they only used a numerical resolution of $128^3$ cells (see the discussion on numerical convergence in Section~\ref{sec:timeevol}).
\end{table*}

Figure~\ref{fig:model_comparison} shows that all the theoretical models exhibit some positive correlation between $\sfrff$ (theory) and $\sfrff$ (simulation). The multi-freefall KM and PN models (right panels) show much better agreement with the simulation data in both the HD and MHD fits, indicated by the smallest \mbox{$\chisqred=1.2$--$1.3$} (see Table~\ref{tab:fits}), than the original KM and PN models (left panels). The HC models exhibit the opposite behavior, i.e., the HC theory gives slightly better fits than the multi-freefall HC theory. This is not surprising because both HC models use the multi-freefall factor, but the HC model additionally includes turbulent support in the estimate of the threshold density (Equations~\ref{eq:rhocrit_th} and~\ref{eq:rhocrit_turb} into Equation~\ref{eq:scrit_hc}), while the multi-freefall HC model only includes thermal support (Equation~\ref{eq:rhocrit_th} only). However, all HC fits exhibit relatively large \mbox{$\chisqred\approx4.9$--6.2}. The reason for this is the choice of the critical density in the HC models and its resulting dependence on the sonic Mach number, $\rhocrit\propto\mach^{-2}$, while all KM and PN models have $\rhocrit\propto\mach^{+2}$, which is (apart from the different choice of fudge factors) the only fundamental difference between the multi-freefall HC and the two multi-freefall KM and PN models (see Table~\ref{tab:theories}). The difference in fudge factors is irrelevant in this comparison because they all enter in the same way for each theory, simply as factors in the critical density, for which the fitting procedure determines the best-fit value automatically. In contrast, the dependence of $\sfrff$ (theory) on $\alphavir$, $\mach$, $b$, and $\beta$ is determined by each analytic theory separately. Table~\ref{tab:theories} gives an overview of the basic similarities and differences between the six theoretical models for the $\sfrff$.

The KM fits also exhibit fairly large $\chisqred=5.3$ and $5.7$ in the HD and MHD fit set, respectively. In contrast, the multi-freefall version of the KM model gives much better fits ($\chisqred=1.3$ for both the HD and MHD fits, respectively). The original PN model already gives fairly good fits ($\chisqred=1.9$ and $1.8$), but again, the multi-freefall PN version gives better fits, in fact the best fits of all analytic theories ($\chisqred=1.3$ for the HD and $\chisqred=1.2$ for the MHD fit). The HD fits for the multi-freefall KM and multi-freefall PN models are identical because in the HD limit the two theories are identical, while in the MHD case, the only difference is the $\beta$-dependence of $\rhocrit$, which is $\rhocrit{_\mathrm{,KM}}\propto1/(1+\beta^{-1})$ for KM (Equation~\ref{eq:scrit_km}), while it is $\rhocrit{_\mathrm{,PN}}\propto f(\beta)$ given by Equation~(\ref{eq:fofbeta}) for the PN theory. However, the difference in $\chisqred$ between multi-ff KM and multi-ff PN is very small, such that both the multi-freefall KM and multi-freefall PN models provide the best match to our set of numerical simulations.

The best-fit MHD theory parameters for the multi-freefall KM and multi-freefall PN models are similar (see Table~\ref{tab:fits}). Taking into account the full range of error margins, we find \mbox{$1/\phit=0.4$--$0.55$}, and \mbox{$\phix=0.15$--$0.21$} and \mbox{$\theta=0.87$--$1.1$}. The multi-ff KM fit thus suggests a close correspondence of the magnetothermal Jeans length~(Equation~\ref{eq:ljmt}) and the magnetosonic scale~(Equation~\ref{eq:lsms}) with a correction of order $\phix=0.18\pm0.03$. The multi-ff PN model fit supports the expected large-scale injection of turbulence, parameterized by $\theta=0.99\pm0.11$ (see Section~\ref{sec:theory}). Moreover, the \mbox{$\chisqred=1.2$--$1.3$} of the multi-ff KM and multi-ff PN fits are similar, but slightly smaller in the MHD fit set than in the HD fit set. This indicates that the magnetic-field dependence in the analytic models provides a good match to the simulation data, and that our extension of the multi-ff KM model to MHD in Section~\ref{sec:sixtheories} is reasonable.

Even though the agreement between $\sfrff$ (theory) and $\sfrff$ (simulation) is very good for the multi-ff KM and multi-ff PN models shown in Figure~\ref{fig:model_comparison}, some numerical simulations only agree within a factor of \mbox{2--3} with the analytic prediction. To distinguish each simulation, we added the simulation numbers of Table~\ref{tab:sims} in each KM box of Figure~\ref{fig:model_comparison}. The values of the measured $\sfrff$ (simulation) and the computed $\sfrff$ (theory) are listed in Table~\ref{tab:sfrffs}. Generally, the multi-ff KM and PN theories agree with the simulation data within a factor of two. The simulation with the largest deviation is model \#30 (GT256mM50), for which the predicted $\sfrff$ by the multi-ff KM and PN models is a factor of 2.9 and 2.7 higher than the measured $\sfrff$ in the simulation. The higher-resolution version of this simulation with $512^3$ cells (\#31: GT512mM50) shows an improvement, such that $\sfrff$ (simulation) is now only a factor of 2.2 higher than $\sfrff$ in both the multi-ff KM and PN theories. A similar trend with increasing resolution is obtained for MHD models \#19 (GT256mM10B3) and \#20 (GT512mM10B3), as well as for \#28 (GT256sM50) and \#29 (GT512sM50), all converging toward the diagonal, solid line in Figure~\ref{fig:model_comparison} for the multi-freefall KM and PN models. This improvement with increasing resolution can be seen best for the $\mach\sim50$, compressive-forcing models \#32 (GT256cM50) with $256^3$, \#33 (GT512cM50) with $512^3$, and \#34 (GT1024cM50) with $1024^3$ resolution in the right panels of Figure~\ref{fig:model_comparison}. The convergence with increasing resolution suggests that the analytic theories give reasonable results and that we have constrained the theory parameters well with our set of numerical simulations.

The overall agreement between the theories and simulations is encouraging. Although some numerical models only agree within a factor of 2--3 at the limited resolution available, we have to keep in mind that the overall agreement holds over two orders of magnitude in SFRs, from $\sfrff\approx0.1$ to 10, as covered by all the numerical simulations with different virial parameters, Mach numbers, forcing, and magnetic-field strengths, combined in Figure~\ref{fig:model_comparison}. All our simulations are fit simultaneously by the multi-ff KM and multi-ff PN models.

\section{Comparison with Observations} \label{sec:obs}
Here we compare the MHD simulation results of the SFR from Section~\ref{sec:model_comparison} with observations of Galactic clouds. Since observed SFRs are usually quoted as SFR column densities, $\sigsfr$, i.e., SFR per unit area, we convert the simulated SFRs to $\sigsfr$ to facilitate the comparison with observations.

\begin{figure*}[t]
\centerline{
\includegraphics[width=0.99\linewidth]{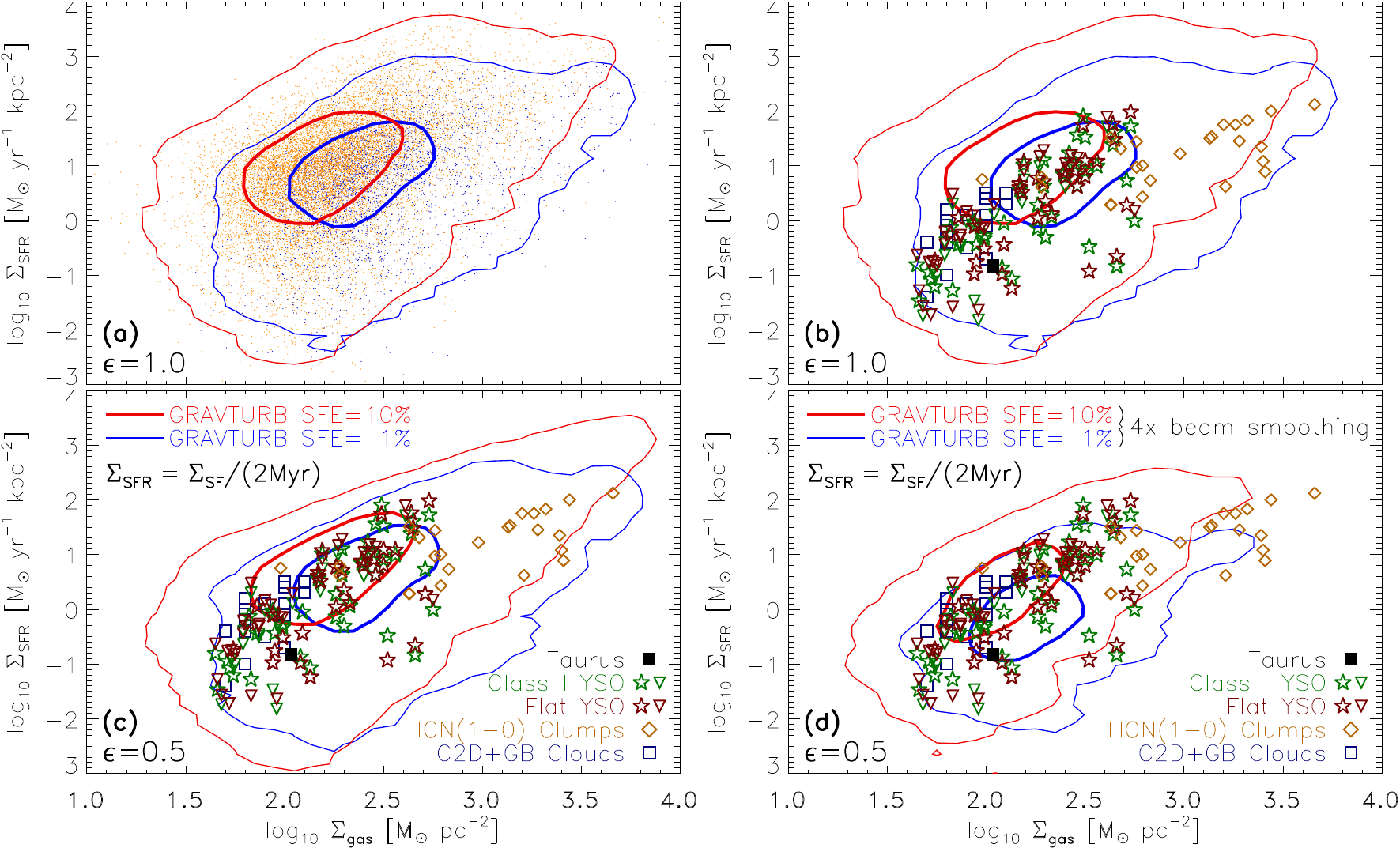}
}
\caption{\textbf{(a)}: Star formation rate column density $\sigsfr$ vs.~gas column density $\siggas$ measured in the GRAVTURB simulations listed in Table~\ref{tab:sims} for a star formation efficiency $\sfe=1\%$ (blue) and $\sfe=10\%$ (red), respectively. Two contour lines for each $\sfe$ are drawn. The thick contours enclose 50\% of all ($\siggas,\,\sigsfr$) simulation pairs, centered on the peak of the distribution, while the thin contours enclose 99\%. \textbf{(b)}: Same as (a), but only the contours of the simulations are drawn, and observational data of Galactic clouds from \citet{HeidermanEtAl2010} are superimposed. The individual data points are labeled in the legend of the bottom panels (Taurus: filled black box, Class I YSOs and Flat YSOs: green and red stars and upper-limits shown as downward-pointing triangles, HCN(1--0) Clumps: golden diamonds, and C2D+GB Clouds: dark blue boxes). The simulation data in panels (a) and (b) are plotted for a local core-formation efficiency $\eps=1$, the value expected without any local feedback from YSOs. \textbf{(c)}: Same as (b), but the simulation data were transformed to $\eps=0.5$ using Equations~(\ref{eq:epsmod}), which changes the GRAVTURB contours compared to (a) and (b). The value $\eps=0.5$ was determined by fitting the simulation data to the observational data using Equation~(\ref{eq:obsfit}), suggesting local efficiencies of \mbox{$\eps\approx0.3$--$0.7$} for an assumed \mbox{$\sfe\approx1\%$--$10\%$} in the observed clouds. \textbf{(d)}: Same as (c), but for the simulation maps smoothed to $4\times$ coarser resolution, demonstrating the effect of observing the simulated clouds with reduced telescope resolution.
}
\label{fig:sfrcoldens}
\end{figure*}

\begin{figure*}[t]
\centerline{
\includegraphics[width=0.99\linewidth]{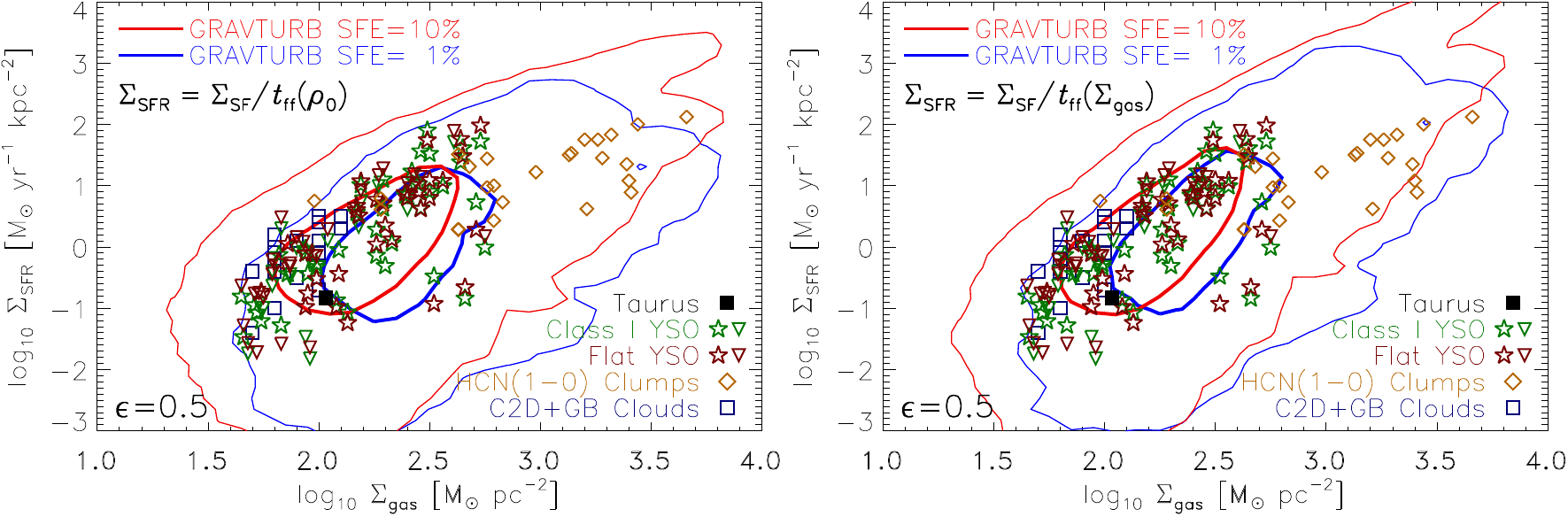}
}
\caption{Same as panel (c) in Figure~\ref{fig:sfrcoldens}, but here we compute the simulation $\sigsfr$ with two other methods, both different from the standard method used in Figure~\ref{fig:sfrcoldens}, where $\sigsfr\equiv\sigsf/(2\,\mathrm{M}\yr)$. Left: $\sigsfr\equiv\sigsf/\tff(\meanrho)$, i.e., the sink particle column density $\sigsf$ is divided by the global freefall time at the mean density $\meanrho$ of the simulation in which the $\sigsf$ pixel was found. Right: $\sigsfr\equiv\sigsf/\tff(\siggas)$, i.e., we divide $\sigsf$ by the local freefall time of the gas for each pixel, $\tff(\siggas)=\sqrt{3\pi L /(32G\siggas)}$ with the line of sight $L$ of the corresponding simulation model. Both $\meanrho$ and $L$ are listed in Table~\ref{tab:sims}. Some minor differences compared to panel (c) in Figure~\ref{fig:sfrcoldens} are apparent, but the overall agreement between simulations and Galactic cloud observations remains good, irrespective of the method used to define $\sigsfr$ in the simulations.}
\label{fig:sfrcoldens_tffs}
\end{figure*}

\subsection{MHD Simulations Converted to $\siggas$ and $\sigsfr$}
We measure $\siggas$ and $\sigsfr$ with a method as close as possible to what observers do to infer $\sigsfr$-to-$\siggas$ relations \citep[see, e.g.,][]{BigielEtAl2008,HeidermanEtAl2010}, including the effects of telescope beam smoothing. For each simulation, we construct two-dimensional projections of the gas column density $\siggas$ and the sink particle column density $\sigsf$ along each coordinate axis: $x$, $y$, $z$. All maps were smoothed to a resolution $N_\mathrm{res}/8$ with the numerical resolution $N_\mathrm{res}$ given in Table~\ref{tab:sims}, such that the size of each pixel in the smoothed maps slightly exceeds the sink particle diameter (which is 5 grid cells; see Section~\ref{sec:sinks}). We also test smoothing to $N_\mathrm{res}/32$ below, which yields similar results. We then search for pixels with a sink particle column density greater than zero, $\sigsf>0$, and extract the corresponding pixel in the gas column density map, which gives $\siggas$ in units of $\msol\,\pc^{-2}$ for that pixel. The SFR column density is computed by taking the sink particle column density $\sigsf$ of the same pixel and dividing it by a characteristic timescale for star formation, $\tsf$, which yields $\sigsfr=\sigsf/\tsf$ in units of $\msol\,\yr^{-1}\,\mathrm{kpc}^{-2}$. The simplest choice for $\tsf$ is a fixed star formation time, $\tsf=2\,\mathrm{M}\yr$, based on an estimate of the elapsed time between star formation and the end of the Class II phase \citep[e.g.,][]{EvansEtAl2009,CoveyEtAl2010}. This is also the $\tsf$ adopted by \citet{LadaLombardiAlves2010} and \citet{HeidermanEtAl2010} to convert young stellar object (YSO) counts into an SFR column density, so we use it here as the standard approach. However, we also experimented with two other choices for $\tsf$ and present a comparison of those choices below, all yielding similar results.

The result of the procedure explained above is plotted in panel (a) of Figure~\ref{fig:sfrcoldens}. It shows a scatter plot of $\sigsfr$ versus $\siggas$ measured in all the maps produced from our simulations listed in Table~\ref{tab:sims} (except for the two low-resolution, $128^3$-simulations) for a star formation efficiency $\sfe=1\%$ (blue) and $\sfe=10\%$ (red). Thus, each pixel shown in panel (a) of Figure~\ref{fig:sfrcoldens} is one pair of ($\siggas,\sigsfr$) extracted for each simulation and each projection direction. By combining all data of maps from the three principal projections in $x$, $y$, and $z$, we increase the statistical sample for each model by about a factor of three on average. A total number of $3.5\times10^3$ and $1.2\times10^4$ simulation pixels for $\sfe=1\%$ and $\sfe=10\%$, respectively, contribute to the scatter plots in Figure~\ref{fig:sfrcoldens}. We also add contours of the ($\siggas,\sigsfr$) distribution, with two contour levels for $\sfe=1\%$ (blue contours) and $\sfe=10\%$ (red contours). The thick contours enclose 50\% of all simulation pixels and the thin contours enclose 99\%. The contours help to easily identify the underlying probability distribution of the scattered data points.

The simulation data have a broad probability distribution with a clear positive correlation between $\sigsfr$ and $\siggas$. The data for $\sfe=10\%$ are shifted to higher $\sigsfr$ and lower $\siggas$ compared to the $\sfe=1\%$ distribution because more gas is accreted by sink particles and thus removed from the gas phase at higher $\sfe$. If we were to fit power laws to the distributions, the slopes would be in the range 1--2, i.e., $\sigsfr\propto\siggas^\textrm{1--2}$ with somewhat flatter slopes at higher $\sfe$.

\subsection{Galactic Observations of $\siggas$ and $\sigsfr$}
To compare the simulation data with observations, we add data of Galactic clouds from \citet{HeidermanEtAl2010} in panel (b) of Figure~\ref{fig:sfrcoldens}, superimposed on the simulation contours. The observational data are from Galactic observations of clouds and YSOs identified in the \emph{Spitzer} Cores-to-Disks (C2D) and GB surveys \citep{EvansEtAl2009} of massive dense clumps \citep{WuEtAl2010}, and of the Taurus molecular cloud \citep{PinedaEtAl2010,RebullEtAl2010}. The simulation data indicated by the same contours of panel (a) fall in the range of the observational data, however, the simulation data show slightly higher $\sigsfr$ than the observational data, on average. This is not surprising, given that our simulations did not include any local feedback from YSOs. It is known, however, that young stars eject a significant amount of accreted material, thereby reducing the overall accretion rate due to feedback from jets, winds, and outflows \citep{WardleKoenigl1993,KoniglPudritz2000,BeutherEtAl2002,PudritzEtAl2007,PetersEtAl2011,SeifriedEtAl2011}. Hence, only a fraction $\eps<1$ of the in-falling gas actually ends up on the protostar.

The local core-formation efficiency is parameterized by the factor $\eps$ in Equation~(\ref{eq:sfrffbasic}), from which all the $\sfrff$-models in Section~\ref{sec:theory} were derived. Since there is no feedback in our simulations, $\eps=1$ by definition. However, we can devise a correction to account for $\eps<1$. For this, we simply have to multiply the original $\sigsfr$ for $\eps=1$ by a given $\eps<1$. To conserve mass, we also have to account for the fact that a fraction $(1-\eps)$ was not accreted and remained in the gas phase due to local feedback. This means we have to increase $\siggas$ according to the reduction of $\sigsfr$, such that $\sigtot=\siggas+\sigsf$ with $\sigsf=\sigsfr\tsf$ is conserved. Given our simulation data $\siggas$ and $\sigsfr$ with $\eps=1$, we can compute values $\sigsfr^\prime$ and $\siggas^\prime$ for $\eps<1$ according to the following equations:
\begin{eqnarray}
\sigsfr^\prime(\eps) & = & \eps\,\sigsfr\,, \nonumber \\
\siggas^\prime(\eps) & = & \siggas + \left(1-\eps\right) \sigsf \label{eq:epsmod}\,.
\end{eqnarray}
Using these expressions, we can correct our simulation data to follow more realistic values of the local efficiency (see also the discussion of $\eps$ in Section~\ref{sec:derivation}).

The \citet{HeidermanEtAl2010} sample of SFR column densities for Galactic clouds shown in panel (b) of Figure~\ref{fig:sfrcoldens} is rather broad and presumably covers different evolutionary stages of the clouds, such that a single $\sfe$ for the whole sample is quite unlikely. However, since we are currently lacking additional information about the $\sfe$ in the observational sample, we can reasonably assume SFEs in the range \mbox{1\%--10\%} in the observational data \citep[][Paper II]{EvansEtAl2009,FederrathKlessen2012paper2}. In order to find the best-fit local efficiency parameter $\eps$, we fit our simulated distribution $p_\mathrm{sim}(\siggas^\prime,\sigsfr^\prime)$ to the observed distribution $p_\mathrm{obs}(\siggas,\sigsfr)$, by applying Equations~(\ref{eq:epsmod}). To do this, we compute the sum of the squared differences $\Delta^2$ between the two distributions, which have both been sampled to the same ($\siggas,\sigsfr$) grid with indexes $i$,
\begin{equation} \label{eq:obsfit}
\Delta^2=\sum_i \left[p_\mathrm{sim}(\siggas^\prime{_{,i}},\,\sigsfr^\prime{_{,i}}) - p_\mathrm{obs}(\siggas{_{,i}},\,\sigsfr{_{,i}})\right]^2,\,
\end{equation}
for $\sfe=1\%$, $3\%$, and $10\%$ and for 21 local efficiencies, $\eps=[0,1]$ in steps of $\deriv\eps=0.05$. For each given $\sfe$, we search for the minimum of $\Delta^2$ as a function of $\eps$. This procedure yields best-fit values of the local efficiency parameter $\eps=0.7$, 0.5, and 0.3 for $\sfe=1\%$, $3\%$, and $10\%$, respectively, in our comparison of simulation data with the \citet{HeidermanEtAl2010} Galactic cloud sample.

The simulation data modified to a local efficiency of $\eps=0.5$ are shown in panel (c) of Figure~\ref{fig:sfrcoldens} together with the original \citet{HeidermanEtAl2010} data. Assuming that the observational data have an $\sfe$ between 1\% and 10\%, the local efficiency parameters would be in between $\eps=0.3$ and $0.7$. This is in good agreement with theoretical models for $\eps$ \citep{MatznerMcKee2000}, with numerical simulations including outflow feedback \citep{WangEtAl2010,SeifriedEtAl2012}, and with observational estimates \citep[][and the discussion on $\eps$ in Section~\ref{sec:derivation}]{BeutherEtAl2002}.

We note that the simulation data in Figure~\ref{fig:sfrcoldens} are furthermore consistent with the Galactic cloud samples in \citet{LadaLombardiAlves2010} and \citet{GutermuthEtAl2011}, showing that $\sigsfr$ can vary by more than an order of magnitude at any given $\siggas$.

Considering the uncertainties in the $\sfe$ from the observations and the uncertainties in the simulations, the overall agreement is encouraging. The HCN(1--0) observational data points of molecular clumps are at the lower end of the distribution, but are still consistent with the simulation data. Possibly, the molecular clumps have a systematically smaller $\sfe$ because they are larger structures compared to the YSOs, such that the molecular clumps fall slightly below the general trend. However, this can only be tested when estimates of the cloud SFEs become available \citep[see Paper II,][]{FederrathKlessen2012paper2}. The Taurus data point as well as a few of the YSO data in the range \mbox{$\log_{10}\siggas\approx1.4$--$2.8$} also lie at the low-$\sigsfr$ end of the distributions obtained in the simulations. This might be caused by an enhanced magnetic-field influence for these objects. For instance, Taurus seems to be trans-Alfv\'enic rather than super-Alfv\'enic \citep{HeyerBrunt2012}, leading to a reduced $\sigsfr$ as discussed in Section~\ref{sec:morphology}. Only one of our MHD simulations approaches this strongly magnetized regime (GT256mM10B10 with $\macha\approx1.3$; see Table~\ref{tab:sims}), where anisotropies induced by the magnetic field start to become important.

\subsection{Influence of Telescope Resolution and Choice of $\tsf$}
We test the effects of telescope beam smoothing in panel (d) of Figure~\ref{fig:sfrcoldens}. Panel (d) is identical to panel (c), except that the simulation data were smoothed to grids with resolution $N_\mathrm{res}/32$, i.e., four times coarser resolution compared to the contours shown in panel (c). The increased beam smoothing results in distributions with somewhat smaller $\siggas$ and $\sigsfr$, best seen by comparing the positions of the thickest contours between panels (c) and (d). However, the overall agreement of the simulation data with the Galactic cloud sample is still good, even when the resolution is decreased by a factor of four.

In Figure~\ref{fig:sfrcoldens_tffs}, we study the influence of different choices for the star formation timescale $\tsf$. The two panels are identical to panel (c) in Figure~\ref{fig:sfrcoldens}, except for the method by which $\sigsfr=\sigsf/\tsf$ was computed in the simulations. The left panel adopts $\tsf=\tff(\meanrho)$, i.e., the sink particle column density $\sigsf$ is divided by the freefall time at the mean density $\meanrho$ of the simulation in which the $\sigsf$ pixel was found. In the right panel, we use $\tsf=\tff(\siggas)=\sqrt{3\pi L /(32G\siggas)}$, i.e., instead of taking the global mean free-fall time, we take the local freefall time of the gas in each pixel. The contours differ slightly between those two last choices and between our standard choice of fixed $\tsf=2\,\mathrm{M}\yr$ in Figure~\ref{fig:sfrcoldens}, but the overall agreement between simulation data and Galactic observations is similar in all three cases.

\subsection{Comparison with Extragalactic Measurements}
Figures~\ref{fig:sfrcoldens} and~\ref{fig:sfrcoldens_tffs} indicate some power-law correlation of the form $\sigsfr\propto\siggas^N$ (albeit with significant scatter), similar in exponents \mbox{$N\approx1$--$2$} to the Kennicutt-Schmidt relation \citep{Schmidt1959,Kennicutt1998} and follow-up measurements for molecular gas \citep[e.g.,][]{WongBlitz2002,GaoSolomon2004,BigielEtAl2008,KennicuttEvans2012}. However, the measured values of $\sigsfr$ in our numerical sample are larger than the extragalactic values of $\sigsfr$ and larger than theoretical estimates for that regime \citep[e.g.,][]{KrumholzMcKeeTumlinson2009} by about 1--2 orders of magnitude. The Galactic measurements of $\sigsfr$ by \citet{HeidermanEtAl2010} in Figure~\ref{fig:sfrcoldens} and by \citet{LadaLombardiAlves2010}, however, also show values of $\sigsfr$ that are 1--2 orders of magnitude above the extragalactic measurements with a scatter of about 1--2 orders of magnitude. \citet{HeidermanEtAl2010} explain this difference between Galactic and extragalactic measurements of $\sigsfr$ with the different telescope resolutions available for both regimes and thus the different areas over which the measurements of $\siggas$ and $\sigsfr$ are averaged. Both disk-averaged and spatially-resolved extragalactic measurements only provide highly smoothed images, mixing both star-forming and non-star-forming gas. Taking these factors into account and correcting for them, \citet{HeidermanEtAl2010} conclude that the extragalactic ($\siggas,\sigsfr$) relations are in agreement with the Galactic measurements. Indeed, decreased telescope resolution (or equivalently observing a region at greater distance) reduces $\sigsfr$, but also $\siggas$, as demonstrated here by comparing panels (c) and (d) of Figure~\ref{fig:sfrcoldens}. \citet{KrumholzDekelMcKee2012} argue that both Galactic and extragalactic measurements are consistent with a local star formation law, correlating $\sigsfr$ with $\siggas/\tsf$, where $\tsf$ ``is the freefall time evaluated at the density averaged over length scales comparable to the outer scale of turbulence, regardless of the mean density of the region being observed''. This seems to be a rather especial definition. Our experiments with three different definitions of $\tsf$ in Figures~\ref{fig:sfrcoldens} and~\ref{fig:sfrcoldens_tffs} do not exclude or prefer any particular choice for $\tsf$ in the Galactic cloud sample studied here. After acceptance of this work, we also learned about a recently submitted paper on a theoretical model for the $\sigsfr$-to-$\siggas$ relation by \citet{RenaudEtAl2012}, which is consistent with our findings for Galactic clouds, favoring a non-universal behavior of the star formation relation.

The simulations and the observational data shown in Figure~\ref{fig:sfrcoldens} are generally in very good agreement. The variations of the observed SFRs in different clouds by up to two orders of magnitude for a given value of $\siggas$ \citep{MooneySolomon1988,LadaLombardiAlves2010,HeidermanEtAl2010} and the different scaling relations of $\sigsfr$ versus $\siggas$ \citep{SuzukiEtAl2010} might thus be a result of different physical conditions in Galactic as well as extragalactic molecular clouds. As shown above, star formation is primarily controlled by the forcing and the sonic Mach number of the turbulence, with the magnetic field having a secondary effect. Molecular clouds cover a range of values for these physical parameters and different combinations of those, providing an explanation for the observed scatter in SFRs.

\section{Discussion and Limitations} \label{sec:limitations}

Here we discuss limitations of the analytic theories for the SFR from Section~\ref{sec:theory}, the numerical simulations from \mbox{Sections~\ref{sec:sims}--\ref{sec:model_comparison}}, and limitations of the comparison of both theory and simulations with observations in Section~\ref{sec:obs}.

\subsection{Analytic Theories}

\subsubsection{Non-log-normal Effects in the Density PDF} \label{sec:pdftails}
One limitation of the current analytic theories for $\sfrff$ is the assumption of a perfect log-normal PDF of the gas density, Equation~(\ref{eq:pdf}), in the derivation of the SFR integral, Equation~(\ref{eq:sfrffbasic}), which affects all six analytic theories (Table~\ref{tab:theories}) similarly.
Even though a log-normal PDF is expected for purely isothermal turbulence \citep{Vazquez1994}, intermittency introduces skewness and kurtosis in the distributions \citep{Klessen2000,KritsukEtAl2007,BurkhartEtAl2009}, which becomes stronger for more compressive forcing \citep{SchmidtEtAl2009,FederrathDuvalKlessenSchmidtMacLow2010} and for higher Mach numbers \citep{KonstandinEtAl2012ApJ}. Temperature variations can also introduce deviations from perfect log normals in the wings of the distributions. This occurs, for instance, if the turbulence is modeled with a polytropic equation of state (EOS), $P\propto\rho^\Gamma$ with $\Gamma$ larger or smaller than unity \citep{PassotVazquez1998,LiKlessenMacLow2003,JappsenEtAl2005}. However, when a detailed, fully coupled, chemical, and radiative cooling and heating model is used instead of a polytropic EOS, the PDF of the main molecular gas component, $\mathrm{H}_2$, follows a log-normal distribution very well \citep{GloverMacLow2007a,GloverFederrathMacLowKlessen2010,ShettyEtAl2011,MicicEtAl2012}. The strongest deviations from log-normal PDF arise, when the gas starts to collapse due to self-gravity, producing power-law tails at high densities \citep{Klessen2000,DibBurkert2005,FederrathGloverKlessenSchmidt2008,VazquezSemadeniEtAl2008,ChoKim2011,KritsukNormanWagner2011,BallesterosEtAl2011,CollinsEtAl2012,SafranekShraderEtAl2012}, which has been observed in the column density PDFs of clouds that have already formed stars \citep{KainulainenEtAl2009,SchneiderEtAl2012}.
One might thus argue that star formation might accelerate over time \citep{ChoKim2011,CollinsEtAl2012}. In our numerical experiments, we see that after an initial transient acceleration of $\sfe(t)$ in \mbox{Figures~\ref{fig:accr_res}--\ref{fig:accr_mag}}, the SFR becomes fairly constant in most of the numerical models for $\sfe\gtrsim4\%$. This taken together with the good fit-quality of $\sfrff$ (theory) to $\sfrff$ (simulation) obtained for the multi-freefall KM and PN models in Figure~\ref{fig:model_comparison} suggests that the development of power-law tails in the density PDF during star formation does not significantly affect star formation itself. Using a log-normal PDF in the analytic theories to estimate $\sfrff$ seems to be a reasonably good approximation. From a certain perspective, we could say that the initial conditions for star formation are basically determined by the log-normal part of the PDF. In regions that form stars, the PDF develops a power-law tail at high densities, which is a result (or a byproduct) of star formation, but does not necessarily affect the process of stellar birth itself. We discuss this further in Paper II \citep{FederrathKlessen2012paper2}, where we present the density PDFs of the simulations, showing the development of power-law tails when star formation sets in, consistent with the assumption that the power-law tails observed in molecular clouds correlate with star formation \citep{KainulainenEtAl2009,SchneiderEtAl2012}.

\subsubsection{Anisotropies in Sub-Alfv\'enic Turbulence}
The present analytic theories only work for super-Alfv\'enic turbulence because Equation~(\ref{eq:sigs}) and~(\ref{eq:sigsmacha}) break down for $\macha\lesssim2$ \citep[see the discussion in][]{MolinaEtAl2012}. All theories assume statistical isotropy, which is only fulfilled in the trans- to super-Alfv\'enic regime of turbulence studied here.

\subsubsection{Virial Parameter} \label{sec:alphavir}
The virial parameter in Equation~(\ref{eq:alphavircirc}) only applies to spherical, uniform-density clouds. In the comparison with numerical simulations (columns 10 and 11 in Table~\ref{tab:sims}), it became clear that the virial parameter, $\alphavir\equiv2E_\mathrm{kin}/|E_\mathrm{grav}|$, Equation~(\ref{eq:alphavir}), based on the spatial gas distribution can be more than an order of magnitude different from the virial parameter estimated by Equation~(\ref{eq:alphavircirc}). This is because turbulent interstellar gas is concentrated in fractal-like structures that differ significantly between solenoidal and compressive forcings, and between different sonic Mach numbers (see Figure~\ref{fig:hdimages}), even when the total mass is identical. However, we also tested using $\alphavircirc$ instead of $\alphavir$ in the \mbox{theory--simulation} comparison of Section~\ref{sec:model_comparison}. Doing so yielded similar fits to the ones shown in Figure~\ref{fig:model_comparison} and listed in Table~\ref{tab:fits}, yet with somewhat larger $\chisqred$ in some cases. We thus preferred to use the direct computation of $\alphavir$ in the simulations, Equation~(\ref{eq:alphavir}), which provides a more meaningful description of the dynamical state of the clouds. In the derivation of the analytic models in Section~\ref{sec:theory}, however, we use the simple definition given by Equation~(\ref{eq:alphavircirc}) because it can be treated analytically.

\subsection{MHD Simulations}

\subsubsection{Approximation of $\sfrff$ as Constant Over Time}
In both the theory and MHD simulations, we approximate $\sfrff$ as constant over time. \mbox{Figures~\ref{fig:accr_res}--\ref{fig:accr_mag}} show that this is a reasonable assumption for $\sfe\gtrsim4\%$, but the initial acceleration of $\sfrff$ when $\sfe\lesssim4\%$ is more complicated and is not accounted for in the present theory and simulations. In real molecular clouds, the $\sfrff$ might also change over time, depending on the evolutionary stage of a cloud, or on environmental parameters.

\subsubsection{Limited Numerical Resolution}
Our numerical resolution studies in Figures~\ref{fig:accr_res} and~\ref{fig:model_comparison} show that $\sfrff$ converges with increasing resolution in the numerical simulations. However, some models still differ by a factor of 2--3 from the best analytic predictions. In particular, the very high Mach number simulations with $\mach\sim50$ are not converged at a resolution of $256^3$ and only marginally resolved with $512^3$ cells. However, the $1024^3$-simulation GT1024cM50 with compressive forcing at $\mach\sim50$ seems reasonably well converged as suggested by Figure~\ref{fig:model_comparison} (model \#34). The lower-Mach number simulations typically agree within a factor of 1.5 with the best analytic theories (see Table~\ref{tab:sfrffs}), which is similar to the typical statistical variation induced by different random realizations of the turbulence (see the comparison of three different random seeds in Figure~\ref{fig:accr_res}).

\subsubsection{Periodic Boundary Conditions}
Our numerical simulations are highly idealized in that the boundary conditions are periodic. Real molecular clouds are embedded in the larger-scale interstellar medium and eventually in galaxies, which sets their boundary conditions.
Our choice of boundaries introduces some uncertainties, e.g., in the virial parameter because the gravitational energy $E_\mathrm{grav}$ entering $\alphavir$ depends on the choice of boundary condition. The other extreme would be to initialize a cloud in isolation as done in related studies \citep[e.g.,][]{BateBonnellBromm2003,ClarkEtAl2005,KrumholzKleinMcKee2007,PriceBate2008,PriceBate2009,SmithClarkBonnell2008,FederrathBanerjeeClarkKlessen2010,WalchEtAl2010,GirichidisEtAl2011}. This is similarly artificial because real clouds are not isolated, but exist in a large-scale interstellar web of filaments and other clouds.

Here, we test the analytic theories introduced in Section~\ref{sec:theory} with such simulations of isolated star formation. For instance, \citet{GirichidisEtAl2011} modeled isolated clouds with different density profiles and an initial turbulent perturbation, i.e., impulsive turbulent forcing. Since the clouds with initial power-law or Bonnor-Ebert profiles already assume a stage of previous evolution that may have led to such a density profile, we prefer to compare the more basic, simple initial condition when the density field is initially uniform. \citet{GirichidisEtAl2011} modeled such a uniform density profile with a mixed ($b=0.4$) turbulent perturbation with two different random seeds, in which the sonic Mach number $\mach=3.3$ for their simulation TH-m-1 and $\mach=3.6$ for TH-m-2. The simulations did not include magnetic fields, so $\beta\to\infty$. The virial parameters are in the range \mbox{$\alphavir=1$--$2$} \citep{GirichidisEtAl2012b}, depending on the time interval and spatial range chosen to determine $\alphavir$, which exhibits some temporal and spatial variation. Using the best-fit multi-freefall PN parameters determined from Figure~\ref{fig:model_comparison} and Table~\ref{tab:fits} ($1/\phit=0.47\pm0.16$ and $\theta=1.0\pm0.3$), an average virial parameter $\alphavir=1.5$, an average Mach number of $\mach=3.45$, and $b=0.4$ for mixed turbulence, we find $\sfrff(\textrm{multi-ff PN})=0.56$ by evaluating Equation~(\ref{eq:sfrff_basicsolution}) with $\scrit{_\mathrm{,PN}}$ from Equation~(\ref{eq:scrit_pn}). Taking the uncertainties in the fit parameters $1/\phit$ and $\theta$, as well as the uncertainty in \mbox{$\alphavir=1$--$2$} and \mbox{$\mach=3.3$--$3.6$} into account, we find the analytic multi-ff PN prediction $\sfrff(\textrm{multi-ff PN})=0.56\pm0.35$ for both TH-m-1 and TH-m-2 simulations by \citet{GirichidisEtAl2011}.  A very similar prediction is obtained using the multi-freefall KM model instead of the multi-freefall PN model with the corresponding parameters listed in Table~\ref{tab:fits}. From a linear fit to the evolution of the total accreted mass versus time in the TH-m-1 and TH-m-2 simulations, we find $\sfrff(\textrm{TH-m-1})\approx0.67$ and $\sfrff(\textrm{TH-m-2})\approx0.61$ \citep[][Figure~7]{GirichidisEtAl2011}, in very good agreement with the analytic model prediction, indicating that different boundary conditions do not severely affect our results and conclusions concerning $\sfrff$.

\subsection{Observations}
Assuming a uniform \mbox{$\sfe=1\%$--$10\%$} in the observed Galactic cloud sample by \citet{HeidermanEtAl2010}, we estimated the local core-formation efficiency parameter \mbox{$\eps=0.3$--$0.7$} with the best-fit value $\eps\approx0.5$, by fitting our numerical simulations to the observed distribution in Figure~\ref{fig:sfrcoldens}. There are three major uncertainties in this comparison of the simulations and observations.

First, the $\sfe$ in the observed sample is not known. We reasonably assumed \mbox{$\sfe=1\%$--$10\%$}, but some of the individual clouds may not fall in this range. Moreover, there could be a systematic correlation of $\sfe$ with gas column density $\siggas$, which is not accounted for. For instance, the HCN(1--0) molecular clump data shown in Figure~\ref{fig:sfrcoldens} has potentially smaller $\sfe$ on average than the YSO data because smaller scales tend to exhibit higher $\sfe$ \citep[][]{McKeeOstriker2007}. For instance, it seems plausible that $\sfe$ approaches the local core-efficiency $\eps$, once scales as small as a single core are considered. In contrast, giant molecular cloud complexes as a whole typically only have SFEs of a few percent at most \citep[see Paper II,][]{FederrathKlessen2012paper2}.

The second uncertainty is the effect of the telescope resolution. Lower resolution (or observation of a very distant region, e.g., a whole galaxy) inevitably means that the observed star-forming regions are smoothed over larger areas compared to a high-resolution observation of the same region. The effect of reducing the beam resolution by a factor of four in our synthetic observations of the simulated clouds is demonstrated by comparing panels (c) and (d) in Figure~\ref{fig:sfrcoldens}, resulting in a relatively weak, but noticeable reduction of $\sigsfr$ and $\siggas$.

The third major uncertainty is the star formation timescale $\tsf$ used to convert a given star formation column density $\sigsf$ into a rate $\sigsfr=\sigsf/\tsf$. In Figure~\ref{fig:sfrcoldens}, we adopted a fixed $\tsf=2\,\mathrm{M}\yr$ as often used by observers \citep[e.g.,][]{HeidermanEtAl2010,LadaLombardiAlves2010}. However, we studied two additional choices of $\tsf$ in Figure~\ref{fig:sfrcoldens_tffs}, one where $\tsf=\tff(\meanrho)$ (division by the global freefall time) and the other where $\tsf=\tff(\siggas)$ (division by the local freefall time). Comparing these three choices for $\tsf$, we find that the resulting $\sigsfr$-to-$\siggas$ correlations change slightly, but the overall effect is rather weak. Given the broad distributions in both the simulation data and in the \citet{HeidermanEtAl2010} Galactic cloud sample, it is hard to decide which method provides better agreement. They all seem to agree reasonably well within the observational range of Galactic clouds.

Finally, we note a fundamental difficulty of estimating actual SFRs or $\sfrff$ in observations. Cloud observations are inevitably limited to a nearly instantaneous snapshot of the state of a cloud with respect to the relevant timescales for star formation, which exceed the lifetime of a human being by orders of magnitude. However, measuring a real SFR requires knowledge about the time evolution of the cloud, which is thus not available. Strictly speaking, a direct measurement of the time derivative of star formation, i.e., the SFR is thus impossible in observations. This is why we can only make meaningful comparisons of star formation in simulations and observations based on the methods explained and applied in Section~\ref{sec:obs} (Figures~\ref{fig:sfrcoldens} and~\ref{fig:sfrcoldens_tffs}), but not the actual SFRs computed from the \emph{time evolution} of star formation.

\section{Summary and Conclusions} \label{sec:conclusions}

We investigated the role of turbulence and magnetic fields for the SFR in molecular clouds. We compared theoretical models for the SFR with a comprehensive set of numerical magnetohydrodynamic simulations of core and star formation, and with observations of Galactic clouds. The main conclusion from this study is that the SFR depends on four parameters: (1) the virial parameter, $\alphavir\equiv 2 E_\mathrm{kin}/|E_\mathrm{grav}|$; (2) the sonic Mach number $\mach$; (3) the turbulent forcing parameter $b$ (solenoidal, mixed, compressive); and (4) the strength of magnetic fields, parameterized by plasma $\beta=2\macha^2/\mach^2$ with the Alfv\'en Mach number $\macha$.

Our simulations are in good agreement with SFR column densities and gas column densities of observed molecular clouds. We suggest that variations of the four basic, dimensionless parameters can explain the scatter in the observations. Given that molecular clouds seem to have an $\alphavir$ of order unity, the most important parameters controlling the SFR are the sonic Mach number $\mach$ and the turbulent forcing of a molecular cloud, with magnetic field having a secondary effect. The turbulent forcing can be parameterized by $b$ in Equation~(\ref{eq:sigs}). It is a measure for the fraction of energy excited in the form of compressive modes in a turbulent cloud. We distinguish solenoidal (divergence-free) forcing ($b=1/3$) from compressive (curl-free) forcing ($b=1$), as well as mixtures of both ($1/3<b<1$). We find that the SFR decreases with increasing magnetic pressure, but only by a factor of two. The sonic Mach number can change the SFR by a factor of 4--5, while $b$ can introduce order-of-magnitude differences in the SFR, emphasizing the role of the turbulent forcing for star formation.

\subsection{Analytic Theories for $\sfrff$}
\begin{enumerate}
\item{In Section~\ref{sec:theory}, we derived six analytic models for the SFR per freefall time, $\sfrff$: the original \citet[][KM]{KrumholzMcKee2005}, \citet[][PN]{PadoanNordlund2011}, and \citet[][HC]{HennebelleChabrier2011} models and the multi-freefall KM, PN, and HC models, which are all based on an integral over the density PDF, Equation~(\ref{eq:pdf}), leading to different analytic solutions for $\sfrff$, summarized in Table~\ref{tab:theories}. They all yield a dimensionless SFR per freefall time, $\sfrff$, based on Equation~(\ref{eq:sfrffbasic}), which can be transformed to a real SFR with units of $\msol\,\yr^{-1}$ by applying Equation~(\ref{eq:sfrffdef}).}
\item{We extended the (multi-freefall) KM and (multi-freefall) HC theories to include magnetic fields by introducing a magnetic-pressure correction given by Equation~(\ref{eq:magextension}), which allows us to replace the sound speed by an effective magnetosonic speed given by Equation~(\ref{eq:csmag}) or~(\ref{eq:machmag}) for super-Alfv\'enic, isothermal turbulence.}
\item{We analyzed the basic dependencies of all six theories on the four parameters listed above. $\sfrff$ decreases with increasing virial parameter $\alphavir$, while it increases with increasing sonic Mach number $\mach$ in the best multi-freefall theories (see Figure~\ref{fig:sfrffmodel_mix}). Varying the forcing parameter $b$ from purely solenoidal forcing ($b=1/3$) to purely compressive forcing ($b=1$) leads to a higher $\sfrff$ by more than an order of magnitude (Figure~\ref{fig:sfrff_on_b}). Stronger magnetic fields parameterized by decreasing plasma $\beta$ (or equivalently decreasing Alfv\'en Mach number $\macha$) lead to decreasing $\sfrff$ (Figure~\ref{fig:sfrff_on_beta}).}
\end{enumerate}

\subsection{Numerical Simulations}
\begin{enumerate}
\item{In Sections~\ref{sec:sims} and~\ref{sec:simresults}, we performed a set of numerical experiments of star formation, covering molecular cloud sizes and masses in the range $L=0.3$ to $200\,\pc$ and $M_c=300$ to $4\times10^6\,\msol$ (see Table~\ref{tab:sims}) with solenoidal, mixed, and compressive forcings of the turbulence (see Section~\ref{sec:forcing} for details of the forcing) to test the analytic models. We also ran super-Alfv\'enic simulations with varying magnetic-field strength to test the influence of magnetic fields on the SFR. All simulations include sink particles to model core and star formation, allowing us to measure $\sfrff$, depending on $\alphavir$, $\mach$, $b$, and $\beta$.}
\item{We computed the virial parameter $\alphavir\equiv 2 E_\mathrm{kin}/|E_\mathrm{grav}|$ based on the uniform-density, spherical approximation given by Equation~(\ref{eq:alphavircirc}), and based on the actual, three-dimensional, inhomogeneous gas distribution in the simulations. Depending on the forcing and Mach number of the turbulence, we find that these two definitions can differ by more than an order of magnitude (compare columns 10 and 11 in Table~\ref{tab:sims}), which means that theoretical and observational estimates of $\alphavir$ based on a uniform-density, spherical approximation must be considered with caution.}
\item{The SFR converges with increasing numerical resolution (Figures~\ref{fig:accr_res} and~\ref{fig:model_comparison}). The statistical uncertainty in $\sfrff$ is about a factor of 1.5, indicated by comparing three different random realizations of the same parameter set (Figure~\ref{fig:accr_res}), similar to the uncertainty introduced by limited numerical resolution.}
\item{We found that for our models with $\mach\sim10$, compressive forcing yields SFRs at least an order of magnitude higher than solenoidal forcing, emphasizing the role of different turbulent energy injection mechanisms for the SFR (Figure~\ref{fig:accr_res}). The cloud morphology also depends strongly on the type of forcing and sonic Mach number (see Figure~\ref{fig:hdimages}). The SFR increases by a factor of about four for compressive forcing between $\mach=3$ and $\mach=50$ (Figure~\ref{fig:accr_mach}).}
\item{Including magnetic fields in simulations with $\mach\sim10$ and mixed turbulent forcing, we found that the magnetic field is amplified in regions of core and cluster formation (Figure~\ref{fig:mhdimagesB}), reducing the $\sfrff$ by a factor of two between purely hydrodynamic turbulence ($\macha\to\infty$) and trans-Alfv\'enic turbulence with $\macha\sim1.3$ (see Figure~\ref{fig:accr_mag}). This is a relatively small change in $\sfrff$ for such a fairly strong magnetic field, compared to the dependence of $\sfrff$ on $\alphavir$, $\mach$, and $b$. However, magnetic fields do affect the morphology of the clouds even on large scales, and they reduce fragmentation (see Figure~\ref{fig:mhdimages}), thus potentially having an important impact on the core and stellar IMF.}
\item{A detailed comparison of $\sfrff$ (simulation) with $\sfrff$ (theory) in Figure~\ref{fig:model_comparison} showed that the multi-freefall analytic theories are generally better than the non-multi-freefall theories. The multi-ff KM and multi-ff PN models give the best fits to our simulation data (see Tables~\ref{tab:fits} and~\ref{tab:sfrffs}) with reasonable best-fit model parameters, $1/\phit\approx0.5$ for both multi-ff KM and PN models, as well as $\phix\approx0.17$ for the multi-ff KM model, and $\theta\approx1$ for the multi-ff PN model, suggesting a close connection between the magnetothermal Jeans scale and the magnetosonic scale, as well as turbulence driven on the outer, largest scales of molecular clouds.}
\item{All numerical simulations agree with the multi-ff KM and PN theories within a factor of three, and come closer to the analytic prediction with increasing numerical resolution. This is an encouraging agreement, given that the modeled SFRs vary over two orders of magnitude in our numerical simulations (see Figure~\ref{fig:model_comparison}).}
\end{enumerate}

\subsection{Comparison with Observations}
\begin{enumerate}
\item{We compared our numerical simulations with observations of the SFR column density $\sigsfr$ as a function of the gas column density $\siggas$, measured in Galactic clouds in Section~\ref{sec:obs} (Figure~\ref{fig:sfrcoldens}). We showed that the simulations slightly overestimate the SFR compared to the observed clouds because we did not include any local radiative and mechanical feedback from young stellar objects, and hence, the local efficiency parameter $\eps=1$ in our simulations, by definition. However, assuming a constant, global star formation efficiency in the observed clouds of \mbox{$\sfe\approx1\%$--10\%} \citep[see Paper II,][]{FederrathKlessen2012paper2}, we can adjust our numerical simulation data with Equations~(\ref{eq:epsmod}) to account for $\eps<1$. Doing so, we found the best-fit local efficiency $\eps\approx0.5$ ($\eps=0.7$, 0.5, and $0.3$ for $\sfe=1\%$, 3\%, and $10\%$, respectively) for the observed Galactic clouds, which is in good agreement with theoretical expectations, independent numerical simulations, and observations of individual protostellar cores.}
\item{We studied the effects of telescope beam smoothing in panels (c) and (d) of Figure~\ref{fig:sfrcoldens}, and the effect of varying the definition of the star formation timescale $\tsf$ to determine $\sigsfr$ in Figure~\ref{fig:sfrcoldens_tffs}. We found that both the telescope beam resolution and the definition of $\tsf$ introduce minor uncertainties in our comparison between simulations and observations.}
\item{The correlation between $\siggas$ and $\sigsfr$ in Figure~\ref{fig:sfrcoldens} is consistent with power laws of the form $\sigsfr\propto\siggas^N$ with exponents \mbox{$N=1$--$2$} (albeit with significant scatter), which is similar to extragalactic measurements of \mbox{$\siggas$--$\sigsfr$} correlations.}
\end{enumerate}

The overall agreement between theory, simulations and observations in Figures~\ref{fig:model_comparison} and~\ref{fig:sfrcoldens} is encouraging, considering the simplifications inherent in the theoretical models, the limitations of the numerical simulations, and the uncertainties in the SFEs of the observed sample of clouds (see Section~\ref{sec:limitations}). We conclude that supersonic, magnetized turbulence is a key process, likely controlling the SFR of molecular clouds in the Milky Way and potentially in other galaxies.

\acknowledgements
We thank Amanda Heiderman for sending us the observed SFR column densities measured in Galactic clouds shown in Figures~\ref{fig:sfrcoldens} and~\ref{fig:sfrcoldens_tffs}, and we thank Patrick Hennebelle, Mark Krumholz, and Paolo Padoan for enlightening discussions and detailed comments on the manuscript. 
We also thank Chris McKee for a timely, detailed, and constructive referee report, which significantly improved this study.
Stimulating discussions with Ben Ayliffe, Christian Baczynski, Robi Banerjee, Chris Brunt, Blakesley Burkhart, Michael Burton, Gilles Chabrier, Paul Clark, David Collins, Benoit Commercon, Timea Csengeri, Maria Cunningham, Bruce Elmegreen, Philipp Girichidis, Karl Glazebrook, Simon Glover, Nathan Goldbaum, Alex Hill, Alexandre Lazarian, Lukas Konstandin, Guillaume Laibe, Mordecai-Mark Mac Low, Faviola Molina, Joe Monaghan, Volker Ossenkopf, Daniel Price, Ralph Pudritz, Chalence Safranek-Shrader, Dominik Schleicher, Wolfram Schmidt, Nicola Schneider-Bontemps, Jennifer Schober, Martin Schr\"on, Rahul Shetty, Rowan Smith, Enrique Vazquez-Semadeni, and Mark Wardle, during the preparation of this study are gratefully acknowledged.
C.~F.~thanks for funding provided by the Australian Research Council under the Discovery Projects scheme (grant DP110102191). R.~S.~K.~acknowledges subsidies from the Baden-W\"urttemberg-Stiftung by contract research Internationale Spitzenforschung (grant P-LS-SPII/18). This work was supported by the Deutsche Forschungsgemeinschaft, priority program 1573 (``Physics of the Interstellar Medium'') and collaborative research project SFB 881 (``The Milky Way system'') in sub-projects B1, B2, and B5. Supercomputing time at the Leibniz Rechenzentrum (project pr32lo) and at the Forschungszentrum J\"ulich (project hhd20) are gratefully acknowledged.
The software used in this work was in part developed by the DOE-supported ASC / Alliance Center for Astrophysical Thermonuclear Flashes at the University of Chicago.

%\bibliographystyle{/Users/chfeder/Documents/Latex-sources/apj.bst}
%\bibliography{/Users/chfeder/Documents/Latex-sources/federrath.bib}

\end{document}